\documentclass[a4paper,UKenglish,cleveref, autoref, thm-restate, final]{lipics-v2021}

\hideLIPIcs  

\usepackage{cite}
\usepackage{amsmath,amssymb,amsfonts}
\usepackage{algorithmic}
\usepackage{graphicx}
\usepackage{textcomp}

\usepackage[T1]{fontenc}

\usepackage[dvipsnames]{xcolor}

\usepackage{float}

\usepackage{amsfonts}

\usepackage{amsmath}

\usepackage{amssymb}

\usepackage{amsthm}

\usepackage{thmtools}
\usepackage{thm-restate}

\usepackage{hyperref}

\usepackage{cleveref}
\crefname{equation}{}{}

\usepackage{mathtools}

\usepackage{mathpartir}

\usepackage{stmaryrd}

\usepackage{pifont}

\usepackage{tikz}
\usetikzlibrary{automata, positioning}

\usepackage{array}

\usepackage{multirow}

\usepackage{booktabs}

\usepackage{arydshln}

\usepackage{xstring}

\usepackage{enumerate}
\usepackage[shortlabels,inline]{enumitem}

\usepackage{listings}

\usepackage[free-standing-units, binary-units]{siunitx}

\usepackage{layouts}

\usepackage{scalerel}

\usepackage{relsize}

\usepackage{marvosym}

\usepackage{lscape}

\usepackage{csquotes}

\usepackage[obeyFinal, prependcaption]{todonotes}

\definecolor{TabBlue}{RGB}{31,119,180}
\definecolor{TabOrange}{RGB}{255,127,14}
\definecolor{TabGreen}{RGB}{44,160,44}
\definecolor{TabRed}{RGB}{214,39,40}
\definecolor{TabPurple}{RGB}{148,103,189}
\definecolor{TabBrown}{RGB}{140,86,75}
\definecolor{TabPink}{RGB}{227,119,194}
\definecolor{TabGray}{RGB}{127,127,127}
\definecolor{TabOlive}{RGB}{188,189,34}
\definecolor{TabCyan}{RGB}{23,190,207}
\definecolor{TabGainsboro}{RGB}{220,220,220}

\definecolor{C0}{RGB}{0,107,164}
\definecolor{C1}{RGB}{255,128,14}
\definecolor{C2}{RGB}{171,171,171}
\definecolor{C3}{RGB}{89,89,89}
\definecolor{C4}{RGB}{95,158,209}
\definecolor{C5}{RGB}{200,82,0}
\definecolor{C6}{RGB}{137,137,137}
\definecolor{C7}{RGB}{162,200,236}
\definecolor{C8}{RGB}{255,188,121}
\definecolor{C9}{RGB}{207,207,207}

\newlength{\commentwidth}
\crefname{figure}{Fig.}{Figs.}
\crefname{example}{Ex.}{Exs.}
\crefname{section}{Sec.}{Secs.}
\crefname{table}{Tbl.}{Tbls.}
\crefname{appendix}{App.}{Apps.}
\crefname{algorithm}{Alg.}{Algs.}
\crefname{ass}{assumption}{assumptions}
\crefname{req}{requirement}{requirements}
\crefname{ln}{line}{lines}
\crefname{prp}{property}{properties}
\crefname{frm}{formula}{formulae}
\crefname{mon}{monitor}{monitors}

\hypersetup{
  colorlinks,
  linkcolor=TabRed,
  citecolor=TabGreen,
  urlcolor=TabBlue
}

\graphicspath{{./figs/}{./graphs/}}

\makeatletter
\newcommand{\pushright}[1]{\ifmeasuring@#1\else\omit\hfill$\displaystyle#1$\fi\ignorespaces}
\newcommand{\pushleft}[1]{\ifmeasuring@#1\else\omit$\displaystyle#1$\hfill\fi\ignorespaces}
\makeatother

\DeclareSIUnit{\nothing}{\relax}

\makeatletter
\newcommand{\chain}[2][,]{%
	\@tempswafalse
	\@for\next:=#2\do
  {\if@tempswa{#1}\else\@tempswatrue\fi{\next}}%
}
\makeatother

\newcommand*{\ifempty}[3]{\if\relax\detokenize{#1}\relax#2\else#3\fi}

\newcommand{\optsub}[2][]{%
  \if\relax\detokenize{#1}\relax{#2}\else{#2}_{#1}\fi%
}

\newcommand{\opttop}[2][]{%
  \if\relax\detokenize{#1}\relax{#2}\else{\overset{#1}{#2}}\fi%
}

\newcommand*{\inline}{inline}
\newcommand{\inlinetok}[1]{
  \def\inlinep{#1} 
  \ifx\inline\inlinep true \else false\fi
}

\makeatletter
\def\@inl{inline}
\def\@flt{float}
\newcommand{\datodo}[2][inline]{%
  \def\@plc{#1}%
  \ifx\@inl\@plc\else\ifx\@flt\@plc\def\@plc{}\else\GenericError{A}{Error: unrecognized option '#1'. Expected 'inline' or 'float'}{}{}\fi\fi%
  \todo[%
    caption={\textit{Duncan}}, 
    linecolor=red,
    backgroundcolor=red!25,
    bordercolor=red,
    shadow,
    size=\footnotesize,
    \@plc]{#2}
}   
\newcommand{\letodo}[2][inline]{%
  \def\@plc{#1}%
  \ifx\@inl\@plc\else\ifx\@flt\@plc\def\@plc{}\else\GenericError{A}{Error: unrecognized option '#1'. Expected 'inline' or 'float'}{}{}\fi\fi%
  \todo[%
    caption={\textit{Leo}}, 
    linecolor=blue,
    backgroundcolor=blue!25,
    bordercolor=blue,
    shadow,
    size=\footnotesize,
    \@plc]{#2}
}   
\makeatother

\newcommand*{\exqed}{\hfill$\blacksquare$}

\newcommand*{\defeq}{\ensuremath{\triangleq}}
\newcommand*{\disjunion}{\uplus}

\newcommand*{\bnfdef}{\Coloneqq}
\newcommand*{\bnfor}{\mid}

\makeatletter
\renewcommand*{\bigsqcap}{%
  \mathop{%
    \mathpalette\@updown\bigsqcup
  }%
}
\newcommand*{\@updown}[2]{%
  \rotatebox[origin=c]{180}{$\m@th#1#2$}%
}
\makeatother

\newcommand*{\N}{\ensuremath{\mathbb{N}}}

\newcommand*{\trace}[1]{\chain[.]{#1}}

\newcommand*{\map}[1]{\ensuremath{[\chain[,]{#1}]}}

\renewcommand*{\frac}[2]{%
  \raisebox{.5ex}{\scriptsize\ensuremath{#1}}%
  \kern-.1em/\kern-.15em%
  \raisebox{-.25ex}{\scriptsize\ensuremath{#2}}
}

\newcommand*{\dvar}{\textsc{DVar}}
\newcommand*{\vdvar}{\ensuremath{x}}
\newcommand*{\vvdvar}{\ensuremath{y}}
\newcommand*{\vvvdvar}{\ensuremath{z}}

\newcommand*{\dat}{\ensuremath{\mathbb{D}}}
\newcommand*{\vdat}{\ensuremath{d}}

\newcommand*{\vlabel}{\ensuremath{a}}
\newcommand*{\alphabet}{\Sigma}

\newcommand*{\vconst}{\ensuremath{c}}
\newcommand*{\const}{C}

\DeclareMathOperator*{\match}{\mathsf{match}}

\renewcommand*{\exp}{\textsc{Exp}}
\newcommand*{\vexp}{\ensuremath{e}}
\newcommand*{\vvexp}{\ensuremath{f}}
\newcommand*{\bexp}{\textsc{BExp}}
\newcommand*{\vbexp}{\ensuremath{b}}
\newcommand*{\vvbexp}{\ensuremath{c}}
\newcommand*{\binder}[1]{\ensuremath{\boldsymbol{#1}}}

\DeclareMathOperator*{\eval}{\Downarrow}

\DeclareMathOperator*{\cand}{\ensuremath{\land}}

\DeclareMathOperator*{\cnot}{\ensuremath{\lnot}}

\newcommand*{\evalrel}[2]{\ensuremath{{#1}\;\eval\;{#2}}}

\newcommand*{\vact}{\ensuremath{\alpha}}

\newcommand{\prf}[2]{\ensuremath{#1\,.\,#2}} 

\newcommand*{\parsum}{\ensuremath{\oplus}}
\newcommand*{\parprod}{\ensuremath{\otimes}}
\newcommand*{\parany}{\ensuremath{\odot}}

\newcommand*{\trc}{\ensuremath{\act^\omega}}
\newcommand*{\vtrc}{\ensuremath{t}}
\newcommand*{\vvtrc}{\ensuremath{u}}
\newcommand*{\vvvtrc}{\ensuremath{w}}

\newcommand*{\vrvar}{\ensuremath{X}}

\newcommand{\rec}[2]{\ensuremath{\mathsf{rec}\:#1.\,#2}}

\newcommand*{\mon}{\textsc{Mon}}
\newcommand*{\vmon}{\ensuremath{m}}
\newcommand*{\vvmon}{\ensuremath{n}}
\newcommand*{\nvmon}{\ensuremath{n}}
\newcommand*{\vvvmon}{\ensuremath{o}}

\newcommand*{\vvrd}{\ensuremath{v}}
\newcommand*{\yes}{\ensuremath{\textsf{yes}}}
\newcommand*{\no}{\ensuremath{\textsf{no}}}
\newcommand*{\inc}{\ensuremath{\textsf{end}}}
\newcommand*{\vcon}{c}
\newcommand*{\vvcon}{c'}
\newcommand*{\vvvcon}{c''}
\newcommand*{\vvvvcon}{c'''}
\newcommand*{\monConfs}{C}

\newcommand*{\under}[2]{\ensuremath{#1,\,#2}}

\newcommand*{\syn}[1]{\ensuremath{\llparenthesis\,#1\,\rrparenthesis}}

\newcommand*{\cl}[2][]{\ensuremath{\mathrm{cl}_{#1}(#2)}}

\newcommand*{\bform}[1]{\vbexp_{#1}}
\newcommand*{\eqorneq}{\bowtie}

\newcommand*{\mvrd}{\textsc{mVrd}}
\newcommand*{\mact}{\textsc{mAct}}
\newcommand*{\mnact}{\textsc{mBlc}}
\newcommand*{\mfork}{\textsc{mFork}}

\newcommand*{\masyl}{\textsc{mAsync\textsubscript{L}}}
\newcommand*{\masyr}{\textsc{mAsync\textsubscript{R}}}
\newcommand*{\msyn}{\textsc{mSyn}}

\newcommand*{\mrec}{\textsc{mRec}}

\newcommand*{\mvrca}{\textsc{mVrC1}}

\newcommand*{\mvrdb}{\textsc{mVrD2}}
\newcommand*{\vrul}{\textsc{r}}

\newcommand*{\trans}[1][]{\ensuremath{\opttop[#1]{\;\longrightarrow\;}}}

\newcommand*{\hml}{\textsc{HML}}
\newcommand*{\rechml}{\ensuremath{\mu}\hml}
\newcommand*{\shml}{\textsc{s}\hml}
\newcommand*{\chml}{\textsc{c}\hml}

\newcommand*{\minhml}{\textsc{min}\hml}

\newcommand*{\withdata}{\ensuremath{^d}}
\newcommand*{\guarded}[1]{\ensuremath{_{\forall_g #1}}}
\newcommand{\guard}{\mathtt{gd}}

\newcommand*{\hmld}{\ensuremath{\hml\withdata}}

\newcommand*{\rechmld}{\ensuremath{\rechml\withdata}}

\newcommand*{\shmld}{\ensuremath{\shml\withdata}}

\newcommand*{\chmld}{\ensuremath{\chml\withdata}}
\newcommand*{\disjhmld}{\ensuremath{\textsc{disj}\hml\withdata}}
\newcommand*{\conjhmld}{\ensuremath{\textsc{conj}\hml\withdata}}

\newcommand*{\minhmld}{\ensuremath{\minhml\withdata}}
\newcommand*{\minhmldg}[1][]{\ensuremath{\minhml\withdata\guarded{#1}}}

\newcommand*{\ltru}{\ensuremath{\mathsf{true}}}
\newcommand*{\lfls}{\ensuremath{\mathsf{false}}}
\newcommand*{\vfvar}{\ensuremath{\varphi}}
\newcommand*{\vvfvar}{\ensuremath{\psi}}

\newcommand{\uni}[2]{\ensuremath{\forall\:#1.#2}}
\newcommand{\exi}[2]{\ensuremath{\exists\:#1.#2}}

\newcommand*{\tru}{\ensuremath{\mathsf{tt}}}
\newcommand*{\fls}{\ensuremath{\mathsf{ff}}}

\newcommand*{\nec}[2]{\ensuremath{\mathsf{[}\,#1\,\mathsf{]}\,#2}}
\newcommand*{\pos}[2]{\ensuremath{\mathsf{\langle}\,#1\,\mathsf{\rangle}\,#2}}

\newcommand*{\vpvar}{\ensuremath{X}}
\newcommand*{\vvpvar}{\ensuremath{Y}}
\newcommand*{\vvvpvar}{\ensuremath{Z}}

\renewcommand*{\max}{\ensuremath{\mathsf{max}}}
\renewcommand*{\min}{\ensuremath{\mathsf{min}}}

\newcommand{\recmin}[2]{\ensuremath{\min\:#1.\,(#2)}}
\newcommand{\recmax}[2]{\ensuremath{\max\:#1.\,(#2)}}
\newcommand{\drecmin}[2]{\ensuremath{\min\:#1.\,\left(#2\right)}}

\newcommand*{\tenv}{\textsc{TEnv}}
\newcommand*{\vtenv}{\ensuremath{\sigma}}

\newcommand*{\vpenv}{\ensuremath{\rho}}
\newcommand*{\denv}{\textsc{DEnv}}
\newcommand*{\vdenv}{\ensuremath{\delta}}

\newcommand*{\sem}[1]{\ensuremath{\llbracket\,#1\,\rrbracket}}

\newcommand*{\fx}[2][]{\ensuremath{\mathtt{fx}_{#1}(#2)}}
\newcommand*{\recform}[2][\vfvar]{\ensuremath{{#1}_{#2}}}
\newcommand*{\dep}[1]{\ensuremath{\stackrel{#1}{\rightarrowtail}}}
\newcommand*{\subform}[1]{\ensuremath{\mathtt{sub}(#1)}}
\newcommand*{\suf}[1]{\ensuremath{\mathtt{suffix}(#1)}}
\newcommand*{\gforall}[3]{\ensuremath{\forall\binder{#3 \leq #1 + #2}}}
\newcommand*{\vguard}{\ensuremath{\gamma}}

\newcommand*{\vars}{\mathsf{vars}}

\newcommand*{\regatm}[1]{A_{#1}}
\newcommand*{\vregatm}{A}
\newcommand*{\lang}{L}
\newcommand*{\locations}{\mathsf{Loc}}
\newcommand*{\exilocs}{\locations_{\exists}}
\newcommand*{\unilocs}{\locations_{\forall}}
\newcommand*{\regcfg}[2]{#1, #2}
\newcommand*{\registers}{R}
\newcommand*{\vreg}{r}
\newcommand*{\iloc}{\ell_0}
\newcommand*{\vloc}{\ell}
\newcommand*{\vvloc}{\vloc'}

\newcommand*{\vdval}{\vdenv}
\newcommand*{\idval}{\vdval_0}
\newcommand*{\flocs}{F}

\newcommand*{\transrel}{\Delta}

\newcommand*{\raguess}{\guess}
\newcommand*{\vrun}{\rho}
\newcommand*{\vvrun}{\vrun'}

\newcommand*{\vdwrd}{w}
\newcommand*{\vvdwrd}{\vdwrd'}
\newcommand*{\vtrs}{t}
\newcommand*{\vtransrel}{\Delta}
\newcommand*{\unravel}{\mathsf{unravel}}
\newcommand*{\vpath}{P}
\newcommand*{\extspath}{\bullet}

\newcommand*{\mrecf}{\mrec \textsc{F}}
\newcommand*{\mrecb}{\mrec \textsc{B}}

\newcommand*{\recmon}[1]{p_{#1}}
\newcommand*{\recedmon}[1]{m_{#1}}
\newcommand*{\submonitors}{\mathsf{sub}}

\newcommand{\ltlNext}[1]{\mathsf{X} #1}
\newcommand{\ltlUntil}[2]{#1 \mathsf{U} #2}

\newcommand*{\partialfunction}{\rightharpoonup}

\newcommand*{\vclass}{\mathcal{L}_1}
\newcommand*{\vvclass}{\mathcal{L}_2}

\newcommand*{\ie}{\emph{i.e.,}}

\newcommand*{\etal}{\textit{et~al.}}

\newcommand*{\wrt}{w.r.t.}

\newcommand*{\resp}{resp.}

\newcommand*{\indata}{\star}
\newcommand*{\guess}[1]{\mathsf{guess}~#1}
\newcommand*{\store}[2]{\mathsf{match}(#1,#2)}

\newcommand*{\mgs}{\textsc{mGs}}
\newcommand*{\data}{\mathbb{D}}

\newcommand*{\DVar}{\textsc{DVar}}
\newcommand*{\tvar}{\textsc{FVar}}
\renewcommand*{\trc}{\textsc{Trc}}
\renewcommand*{\vtenv}{\rho}

\newcommand*{\acc}{\textbf{acc}}
\newcommand*{\rej}{\textbf{rej}}
\newcommand*{\idenv}{\vdenv_0}

\newcommand*{\vvdenv}{\vdenv'}

\newcommand*{\vvvdenv}{\vdenv''}

\newcommand*{\vvvvmon}{r}
\newcommand*{\vvvfvar}{\chi}
\newcommand*{\vvvvfvar}{\omega}
\newcommand*{\vvdat}{\vdat'}
\newcommand*{\vftrc}{w}
\newcommand*{\vvftrc}{y}
\newcommand*{\vvvftrc}{z}
\newcommand*{\vfnftrc}{u}
\newcommand*{\vvfnftrc}{v}
\newcommand*{\ftrc}{\textsc{FTrc}}
\newcommand*{\dint}{\textsc{DInt}}

\DeclarePairedDelimiter{\length}{\lvert}{\rvert}

\newcommand*{\vstrc}{T}

\newcommand*{\goodprefixes}{G}
\newcommand*{\badprefixes}{B}
\newcommand*{\vren}{\sigma}

\newcommand*{\vset}{S}
\newcommand*{\vword}{w}
\newcommand*{\vvword}{x}
\newcommand*{\type}{\mathsf{type}}
\newcommand*{\eqrel}[2][]{\sim_{#2}^{#1}}
\newcommand*{\vtype}{\tau}
\newcommand*{\vvtype}{\vtype'}
\newcommand*{\vvvtype}{\vtype''}
\newcommand*{\typeGraph}{\mathcal{G}}

\newcommand*{\headOn}[1]{\stackrel{\downarrow}{#1}}
\newcommand*{\encoding}{\mathsf{enc}}

\newcommand*{\vgvar}{\gamma}

\newcommand*{\vanv}{a}

\newcommand*{\langfsefblocks}{L_{\forall \# \exists \$}}

\newcommand*{\formfsequal}{\vfvar_{\ref{eq:first_second_equal}}}
\newcommand*{\formfrepeats}{\vfvar_{\mathrm{leak}}}
\newcommand*{\monfrepeats}{\vmon_{\mathrm{leak}}}
\newcommand*{\formrepeats}{\vfvar_{\ref{eq:repeats}}}
\newcommand*{\formpairwisedistinct}{\vfvar_{\ref{eq:pairwise_distinct}}}
\newcommand*{\formneverappears}{\vfvar_{\ref{eq:never_appears}}}
\newcommand*{\formfirstrepeatspairwisedistinct}{\vfvar_{\mathrm{dist}}}

\newcommand*{\formguardedfsefblocks}{\vfvar_{\forall \# \exists \$}}
\newcommand{\parc}{\mathsf{parc}}
\newcommand{\psub}{\mathsf{sub}_p}

\DeclareFontFamily{U}{mathb}{\hyphenchar\font45}
\DeclareFontShape{U}{mathb}{m}{n}{
      <5> <6> <7> <8> <9> <10> gen * mathb
      <10.95> mathb10 <12> <14.4> <17.28> <20.74> <24.88> mathb12
      }{}
\DeclareSymbolFont{mathb}{U}{mathb}{m}{n}

\DeclareMathSymbol{\sqsubsetneq}{3}{mathb}{"88}

\usepackage{ifdraft}

\title{Monitorability for the Modal Mu-Calculus over Systems with Data: From Practice to Theory}

\titlerunning{Monitorability of Systems with Data}

\author{Luca Aceto}{Dept. of Computer Science, Reykjavik University, Iceland \and Gran Sasso Science Institute, L'Aquila, Italy \and \url{https://staff.ru.is/luca/}}{luca@ru.is}{https://orcid.org/0000-0002-2197-3018}{}
\author{Antonis Achilleos}{Dept. of Computer Science, Reykjavik University, Iceland \and \url{https://sites.google.com/view/antonisachilleos/}}{antonios@ru.is}{https://orcid.org/0000-0002-1314-333X}{}
\author{Duncan Paul Attard}{University of Malta, Msida, Malta \and \url{https://duncanatt.github.io/}}{duncan.attard.01@um.edu.mt}{https://orcid.org/0000-0002-2448-5394}{}
\author{L{\'e}o Exibard}{LIGM, CNRS, Univ Gustave Eiffel, F77454 Marne-la-Vallée, France \and \url{https://igm.univ-mlv.fr/~exibard/}}{leo.exibard@univ-eiffel.fr}{https://orcid.org/0000-0003-0318-1217}{}
\author{Adrian Francalanza}{University of Malta, Msida, Malta \and \url{https://staff.um.edu.mt/afra1/}}{afra1@um.edu.mt}{https://orcid.org/0000-0003-3829-7391}{}
\author{Anna Ing{\'o}lfsd{\'o}ttir}{Dept. of Computer Science, Reykjavik University, Iceland}{annai@ru.is}{https://orcid.org/0000-0001-8362-3075}{}
\author{Karoliina Lehtinen}{CNRS, Aix-Marseille University, LIS, Marseille, France \and \url{https://lehtinenkaroliina.wordpress.com/}}{lehtinen@lis-lab.fr}{https://orcid.org/0000-0003-1171-8790}{}

\authorrunning{L.~Aceto, A.~Achilleos, D.~P.~Attard, L.~Exibard, A.~Ing{\'o}lfsd{\'o}ttir and K.~Lehtinen}

\Copyright{Luca~Aceto, Antonis~Achilleos, Duncan~Paul~Attard, L\'{e}o~Exibard, Anna~Ing{\'o}lfsd{\'o}ttir and Karoliina~Lehtinen}



\category{} 

\relatedversion{} 

\nolinenumbers 

\EventEditors{John Q. Open and Joan R. Access}
\EventNoEds{2}
\EventLongTitle{42nd Conference on Very Important Topics (CVIT 2016)}
\EventShortTitle{CVIT 2016}
\EventAcronym{CVIT}
\EventYear{2016}
\EventDate{December 24--27, 2016}
\EventLocation{Little Whinging, United Kingdom}
\EventLogo{}
\SeriesVolume{42}
\ArticleNo{23}

\begin{document}

\maketitle


\begin{abstract}
  Runtime verification consists in checking whether a system satisfies a given specification by observing the execution trace it produces. In the regular setting, the modal $\mu$-calculus provides a versatile formalism for expressing specifications of the control flow of the system. This paper focuses on the \emph{data} flow and studies an extension of that logic that allows it to express data-dependent properties, identifying fragments that can be verified at runtime and with what correctness guarantees. The logic studied here is closely related with register automata with \emph{guessing}. That correspondence yields a monitor synthesis algorithm, and a strict hierarchy among the various fragments of the logic, in contrast to the regular setting. We then exhibit a fragment of the logic that can express all monitorable formulae in the logic without greatest fixed-points but not in the full logic, and show this is the best we can get.
\end{abstract}

\ifoptionfinal{}{\bigskip {\LARGE \textcolor{red}{Number of pages: \pageref{references}}}}

\section{Introduction}
Runtime verification is an increasingly important lightweight validation technique that consists in checking a specification by observing an execution trace at runtime~\cite{BartocciFFR18}. Not all system properties can be verified this way, e.g. those that mention behaviours that are not observed in the given trace, or limit behaviours such as ``every request is always eventually granted''. However, it can check properties for which an exhaustive state-space exploration is impractical, and verify systems whose model is unavailable, e.g. closed source code.

In the classical setting, system properties are typically expressed through formalisms whose models are ($\omega$-)words or ($\omega$-)trees, e.g. linear-time temporal logic (LTL), computation tree logic (CTL/CTL$^*$) or the modal $\mu$-calculus (equivalently, Hennessy-Milner logic with recursion), all falling within the realm of ($\omega$-)regular behaviours~\cite{Emerson90}. While this setting enjoys numerous desirable properties (reasonable computational complexity, closure properties, correspondence with automata models, etc.), it falls short of capturing properties of the \emph{data flow} of the system---what information it manipulates and how---since the alphabet of the traces or computation trees is assumed to be finite and typically small, corresponding to a focus on the \emph{control} flow of the system---which signals it emits and when. The data flow has typically higher complexity, due to its unbounded nature, making it seem out of reach. However, due to the ubiquity of data manipulation and the increasing availability of computational power, numerous formal methods have shifted the focus to data, in runtime verification~\cite{DBLP:conf/amast/GrooteM98, Havelund2018}, model-checking~\cite{10.1007/978-3-642-28729-9_26} and reactive synthesis~\cite{DBLP:conf/vmcai/EhlersSK14, DBLP:journals/lmcs/ExibardFR21}.

In the field of runtime verification, tools supporting monitoring of data-dependent properties of systems have been available for some time and have been applied in a variety of settings~\cite{BGHS04, AAC05, BRH07, CR09, BKM10, BFHRR12, MJGCR12, GDPT13, BKV13, DLT16, DBLP:journals/scp/AcetoAAEFI24} (see also the surveys~\cite{Havelund2018, DBLP:journals/sttt/FKRT21}).
However, to our mind, the systematic development of their theoretical underpinnings
has lagged behind their practice.  To quote Milner in~\cite{DBLP:journals/jitech/Milner87} ``the design of computing systems can only properly succeed if it is well grounded in theory''. That quote motivates us to study the theoretical foundations of runtime monitoring for properties of data-dependent systems and to provide a systematic analysis of which properties can be monitored at runtime and with what correctness guarantees, paralleling our analysis in the regular setting~\cite{DBLP:journals/pacmpl/AcetoAFIL19}.

To represent systems with data, we use data words and trees, whose elements are pairs of a letter from a finite alphabet and a value from an infinite domain, structured by a set of predicates to compare data values. 
In~\cite{DBLP:conf/amast/GrooteM98}, the authors introduce a modal $\mu$-calculus based formalism to express data properties that can be monitored at runtime. In~\cite{DBLP:journals/scp/AcetoAAEFI24}, we introduced a variant with only the equality predicate. We provide an in-depth study of this logic by studying its expressiveness and monitorability. This reveals an intricate landscape where, in contrast to the $\omega$-regular setting, most variations on the notion of monitor are \emph{not} equivalent, demonstrating the need for distinct monitor models dependant on the property being monitored for. We also uncover a mistake in~\cite[Theorem~18]{AcetoCFI18}, since what was believed to be a normal form is actually less expressive than the full monitor model. This observation may trigger development in the corresponding tool.
The main contributions of the paper are:
\begin{itemize}
\item A formulation of the Hennessy-Milner logic with recursion over data words (\Cref{sec:rechmld}), inspired from~\cite{DBLP:conf/amast/GrooteM98}, with the equality predicate only (and hence without functions).
\item The delineation of the \hmld{} fragment, capturing all completely monitorable properties expressible over data domains with only equality (\Cref{sec:complete_monitorability} and \Cref{thm:hmld_complete_monitorability}).
\item The delineation of the \chmld{} fragment, which we show is monitorable for satisfactions by a natural extension of the monitor model in~\cite{DBLP:journals/pacmpl/AcetoAFIL19}, along with its compositional synthesis algorithm (\Cref{sec:satisfaction_complete_monitorability} and \Cref{thm:mon_satisfactions_cHMLd}). This monitor model is moreover as expressive as alternating register automata with existential guessing~\cite{DBLP:journals/corr/abs-1202-3957, DBLP:phd/hal/Figueira10} (\Cref{thm:monitors_register_automata}).
\item We establish that, contrary to the $\omega$-regular setting~\cite{DBLP:journals/pacmpl/AcetoAFIL19}, this fragment is not maximal (\Cref{prop:langfsefblocks_not_chmld}), and delineate a fragment (\Cref{sec:candidate_maximal_fragment}) that is maximal among properties without greatest fixed points (\Cref{cor:maximality-of-guarded-in-min}). We show that it is not maximal in general, and that there is no maximal fragment whose membership is decidable (\Cref{cor:undec-monit}).
\end{itemize}

\subparagraph*{Related work}
\label{sec:related_works}
Runtime verification tools often integrate some data capabilities. Indeed, according to Falcone \etal~\cite{DBLP:journals/sttt/FKRT21}, 13 of the 20 tools surveyed have some data in the input specification.
Among tools with data support, we mention AspectJ~\cite{AAC05}, with data included in regular expression matching, the MOP Framework, which integrates runtime verification with data-handling capabilities into the software development cycle~\cite{MJGCR12}. Rule-based monitor Ruler~\cite{BRH07} and the corresponding logic Eagle~\cite{BGHS04} have both been extended with data parameters. The work \cite{DLT16} uses SMT solvers to handle data added to the (potentially infinite state) monitor directly. Trace slicing reduces the problem to checking projections of traces onto a finite set of values~\cite{CR09} while quantified event automata allow for initial quantification over the domain and then spawn copies of the automaton for all possible values~\cite{BFHRR12}.

Another approach is to add data to the logic and monitor fragments thereof. The study in \cite{BKM10} proposes monitors for security policies expressed in metric first order temporal logic. Temporal Object Property Language is a high level logic designed for Java developers, with register automata as a backend formalism~\cite{GDPT13}, bridging the programmer--automata gap.

On the theoretical side, in\cite{BKV13} Bauer et al. study the monitorability of $LTL^{FO}$, LTL with first order quantification over data. The prefix problem is undecidable, so there is no hope of computing complete monitors but the authors establish a hierarchy based on how much of the trace must be stored.
Regarding specifications, the relations between the many logics and automata handling data~\cite{DBLP:journals/corr/abs-1208-5980} remain largely unmapped, and most models are not equivalent.
Among automata models, register automata are well studied~\cite{Bojanczyk19}. Pebble automata~\cite{DBLP:journals/tocl/NevenSV04} are closer to logic, but at the cost of decidability. Class memory automata and data automata coincide~\cite{DBLP:journals/tcs/BjorklundS10}.
Among logics, LTL has been extended to data domains in various ways~\cite{DBLP:journals/corr/abs-1208-5980}. In particular, \emph{freeze} LTL is recognisable by alternating register automata~\cite{DBLP:journals/tocl/DemriL09}, which are also closely related to an extension of the modal $\mu$-calculus~\cite{DBLP:conf/lics/JurdzinskiL07}.
We also mention the Logic of Repeating Values~\cite{DBLP:journals/lmcs/FigueiraMP20} and first-order two-variable logic~\cite{DBLP:journals/tocl/BojanczykDMSS11} which both have promising algorithmic properties.
Beyond the equality predicate, some logics also handle richer domains such as $(\mathbb{Q}, <)$~\cite{DBLP:journals/mscs/FigueiraHL16} and uninterpreted functions~\cite{DBLP:conf/cav/Finkbeiner0PS19}.

\section{The Logic \texorpdfstring{\rechmld}{recHMLd}}
\label{sec:rechmld}

In this section, we define \rechmld, an extension of \rechml{} tailored to express properties of traces of system executions that contain data values. Note that \rechml{} comes in two flavours: \emph{branching time} (the logic describes possible executions of the system) and \emph{linear time} (it describes actual traces). Here, we are concerned with the linear-time setting.

\subsection{Data Words and Traces}
\label{sec:data_words}

In formal methods, data words and trees constitute popular formalisms to model respectively the traces and possible executions of systems~\cite{DBLP:conf/csl/Segoufin06}. Since we consider linear-time properties, we model execution \emph{traces} as data $\omega$-words. They consist in infinite words whose elements are pairs of a letter from a finite alphabet and of a \emph{data value} from an infinite domain. The finite alphabet plays no role here and can be simulated, so we omit it for simplicity~\cite{DBLP:journals/tocl/NevenSV04} (see also \Cref{sec:rechmld_extensions}). A data word is thus an infinite sequence of values from an infinite domain.

For the rest of the paper, we fix a countably infinite \emph{data domain} $\dat$, whose only predicate is `$=$' and is decidable. An \emph{action} is modelled as a data value $\vdat \in \dat$. An infinite (respectively, finite) \emph{trace} is a data word, \ie{} an infinite (resp., finite) sequence $\vtrc \in \dat^\omega$ (resp., $\vftrc \in \dat^n$ for some $n \in \N$); the set of all infinite traces is denoted $\trc = \dat^\omega$ (resp., $\ftrc = \dat^*$ for finite traces). For $\vftrc = \vftrc_0 \dots \vftrc_n \in \ftrc$ and $\vfnftrc = \vfnftrc_0 \vfnftrc_1 \dots  \in \ftrc \cup \trc$,
the \emph{concatenation} of $\vftrc$ and $\vfnftrc$ is $\vftrc \cdot \vfnftrc = \vftrc_0 \dots \vftrc_n \vfnftrc_0 \dots$ (we may omit the $\cdot$).
When $\vfnftrc = \vvftrc \cdot \vvfnftrc$, $\vvftrc$ is a \emph{prefix} of $\vfnftrc$, and $\vvfnftrc$ is a \emph{suffix} of $\vfnftrc$.
The set of suffixes of $\vfnftrc$ is denoted $\suf{\vfnftrc}$.

\subsection{Syntax and Semantics}
\label{sec:syntax_semantics}

\begin{figure}
  \subparagraph*{\rechmld{} Syntax}
  $
  \begin{aligned}[t]
    \vfvar, \vvfvar \in \rechmld{}~
    & \bnfdef{}~
    \tru ~\mid~  \fls ~\mid~
     \pos{\vbexp}{\vfvar} ~\mid~
     \nec{\vbexp}{\vfvar} ~\mid~
     \exi{\binder{\vdvar}}{\vfvar} ~\mid~
     \uni{\binder{\vdvar}}{\vfvar} ~\mid~
     \vfvar \lor \vvfvar ~\mid~
      \vfvar \land \vvfvar \\
    &~\mid~ \recmin{\vpvar}{\vfvar}~\mid~
     \recmax{\vpvar}{\vfvar}~\mid~
     \vpvar
  \end{aligned}
  $
  \medskip
  \subparagraph*{Fragments}
  $
  \begin{aligned}[t]
    \vfvar, \vvfvar \in \chmld
    & \bnfdef
    \tru ~\mid~
     \pos{\vbexp}{\vfvar} ~\mid~
     \exi{\binder{\vdvar}}{\vfvar} ~\mid~
     \vfvar \lor \vvfvar ~\mid~
     \vfvar \land \vvfvar ~\mid~
     \recmin{\vpvar}{\vfvar}~\mid~
      \vpvar \\
    \vfvar, \vvfvar \in \shmld
    & \bnfdef
    \fls ~\mid~
     \nec{\vbexp}{\vfvar} ~\mid~
     \uni{\binder{\vdvar}}{\vfvar} ~\mid~
     \vfvar \lor \vvfvar ~\mid~
     \vfvar \land \vvfvar ~\mid~
     \recmax{\vpvar}{\vfvar}~\mid~
     \vpvar \\
    \vfvar, \vvfvar \in \disjhmld
    & \bnfdef
    \tru ~\mid~
     \pos{\vbexp}{\vfvar} ~\mid~
     \exi{\binder{\vdvar}}{\vfvar} ~\mid~
     \vfvar \lor \vvfvar ~\mid~
     \recmin{\vpvar}{\vfvar}~\mid~
      \vpvar \\
    \vfvar, \vvfvar \in \hmld
    & \bnfdef
    \tru ~\mid~  \fls ~\mid~
     \pos{\vbexp}{\vfvar} ~\mid~
     \nec{\vbexp}{\vfvar} ~\mid~
     \exi{\binder{\vdvar}}{\vfvar} ~\mid~
     \uni{\binder{\vdvar}}{\vfvar} ~\mid~
     \vfvar \lor \vvfvar ~\mid~
     \vfvar \land \vvfvar
  \end{aligned}
$
  \medskip
  \subparagraph*{Semantics}  \hspace{-3em}
  $
  \begin{aligned}[t]
    \sem{\tru, \vpenv, \vdenv} &\defeq \trc
    \qquad
    \sem{\fls, \vpenv, \vdenv} \defeq \varnothing
    \qquad
    \sem{\vpvar, \vpenv, \vdenv} \defeq \big(\vpenv(\vpvar)\big)(\vdenv)
    \\
    \sem{\pos{\vbexp}{\vfvar}, \vpenv, \vdenv} &\defeq \big\{\vtrc \mid (\exists\vvtrc,\vdat. \vtrc = \vdat \vvtrc \text{~and~} \evalrel{\vbexp\vdenv[\star\mapsto\vdat]}{\ltru} \text{~and~} \vvtrc \in \sem{\vfvar, \vpenv, \vdenv})\big\} \\
    \sem{\nec{\vbexp}{\vfvar}, \vpenv, \vdenv} &\defeq \big\{\vtrc \mid (\forall\vvtrc,\vdat. (\vtrc = \vdat\vvtrc \text{~and~} \evalrel{\vbexp\vdenv[\star\mapsto\vdat]}{\ltru}) \text{~implies~} \vvtrc \in \sem{\vfvar, \vpenv, \vdenv})\big\} \\
    \sem{\exi{\binder{\vdvar}}{\vfvar}, \vpenv, \vdenv} &\defeq
    \bigcup_{\vdat \in \dat} \sem{\vfvar,\vpenv,\vdenv[\vdvar\mapsto\vdat]} \qquad
    \sem{\uni{\binder{\vdvar}}{\vfvar}, \vpenv, \vdenv} \defeq
    \bigcap_{\vdat \in \dat} \sem{\vfvar,\vpenv,\vdenv[\vdvar\mapsto\vdat]}\\
    \sem{\vfvar \lor \vvfvar, \vpenv, \vdenv} &\defeq \sem{\vfvar, \vpenv, \vdenv} \cup \sem{\vvfvar, \vpenv, \vdenv} \qquad
    \sem{\vfvar \land \vvfvar, \vpenv, \vdenv} \defeq \sem{\vfvar, \vpenv, \vdenv} \cap \sem{\vvfvar, \vpenv, \vdenv}\\
    \sem{\recmin{\vpvar}{\vfvar}, \vpenv, \vdenv} &\defeq \big(\bigsqcap\big\{F \mid \lambda\vdenv'.\sem{\vfvar, \vpenv[\vpvar \mapsto F], \vdenv'} \sqsubseteq F \big\}\big)(\vdenv)\\
    \sem{\recmax{\vpvar}{\vfvar}, \vpenv, \vdenv} &\defeq \big(\bigsqcup\big\{F \mid F \sqsubseteq \lambda\vdenv'.\sem{\vfvar, \vpenv[\vpvar \mapsto F], \vdenv'} \big\}\big)(\vdenv)
  \end{aligned}
  $
  \medskip

  \subparagraph*{Expressions}
  $
  \begin{aligned}
    \vbexp, \vvbexp \in \bexp &\bnfdef
    \ltru ~\mid~
    \vexp = \vvexp ~\mid~
    \cnot\vbexp ~\mid~
    \vbexp \cand \vvbexp &
    \vexp, \vvexp \in \exp &\bnfdef
    \vdvar \in \dvar ~\mid~
    \star
  \end{aligned}
$
\caption{Syntax and linear-time semantics of \rechmld}\label{fig:rechml_LT}
\end{figure}

We define an extension of \rechml, called \rechmld. Its syntax and semantics are described in \Cref{fig:rechml_LT}.
Formulae are built from a countable set of formula variables, $\vpvar, \vvpvar \in \tvar$, and data variables, $\vdvar, \vvdvar \in \dvar$, ranging over an infinite domain of data values, $\vdat \in \dat$.
In addition to the standard Boolean constructs, \rechmld\ can express recursive properties as least ($\recmin{\vpvar}{\vfvar}$) and greatest ($\recmax{\vpvar}{\vfvar}$) fixed-point formulae that bind the free occurrences of $\vpvar$ in $\vfvar$.
The logic includes the \emph{possibility} ($\pos{\vbexp}{\vfvar}$) and \emph{necessity} ($\nec{\vbexp}{\vfvar}$) modal constructs.
To reason on the data carried by process actions,
modalities are augmented with decidable, quantifier-free Boolean \emph{constraint expressions}, $\vbexp, \vvbexp \in \bexp$,
defined over $\dat$ and $\dvar \cup \{\star\}$, where $\star\notin\dvar$ is a placeholder variable for the current action $\vdat \in \dat$.
The free data variables $\vdvar \in \dvar$ that appear in $\vbexp$ are bound by existential and universal quantification constructs $\exi{\binder{\vdvar}}{\vfvar}$ and $\uni{\binder{\vdvar}}{\vfvar}$.
%

In what follows, the standard notions of open and closed expressions, formula equivalence up to alpha-conversion and variable substitution are used.
%
%
We assume wlog that every occurrence of each fixed-point variable is within the scope of a modal operator in its defining fixed-point formula. This is the case e.g. of $\recmax{\vpvar}{\nec{\vbexp}{\vrvar}}$, but not of $\recmax{\vpvar}{\vrvar \land \nec{\vbexp}{\vrvar}}$.

We define the domains $\denv = \dvar \partialfunction \dat$ of data environments, $\dint = \denv \partialfunction 2^\trc$ of data interpretations, and $\tenv = \tvar \partialfunction \dint$ of trace environments (where $A \partialfunction B$ denotes the set of partial functions from set $A$ to set $B$).
A \emph{data environment}, $\vdenv \in \denv$, is a partial function with a finite domain mapping data variables to values from $\dat$; analogously, a \emph{trace environment}, $\vtenv \in \tenv$, maps formula variables to data interpretations $F, G \in \dint$, that given $\vdenv$, return a set of traces, whose intended meaning is the interpretation of the formula variable in the data environment $\vdenv$.

The \emph{linear-time} semantics of $\rechmld$ is given by the denotational semantic function, $\sem{-}$, defined inductively in \cref{fig:rechml_LT}.
In $\sem{-}$, formulae are interpreted \wrt\ a trace environment $\vtenv$ that gives meaning to formula variables, and a data environment $\vdenv$ that assigns values to data variables in Boolean constraint expressions.
An expression $\vbexp$ defines a set of \emph{external} system actions.
An action $\vdat$ is in this set when the data value it carries satisfies $\vbexp$ with regards to the data environment $\vdenv$, \ie\ $\evalrel{\vbexp\vdenv[\star\mapsto\vdat]}{\ltru}$.
Possibility formulae $\pos{\vbexp}{\vfvar}$ denote all the traces $\vtrc = \vdat \vvtrc$ that begin with an action $\vdat$ that is in the action set described by $\vbexp\vdenv$
\emph{and} whose tail $\vvtrc$ satisfies the continuation formula $\vfvar$.
Dually, necessity formulae $\nec{\vbexp}{\vfvar}$ describe all the traces that, \emph{whenever} they begin with such an action $\vdat$,
continue with a trace that satisfies $\vfvar$.
Note that in the linear-time setting, necessity can be expressed as possibility: $\nec{\vbexp}{\vfvar} \equiv \pos{\neg \vbexp}{\tru} \lor \pos{\vbexp}{\vfvar}$, and dually $\pos{\vbexp}{\vfvar} = \nec{\neg \vbexp}{\fls} \land \nec{\vbexp}{\vfvar}$.
The existential quantifier $\exi{\binder{\vdvar}}{\vfvar}$ is interpreted as the set of traces that satisfy $\vfvar$ by assigning \emph{some} $\vdat \in \dat$ to $\vdvar$; the universal quantifier $\uni{\binder{\vdvar}}{\vfvar}$ is the set of traces satisfying $\vfvar$ under \emph{all} such assignments. Formulae are only interpreted with regards to data environments whose domain includes the set of free data variables occurring in them. Note that existential quantification cannot be expressed using universal quantification, except using negation, which is not allowed in the syntax outside modalities.

Since the logic does not have an explicit negation operator, for all $\vfvar$ the semantic function $\sem{\vfvar, \vtenv, \vdenv}$ is monotonic in $\vtenv$ over the complete lattice $(\dint, \sqsubseteq)$, where the partial order $\sqsubseteq$ corresponds to graph inclusion. Formally, it is defined, for all $F,G \in \dint$, as $F \sqsubseteq G$ whenever $\forall \vdenv \in \denv. F(\vdenv) \subseteq G(\vdenv)$.
%
%
As is standard in the modal $\mu$-calculus, recursion is interpreted through fixed points: by the Knaster-Tarski theorem~\cite{pjm/1103044538}, $\recmin{\vpvar}{\vfvar}$ and $\recmax{\vpvar}{\vfvar}$ respectively correspond to the least and greatest fixed point of the operator that maps a data interpretation $F: \denv \rightarrow 2^{\trc}$ to the data interpretation $\vdenv \mapsto \sem{\vfvar, \vpenv\map{\vrvar \leftarrow F}, \vdenv}$. This is the analogue of the operator used to define the semantics of the modal $\mu$-calculus over traces, lifted to the case of infinite alphabets by parameterising the interpretation by a data environment, in the spirit of~\cite{DBLP:conf/amast/GrooteM98}.
%
To obtain the sought interpretation for $\recmin{\vpvar}{\vfvar}$ and $\recmax{\vpvar}{\vfvar}$, one then applies the least (resp., greatest) fixed point of this operator (which is a function from data environments to sets of traces) to the current data environment $\vdenv$.
%
By construction, they both satisfy the following fixed-point equations:
\begin{proposition}
  \label{prop:fixed_point_equation}
  For all formulae $\vfvar$, all trace environments $\vpvar$, all data environments $\vdenv$,
  $
    \sem{\recmin{\vpvar}{\vfvar}, \vpenv, \vdenv} = \sem{\vfvar\map{\frac{\recmin{\vpvar}{\vfvar}}{\vrvar}}, \vpenv, \vdenv}$ and
    $
    \sem{\recmax{\vpvar}{\vfvar}, \vpenv, \vdenv} = \sem{\vfvar\map{\frac{\recmax{\vpvar}{\vfvar}}{\vrvar}}, \vpenv, \vdenv}
    $.
\end{proposition}

When a formula is closed with regards to recursion variables (respectively, data variables), its interpretation does not depend on the trace environment $\vpenv$ (resp., the data environment $\vdenv$) and we write $\sem{\vfvar, \vdenv}$ (resp., $\sem{\vfvar, \vpenv}$) in lieu of $\sem{\vfvar, \vpenv, \vdenv}$.
For closed formulae, we drop both and write $\sem{\vfvar}$ in lieu of $\sem{\vfvar,\vpenv,\vdenv}$ for clarity.
We say that a trace $\vtrc$ satisfies a closed formula $\vfvar$ if $\vtrc \in \sem{\vfvar}$, and violates $\vfvar$ if $\vtrc \notin \sem{\vfvar}$.
In the following, in all closed formulae $\vfvar$ we assume that each recursion variable $\vpvar$ appears in a unique fixed-point formula $\fx[\vfvar]{\vpvar}$,
which is either of the form $\recmin{\vpvar}{\recform[\vfvar]{\vpvar}}$ or $\recmax{\vpvar}{\recform[\vfvar]{\vpvar}}$.
If $\fx{\vpvar}$ is $\recmin{\vpvar}{\vfvar_\vpvar}$, then $\vpvar$ is called an \emph{lfp variable}; otherwise, $\vpvar$ is called a \emph{gfp variable}.
We write $\vpvar \leq \vvpvar$ when $\vfvar_\vpvar$ is a subformula of $\vfvar_\vvpvar$, $\vpvar < \vvpvar$ when moreover $\vpvar \neq \vvpvar$, and denote by $\subform{\vfvar}$ the set of subformulae of $\vfvar$.

\begin{example}
  \label{ex:first_repeats_pairwise_distinct}
The following formula belongs to the \chmld{} fragment. It states that the first data value eventually repeats, and in between all data values are pairwise distinct:
$
      \formfirstrepeatspairwisedistinct \defeq~ \exi{\binder{\vdvar}}{\pos{\star = \vdvar}{\drecmin{\vpvar}{\pos{\vdvar = \star}{\tru} \lor \left(\exi{\binder{\vvdvar}}{\pos{\star = \vvdvar}{\recmin{\vvpvar}{\pos{\star = \vdvar}{\tru} \lor \pos{\star \neq \vdvar, \vvdvar}{\vvpvar}}}} \land \pos{\star \neq \vdvar}{\vpvar}}\right)}}
$.
Intuitively, a $\mathsf{min}$ construct is satisfied whenever $\tru$ can be reached by unfolding it finitely many times using \Cref{prop:fixed_point_equation}. The first diamond $\pos{.}{}$ implies that $\vdvar$ is bound to the first data value. Then, the first $\mathsf{min}$ is satisfied when $\vdvar$ occurs again, or when the rhs of the first disjunction is satisfied.
This happens when the current data value (bound to $\vvdvar$ thanks to the $\pos{\star = \vvdvar}$ diamond) does not appear before $\vdvar$ is found, as checked by the $\min.\,\vvpvar$, \emph{and} that the overall property is true at next step. Other examples can be found in \Cref{app:formulae_examples}.
\end{example}

\subsection{Satisfiability and Validity}
\label{sec:rechmld_satisfiability_validity}

Over data words, the infinity of the domain implies that compromises have to be made between expressiveness, closure properties and decidability~\cite{DBLP:journals/tcs/BjorklundS10, DBLP:journals/corr/abs-1208-5980}. By adapting the classical encoding~\cite[Section~12]{esslli1994-Dam}, one can observe that \rechmld{} is strictly more expressive than LTL with freeze~\cite{DBLP:journals/iandc/DemriLN07} (see \Cref{prop:ltl_freeze_rechmld} in \Cref{app:proof:prop:ltl_freeze_rechmld}). Thus, in our setting, decidability fails: the satisfiability and validity problems of \rechmld{} are undecidable, in contrast with the finite alphabet case (\rechml{})~\cite{DBLP:conf/popl/Vardi88}.
By adapting the reduction of~\cite[Theorem~18]{DBLP:journals/tocl/NevenSV04}, we can sharpen the undecidability result thus (see
\Cref{app:proof_thm_validity_disjhmld} and \Cref{app:proof_thm_satisfiability_chmld}):
\begin{restatable}{theorem}{validityDisjHMLd}
  \label{thm:validity_disjhmld}
  The validity problem for \disjhmld{} is undecidable.
\end{restatable}
\begin{restatable}{theorem}{satisfiabilitycHMLd}
  \label{thm:satisfiability_chmld}
  The satisfiability problem for \chmld{} is undecidable.
\end{restatable}
The decidability picture for \rechmld{} is quite grim, but fortunately, as we will see in \Cref{sec:complete_sat_complete_fragments}, this does not prevent us from delineating monitorable fragments of that logic.

\subsection{Annotation Semantics}
\label{sec:annotations}

We introduce an alternative semantics for formulae in \rechmld, to better argue about the monitorability of a formula. Annotations are analogous to choice functions~\cite[Section~4]{StreettE89} (see also~\cite[Theorem~2.1]{DAM199477}), and consist in (possibly infinite) witnesses that a formula holds.

\begin{definition}
  \label{def:annotation}
An \emph{annotation} is a graph $(A, \dep{})$, where $A \subseteq \rechmld \times \denv \times \trc$, and:
\begin{itemize}
    \item it is not the case that $(\fls, \vdenv, \vtrc) \in A$  for any $\vtrc \in \trc$ and $\vdenv \in \denv$;
    \item if $(\pos{\vbexp}{\vfvar}, \vdenv, \vdat\vtrc) \in A$, then $\evalrel{\vbexp\vdenv[\star\mapsto\vdat]}{\ltru}$, $(\vfvar, \vdenv, \vtrc) \in A$, and $(\pos{\vbexp}{\vfvar}, \vdenv, \vdat\vtrc) \dep{} (\vfvar, \vdenv, \vtrc)$;
    \item if $(\nec{\vbexp}{\vfvar}, \vdenv, \vdat\vtrc) \in A$, and  $\evalrel{\vbexp\vdenv[\star\mapsto\vdat]}{\ltru}$, then $(\vfvar, \vdenv, \vtrc) \in A$, and $(\nec{\vbexp}{\vfvar}, \vdenv, \vdat\vtrc) \dep{} (\vfvar, \vdenv, \vtrc)$;
    \item if $(\exi{\binder{\vdvar}}{\vfvar}, \vdenv, \vtrc) \in A$, then $(\vfvar, \vdenv[\vdvar \mapsto \vdat], \vtrc) \in A$ and $(\exi{\binder{\vdvar}}{\vfvar}, \vdenv, \vtrc) \dep{} (\vfvar, \vdenv[\vdvar \mapsto \vdat], \vtrc)$ for some $\vdat \in \dat$;
    \item if $(\uni{\binder{\vdvar}}{\vfvar}, \vdenv, \vtrc) \in A$, then $(\vfvar, \vdenv[\vdvar \mapsto \vdat], \vtrc) \in A$ and $(\uni{\binder{\vdvar}}{\vfvar}, \vdenv, \vtrc) \dep{} (\vfvar, \vdenv[\vdvar \mapsto \vdat], \vtrc)$ for all $\vdat \in \dat$;
    \item if $(\vfvar \lor \vvfvar, \vdenv, \vtrc) \in A$, then $(\vfvar, \vdenv, \vtrc) \in A$ and $(\vfvar \lor \vvfvar, \vdenv, \vtrc) \dep{} (\vfvar, \vdenv, \vtrc)$, or $(\vvfvar, \vdenv, \vtrc) \in A$ and $(\vfvar \lor \vvfvar, \vdenv, \vtrc) \dep{} (\vvfvar, \vdenv, \vtrc)$;
    \item if $(\vfvar \land \vvfvar, \vdenv, \vtrc) \in A$, then $(\vfvar, \vdenv, \vtrc) \in A$, $(\vvfvar, \vdenv, \vtrc) \in A$, $(\vfvar \land \vvfvar, \vdenv, \vtrc) \dep{} (\vfvar, \vdenv, \vtrc)$, and $(\vfvar \land \vvfvar, \vdenv, \vtrc) \dep{} (\vvfvar, \vdenv, \vtrc)$;
    \item if $(\recmax{\vpvar}{\vfvar_\vpvar}, \vdenv, \vtrc) \in A$, then $(\vfvar_\vpvar, \vdenv, \vtrc) \in A$ and $(\recmax{\vpvar}{\vfvar_\vpvar}, \vdenv, \vtrc) \dep{} (\vfvar_\vpvar, \vdenv, \vtrc)$;
    \item if $(\recmin{\vpvar}{\vfvar_\vpvar}, \vdenv, \vtrc) \in A$, then $(\vfvar_\vpvar, \vdenv, \vtrc) \in A$  and $(\recmin{\vpvar}{\vfvar_{\vpvar}}, \vdenv, \vtrc) \dep{} (\vfvar_\vpvar, \vdenv, \vtrc)$; and
    \item if $(\vpvar, \vdenv, \vtrc) \in A$, then $(\vfvar_\vpvar, \vdenv, \vtrc) \in A$ and $(\vpvar, \vdenv, \vtrc) \dep{} (\vfvar_\vpvar, \vdenv, \vtrc)$.
\end{itemize}

Given an annotation $(A,\dep{})$ and $\vpvar \in \tvar$, we define the relation $\dep{\vpvar} \subseteq \dep{}$ thus:
$(\vfvar, \vdenv, \vtrc) \dep{\vpvar} (\vvfvar, \vvdenv, \vvtrc)$ if and only if $(\vfvar, \vdenv, \vtrc) \dep{} (\vvfvar, \vvdenv, \vvtrc)$ and $\vvfvar \neq \vvpvar$ for any $\vvpvar \in \tvar$ such that $\vpvar < \vvpvar$.
We say that $(A,\dep{})$ is \emph{lfp-consistent} if there is no lfp variable $\vpvar$ that appears infinitely often on a $\dep{\vpvar}$-path.
For a formula $\vfvar$, a data valuation $\vdenv \in \denv$ and trace $\vtrc$, we say that $(A,\dep{})$ is an annotation for $\vfvar, \vdenv$ on $\vtrc$ (equivalently, $\vfvar, \vdenv$ have annotation $(A,\dep{})$ on $\vtrc$) if
\begin{enumerate*}
    \item $A \subseteq \subform{\vfvar} \times \denv \times \suf{\vtrc}$;
    \item $(\vfvar, \vdenv, \vtrc) \in A$; and
    \item $(A,\dep{})$ is lfp-consistent.
\end{enumerate*}
We say that $(A,\dep{})$ is a \emph{finite annotation} for $\vfvar,\vdenv$ on $\vtrc$ when $A$ is finite and $(A,\dep{})$ is acyclic.
\end{definition}

The following result shows that annotations do yield an alternative semantics for $\rechmld{}$ (see \Cref{app:example_annotation} for an example of an annotation, and \Cref{app:proof:prop:annotation-semantics} for a detailed proof):
\begin{restatable}{proposition}{annotationSemantics}
  \label{prop:annotation-semantics}
    For every closed $\vfvar \in \rechmld$, all $\vdenv \in \denv$ and all $\vtrc \in \trc$:
    $\vfvar, \vdenv$ have an annotation on $\vtrc$ if and only if
    $ \vtrc \in \sem{\vfvar,\vdenv}$.
  \end{restatable}
  \begin{proof}[Proof sketch]
    The left-to-right direction follows from the definition, except for fixed points that need to be inductively unfolded using \Cref{prop:fixed_point_equation}.
    The lfp-consistency of the annotation
    allows us to use well-founded induction on the annotation.
    Conversely, if a trace satisfies a formula, one reconstructs an annotation using the iterative characterisation of fixed points~\cite[Section~3]{DBLP:books/el/07/BradfieldS07}, in the same spirit as~\cite[Lemma~2.12 and Theorem~2.13]{aceto2024complexity}, using transfinite induction to ensure that the constructed annotation is lfp-consistent.
  \end{proof}
The annotations used in the proof of \Cref{prop:annotation-semantics} may be infinite. However, the only rule that induces cycles or infinite unfolding is $\mathsf{max}$, and only $\forall$ requires infinite branching (and indeed, the below proposition fails for them). Thus, closed formulae in \chmld{} that have an annotation on some trace $\vftrc$ always admit a finite one (details are in \Cref{app:proof:prop:annotation-chmld}). Overall,
  \begin{proposition}
    \label{prop:chmld_finite_annotation}
    Let $\vfvar \in \chmld$, $\vdenv \in \denv$ and $\vtrc \in \trc$. Then,
    $\vtrc \in \sem{\vfvar, \vdenv}$ \emph{if and only if} $\vfvar, \vdenv$ have a finite annotation on $\vtrc$.
  \end{proposition}

\begin{corollary}\label{cor:sat-chml-re}
    The satisfiability problem for $\chmld$ is recursively enumerable.
\end{corollary}

\section{Complete and Satisfaction-Complete Fragments}
\label{sec:complete_sat_complete_fragments}

\subsection{Monitorability}
\label{sec:monitorability}
The goal of this paper is to determine which properties can be verified at runtime. Informally, runtime verification is conducted as follows: along its execution, the \emph{system under scrutiny} produces a trace, whose elements carry information about its operations; it can be thought of as a dynamically produced log file. We do not assume that the system terminates, so the trace is infinite, but termination can obviously be modelled e.g. by a termination symbol.

In parallel with the execution of the system, a program, called \emph{monitor}, passively reads each element of the execution trace in an on-line manner. At any time, the monitor can emit a $\yes$ (respectively, a $\no$) verdict, meaning that it considers that the system under scrutiny satisfies (resp., violates) a given specification. We then say that the monitor \emph{accepts} (respectively, \emph{rejects}) the trace. Note, however, that a monitor may never emit a verdict on reading an execution trace, e.g. if it does not have enough information to conclude. In this paper, we focus on \emph{irrevocable} verdicts, meaning that once a verdict is emitted, the monitor cannot change its mind. Thus, for now, we can define a monitor through its acceptance and rejectance predicates (which will later on be defined through an operational model):
\begin{definition}
  \label{def:monitors}
  A monitor is an object $\vmon$ on which two predicates $\acc(\vmon, \vftrc)$ and $\rej(\vmon, \vftrc)$ are defined for all finite traces $\vftrc \in \ftrc$, which satisfy the following properties:
  \begin{description}
  \item[(consistency)] There is no finite trace $\vftrc$ such that both $\acc(\vmon, \vftrc)$ and $\rej(\vmon, \vftrc)$ hold;
  \item[(irrevocability)] For all $\vftrc, \vvftrc \in \ftrc$ such that $\vftrc \preceq \vvftrc$, $\acc(\vmon, \vftrc) \Rightarrow \acc(\vmon, \vvftrc)$ and $\rej(\vmon, \vftrc) \Rightarrow \rej(\vmon, \vvftrc)$.
  \end{description}
  We extend the definitions to infinite traces: for all $\vtrc \in \trc$, $\acc(\vmon, \vtrc)$ iff there exists some $\vftrc \prec \vtrc$ such that $\acc(\vmon, \vftrc)$, and similarly for $\rej(\vmon, \vtrc)$.
  Finally, a (finite or infinite) trace $\vfnftrc \in \ftrc \cup \trc$ is \emph{accepted} (respectively, \emph{rejected}) by $\vmon$ whenever $\acc(\vmon, \vfnftrc)$ (resp., $\rej(\vmon, \vfnftrc)$).
\end{definition}
Note that the irrevocability criterion implies that monitors only recognise suffix-closed languages, in the sense that both the sets of accepted and rejected traces of a monitor are suffix-closed.
We can now relate monitors and properties:
\begin{definition}
  \label{def:monitors_soundness_completeness}
  Let $\vstrc \subseteq \trc$ be a property of traces, and $\vmon$ be a monitor.

  We say that $\vmon$ is \emph{sound} for satisfactions (respectively, for violations) for $\vstrc$ if for all $\vtrc \in \trc$, $\acc(\vmon, \vtrc) \Rightarrow \vtrc \in \vstrc$ (respectively, $\rej(\vmon, \vtrc) \Rightarrow \vtrc \notin \vstrc$). We say that $\vmon$ is \emph{sound} when it is sound for both satisfactions and violations.

  Conversely, we say that $\vmon$ is \emph{complete} for satisfactions (resp., violations) for $\vstrc$ if the converse holds, \ie{} for all $\vtrc \in \trc$, $\vtrc \in \vstrc \Rightarrow \acc(\vmon, \vtrc)$ (resp., $\vtrc \notin \vstrc \Rightarrow \rej(\vmon, \vtrc)$). We then say that $\vstrc$ is \emph{completely monitorable for satisfactions} (resp., \emph{for violations}).
  We say that $\vmon$ is \emph{complete} if it is complete for both, and correspondingly that $\vstrc$ is \emph{completely monitorable}.

  We say that the above are \emph{effective} when $\vmon$ can be computed by a Turing machine (if needed, the definition is spelled out in \Cref{app:effective_monitorability}).

  We extend those definitions to any formula $\vfvar \in \rechmld$ by considering $\vstrc = \sem{\vfvar}$.
\end{definition}
In plain words, a property is completely monitorable if there exists a monitor that detects all its satisfactions and violations. Monitorability is thus defined relative to a monitor model, since it depends on the computational power of the monitoring program. As a first step, we consider monitors with arbitrary power; we do not even assume that they are computable. This very strong definition is to be thought of as an overapproximation. However, as witnessed by \Cref{thm:hmld_complete_monitorability}, a quite weak monitor model suffices, since completely monitorable properties turn out to be very simple. This motivates the study of satisfaction-completeness in \Cref{sec:satisfaction_complete_monitorability,sec:beyond_chmld}, as well as that of optimal monitors (\Cref{def:optimal_monitors} below) in \Cref{sec:optimal_monitors}.

For now, a monitor is to be conceived simply as a machine (possibly with access to arbitrary oracles) $\vmon$ which processes a trace and possibly eventually raises a $\yes$ or a $\no$ verdict. When monitoring for some formula $\vfvar \in \rechmld$, if $\vmon$ emits $\yes$ upon reading a finite trace $\vftrc \in \ftrc$, it means that any continuation $\vftrc \vtrc \in \dat^\omega$ belongs to $\sem{\vfvar}$. Conversely, if it emits a $\no$, it means that $\vftrc\dat^\omega \cap \sem{\vfvar} = \varnothing$. Thus, as long as we are not concerned with the way it executes, a monitor $\vmon$ is fully described by the set of prefixes for which it emits a verdict. Observe that $\vstrc$ is completely monitorable iff there exist two sets $\goodprefixes, \badprefixes \subseteq \ftrc$ such that $\vstrc = \goodprefixes \dat^\omega = \trc \backslash (\badprefixes \dat^\omega)$, \ie{} it is characterised by its good and bad prefixes.
\begin{definition}[{\cite{DBLP:journals/dc/AlpernS87, DBLP:conf/cav/dAmorimR05}}]
  Let $\vstrc \subseteq \trc$ be a
  set of traces.
  We say that $\vftrc \in \ftrc$ is a \emph{good} (respectively, a \emph{bad}) prefix for $\vstrc$ when $\vftrc \dat^\omega \subseteq \vstrc$ (respectively, $\vftrc \dat^\omega \cap \vstrc = \varnothing$).
\end{definition}


\begin{definition}[{\cite[Definition~10]{DBLP:conf/csl/AcetoAFIL21}}]
  \label{def:optimal_monitors}
  Let $\vstrc \subseteq \trc$ be a property of traces and $\mon$ be a set of monitors. A monitor $\vmon \in \mon$ is \emph{optimal for violations} (respectively, for satisfactions) in $\mon$ for $\vstrc$ if for each monitor $\vmon' \in \mon$ that is sound for violations (resp., for satisfactions) for $\vstrc$ and each $\vtrc \in \trc$, if $\rej(\vmon', \vtrc)$ then $\rej(\vmon, \vtrc)$ (resp., if $\acc(\vmon', \vtrc)$ then $\acc(\vmon, \vtrc)$).

If one considers arbitrary monitors (including non-computable ones), we can then say that a monitor $\vmon$ is \emph{violation-optimal} for $\vstrc$ if for all $\vftrc \in \ftrc$, if $\vftrc$ is a bad prefix for $\vstrc$ then $\rej(\vmon, \vftrc)$, and dually for satisfaction-optimality.
\end{definition}

\subsection{The Complete Fragment: \texorpdfstring{\hmld{}}{HMLd}}
\label{sec:complete_monitorability}

In the finite alphabet case, all completely monitorable formulae can be expressed in the fragment \hml{}, which consists in formulae of \rechml{} without recursion~\cite[Theorem~4.8]{DBLP:journals/pacmpl/AcetoAFIL19}.
The proof of that result can be adapted to establish the following theorem, taming the infinity of the domain by quotienting finite traces by bijections over $\dat$ (\hmld{} is defined in \Cref{fig:rechml_LT}).
\begin{theorem}
  \label{thm:hmld_complete_monitorability}
  Let $\vstrc \subseteq \trc$ be a set of traces that is stable under renamings (\ie{} for all bijections $\vren: \dat \rightarrow \dat$, $\vren(\vstrc) = \vstrc$).
  $\vstrc$ is completely monitorable \emph{iff} it can be expressed in \hmld{}.
\end{theorem}
\begin{remark}
  \label{rem:richer_domains}
This result does not hold if we consider the domain $(\mathbb{N}, <)$. Indeed, there, one can define the set $D = \{d_0 d_1 \dots d_n \# w \mid \forall i < j, d_i > d_j\}$, which is completely monitorable, since $n$ is bounded by $d_0$, but cannot be expressed in \hmld{} since $n$ depends on $d_0$.
\end{remark}

\subsection{A Satisfaction-Complete Fragment: \chmld{}}
\label{sec:satisfaction_complete_monitorability}

Having to detect \emph{all} satisfactions and \emph{all} violations prevents us from monitoring for behaviours that can happen after an unbounded number of steps in a system execution, restricting us to the tiny fragment \hmld{}, where properties cannot be recursive. In the following, we relax our notion of completeness and focus on detecting only one kind of verdict, \ie{} single-verdict monitors. We also consider the richer setting of optimal monitors in \Cref{sec:optimal_monitors}, but most results are unfortunately negative.
Without loss of generality, we consider satisfaction-completeness, the ability to detect all satisfactions of a property. In practice, as reflected in the literature, runtime verification is more focussed on detecting violations, which are often more critical. Since this adds one level of negation and hence of technicality, we work with satisfactions and results about violation-completeness are obtained by duality.

\subparagraph*{Monitor Synthesis}

In Figure~\ref{fig:monitors_guessing}, we introduce a model of monitors, along with a synthesis procedure. We now show that it yields sound and satisfaction-complete monitors for formulae in \chmld{} (defined in \Cref{fig:rechml_LT} on page~\pageref{fig:rechml_LT}). Note that this fragment includes conjunctions, and it can express $\fls \equiv \pos{\bot}{\tru}$ (where $\bot$ stands e.g. for $\vdvar \neq \vdvar$) and (linear-time) necessity.

\begin{figure}[t]
{\normalsize
\subparagraph{Syntax}
  $
    \vmon, \vvmon \in \mon \bnfdef
    \yes ~\mid~ \inc ~\mid~
    \prf{(\vbexp)}{\vmon} ~\mid~
    \prf{\guess{\vdvar}}{\vmon} ~\mid~
     \vmon\parsum\vvmon ~\mid~
     \vmon\parprod\vvmon ~\mid~
    \rec{\vrvar}{\vmon} ~\mid~
    \vrvar
$

\medskip

  \subparagraph{Configurations}
  $\displaystyle \vcon \in \monConfs \bnfdef (\vmon,\vdenv)  \bnfor \vcon{} \parany \vcon$,
where $\vmon \in \mon$ is a monitor, $\vdenv \in \denv$ is a data environment and $\parany$ is either $\parsum$ (parallel OR) or $\parprod$ (parallel AND).

\medskip

  \subparagraph{Small-Step Semantics}
  {\small
  \begin{mathpar}
    \inferrule*[left=\mvrd]
      {\vvrd \in \{\yes, \inc\}}
      {\under{\vvrd}{\vdenv} \trans[\vdat] \under{\vvrd}{\vdenv}}\qquad
      \inferrule*[left=\mact,right=$\vdat \in \dat$]
      {\evalrel{\vbexp\vdenv[\star\mapsto\vdat]}{\ltru}}
      {\under{\prf{(\vbexp)}{\vmon}}{\vdenv} \trans[\vdat] \under{\vmon}{\vdenv}}
      \qquad
      \inferrule*[left=\mnact,right=$\vdat \in \dat$]
      {\evalrel{\vbexp\vdenv[\star\mapsto\vdat]}{\lfls}}
      {\under{\prf{(\vbexp)}{\vmon}}{\vdenv} \trans[\vdat] \under{\inc}{\vdenv}}

      \inferrule*[left=\mgs, right=$\vdat\in\dat$]
      {~}
      {\under{\prf{\guess{\vdvar}}{\vmon}}{\vdenv} \trans[\tau] \under{\vmon}{\vdenv[\vdvar\mapsto\vdat]}}\qquad
      \inferrule*[left=\mrec]
      {~}
      {\under{\rec{\vrvar}{\vmon}}{\vdenv} \trans[\tau] \under{\vmon\map{\frac{\rec{\vrvar}{\vmon}}{\vrvar}}}{\vdenv}}


      \inferrule*[left=\mfork]
      {~}
      {\under{\vmon \parany \nvmon}{\vdenv} \trans[\tau] \under{\vmon}{\vdenv} \parany \under{\nvmon}{\vdenv}}
      \qquad
      \inferrule*[left=\msyn]
      {\vcon_1 \trans[\vdat] \vcon_1' \\ \vcon_2 \trans[\vdat] \vcon_2'}
      {\vcon_1 \parany \vcon_2 \trans[\vdat] \vcon_1' \parany \vcon_2'}

      \inferrule*[left=\masyl]
      {\vcon_1 \trans[\tau] \vcon_1'}
      {\vcon_1 \parany \vcon_2 \trans[\tau] \vcon_1' \parany \vcon_2}
      \qquad
      \inferrule*[left=\masyr]
      {\vcon_2 \trans[\tau] \vcon_2'}
      {\vcon_1 \parany \vcon_2 \trans[\tau] \vcon_1 \parany \vcon_2'}

      \inferrule*[left=\mvrca]
      {~}
      {\under{\yes}{\vdenv} \parprod \vcon \trans[\tau] \vcon} \qquad
      \qquad
      \inferrule*[left=\mvrdb]
      {~}
      {\under{\yes}{\vdenv} \parsum \vcon \trans[\tau] \under{\yes}{\vdenv}}

      \end{mathpar}
      }
      \subparagraph{Synthesis}\hspace{-1.5em}
      $
  \begin{aligned}[t]
    \syn{\tru} &= \yes &
    %
    %
    \syn{\exi{\binder{\vdvar}}{\vfvar}} &= \prf{\guess{\vdvar}}{\syn{\vfvar}}
    & \syn{\pos{\vbexp}{\vfvar}} &= \prf{(\vbexp)}{\syn{\vfvar}} \\
    \syn{\vfvar \lor \vvfvar} &= \syn{\vfvar} \parsum \syn{\vvfvar}
    & \syn{\vfvar \land \vvfvar} &= \syn{\vfvar} \parprod \syn{\vvfvar} &
    \syn{\recmin{\vpvar}{\vfvar}} &= \rec{\vpvar}{\syn{\vfvar}}
    & \syn{\vpvar} &= \vrvar
  \end{aligned}
  $
}
  \caption{Syntax, small-step semantics and synthesis of monitors\label{fig:monitors_guessing}}
\end{figure}

To keep track of the value of each data variable, a monitor $\vmon \in \mon$ is equipped with a data environment $\vdenv \in \denv$ forming a pair $(\vmon, \vdenv)$. It begins its execution in the context of an initial data environment $\idenv$, as a single component $(\vmon, \idenv)$. Unless otherwise stated, $\idenv = \varnothing$. Note that for closed monitors, the semantics do not depend on $\vdenv$.
Along its execution, a monitor might fork into parallel components. On forking, each component receives a local copy of the parent monitor's data environment (rule $\mfork$) and then evolves independently. The only way to recombine components is when (at least) one has raised a verdict. The verdict is then aggregated with the other components following the usual rules of propositional logic, where $\yes$ corresponds to $\top$, $\parsum$ to $\lor$ and $\parprod$ to $\land$ (rules $\textsc{mVr}$).
To simulate existential quantification, a monitor can non-deterministically guess the value of a data variable and store it in its data environment (rule $\mgs$). This overwrites a previous valuation if any.

There are two kinds of transitions. Ones of the form $\vcon \xrightarrow{\tau} \vvcon$ are called $\tau$-transitions, and correspond to internal moves of the monitor, that happen without reading any trace elements. Correspondingly, $\tau$ is such that for all (finite or infinite) traces $\vfnftrc \in \ftrc \cup \trc$, $\tau \vfnftrc = \vfnftrc$. Those of the form $\vcon \xrightarrow{\vdat} \vvcon$, for $\vdat \in \dat$, are transitions that \emph{process} an element from the trace.

For two configurations $\vcon, \vvcon$ and a data value $d$, we write $\vcon \xRightarrow{d} \vvcon$ whenever $\vcon \xrightarrow{\tau}^* \vvvcon \xrightarrow{d} \vvvvcon  \xrightarrow{\tau}^* \vvcon$ for some configurations $\vvvcon$ and $\vvvvcon$. For a finite trace $\vftrc = d_0 d_1 \dots d_l$, we then write $\vcon \xRightarrow{\vftrc} \vvcon$ whenever $\vcon \xRightarrow{d_0} \vcon_1 \xRightarrow{d_1} \vcon_2 \dots \vcon_l \xRightarrow{d_l} \vvcon$. By a slight abuse of notation, for all $\vtrc \in \trc$, we define $\acc(\vcon, \vtrc)$ as $\vcon \xRightarrow{\vftrc} \under{\yes}{\vvdenv}$ for some $\vvdenv \in \denv$ and some $\vftrc \prec \vtrc$.

\begin{example}
  \label{ex:token_server}
  Consider a server that issues identifier tokens. Assume that the first token it issues is its own and should not be leaked, \ie{} that the server \emph{does not} satisfy the formula $\formfrepeats \defeq \exi{\binder{\vdvar}}{\pos{\vdvar = \star}{\recmin{\vpvar}{\pos{\vdvar = \star}{\tru} \lor \pos{\vdvar \neq \star}{\vpvar}}}}$.
The procedure of \cref{fig:monitors_guessing} yields
$
    \monfrepeats \defeq \syn{\formfrepeats}=
    \prf{
      \guess{\vdvar}
    }{
      \prf{
      (\star=\vdvar)
      }{
        \rec{\vpvar}{\left(
          \prf{(\star=\vdvar)}{\yes}
          \parsum
          \prf{(\star\neq\vdvar)}{\vpvar}
      }
      \right)}
      }
    $.
  %

  Consider an erroneous execution `\trace{1,0,1,\ldots}' exhibited by the server. $\monfrepeats$ starts in configuration
  $
    \under{
    \prf{
      \guess{\vdvar}
    }{
      \prf{
      (\star=\vdvar)
      }{
        \rec{\vpvar}{\left(
          \prf{(\star=\vdvar)}{\yes}
          \parsum
          \prf{(\star\neq\vdvar)}{\vpvar}
      }
      \right)}
      }
  }{\varnothing}
$.
Following rule $\mgs$, $\monfrepeats$ \emph{internally} selects a concrete value $\vdat\in\dat$ for $\vdvar$. Note that such a value is selected over a possibly infinite domain, reminiscent of~\cite{DBLP:journals/jacm/AptP86}.
Assume it chooses the value $0$ for $\vdvar$. On the next step, the system emits $1$ and the monitor checks for the guard $(\star=\vvdvar)$, which does not hold. Following rule $\mnact$, it transitions to the inconclusive verdict $\under{\inc}{\vdvar\mapsto 1}$, where it stays forever.
Assume instead that the monitor picks $\vdvar = 1$. Then, we have:
$
  \under{\monfrepeats}{\varnothing} \xrightarrow[\mgs]{\tau}{}
    \under{
      \prf{
      (\star=\vdvar)
      }{
        \rec{\vpvar}{\left(
          \prf{(\star=\vdvar)}{\yes}
          \parsum
          \prf{(\star\neq\vdvar)}{\vpvar}
      }
      \right)}
  }{\vdvar\mapsto 1}
$.

The execution of the monitor continues and it eventually raises a $\yes$ verdict (a comprehensive execution is provided in \Cref{app:ex:token_server}).
Thus, the trace is accepted by the monitor: it recognises that the system repeats its first action, and hence violates its specification. Note the importance of the non-deterministic choice of a value for $\vdvar$ using rule $\mgs$.
  \exqed
\end{example}
It would not be difficult to establish that $\monfrepeats$ is a sound and satisfaction-complete monitor for $\formfrepeats$. This is more generally the case for \chmld{}, and dually for \shmld{}:
\begin{theorem}
  \label{thm:mon_satisfactions_cHMLd}
  \chmld{} (respectively, \shmld{}) is completely monitorable for satisfactions (resp., violations).
\end{theorem}
\begin{proof}[Proof sketch]
Soundness of the synthesis procedure of \Cref{fig:monitors_guessing} is proven similarly to~\cite[Prop.~4.15]{DBLP:journals/pacmpl/AcetoAFIL19}. The proof is written for violations but is easily adapted, and data variables do not interfere: they play the same role in monitors as in formulae (full proof in \Cref{sec:proof:sec:soundness}).

To prove satisfaction-completeness, we use annotation semantics: in essence, monitors compute annotations of \chmld{} formulae, so from an annotation of $\vfvar \in \chmld{}$, one can build an accepting run of $\syn{\vfvar}$. The full proof is in \Cref{sec:proof:sec:partial_completeness}.
\end{proof}

\subparagraph*{Monitors and Register Automata}

We conclude by observing that our model of monitors for \chmld{} is equivalent to a model of register automata. This result echoes the equivalence between alternation-free modal $\mu$-calculus and register tree automata in~\cite[Theorems~3 and~7]{DBLP:conf/lics/JurdzinskiL07}. The proof (in \Cref{sec:app:register_automata_monitors}) uses the same ingredients to adapt the one in~\cite[Section~4.2]{DBLP:journals/jlap/AcetoAFIK20}.

Register automata were introduced in~\cite{DBLP:journals/tcs/KaminskiF94} as \emph{finite-memory automata}. They consist in a finite-state automaton equipped with a finite set of \emph{registers}, that can store values from an infinite domain (here, $\dat$). It is able to compare the value it reads with the content of its registers, and transition accordingly.
They are formally defined in \Cref{def:register_automaton} in \Cref{sec:app:register_automata_monitors}, which is equivalent to that of~\cite[Section~1.3]{Bojanczyk19}, omitting labels which play no role here.
\begin{restatable}{theorem}{monitorsRegisterAutomata}
  \label{thm:monitors_register_automata}
  Let $L \subseteq \dat^*$ be a suffix-closed language.
  There exists an alternating register automaton with existential guessing that recognises $L$ \emph{if and only if} there exists a monitor that accepts exactly the traces in $L$.
\end{restatable}
The correspondence also holds between register automata with no universal (respectively, existential) states and monitors with no $\otimes$ (resp., $\oplus$). Moreover, if one defines $\store{\vreg}{\vbexp} \defeq \prf{\guess{\vreg}}{(\vbexp \land \vreg = \star)}$,
all the above correspondences hold for register automata with no guessing and monitors whose $\guess{\vreg}$ construct is replaced with the $\store{\vreg}{\vbexp}$ one.

Since all those classes of register automata are inequivalent~\cite[Section~1.5]{Bojanczyk19}, we know that all those variants of monitors correspond to different classes of properties. Thus, in \chmld{} and \shmld{}, removing conjunctions or disjunctions reduces expressiveness, and the same holds when replacing existential quantification with a $\match(\vreg,\vbexp)$ construct (as defined in~\cite{AcetoCFI18}). This also shows that deterministic monitors (defined as the counterpart of deterministic register automata) are strictly less expressive than non-deterministic or alternating ones, which invalidates~\cite[Theorem~18]{AcetoCFI18}. Those facts are formally stated in \Cref{app:sec:comparing_expressiveness}.

\subsection{Optimal Monitors}
\label{sec:optimal_monitors}
The main obstacle to complete monitorability is that of behaviours that happen in the limit, which obviously cannot be monitored for at runtime. For instance, no monitor can ever detect that there are only finitely many occurrences of a given data value. In the spirit of~\cite{DBLP:conf/csl/AcetoAFIL21}, we thus consider \emph{optimal monitors}, that are only required to flag all violations or satisfactions that may be detected by some monitor. The proofs of the following results are in \Cref{app:proof:sec:optimal_monitors}.

First, \disjhmld{} (as defined in \Cref{fig:rechml_LT}) is equivalent to non-deterministic register automata. Their emptiness problem is decidable~\cite[Theorem~1]{DBLP:journals/tcs/KaminskiF94}, so we can build optimal monitors:
\begin{restatable}{theorem}{optimalMonitorsDisjhmld}
\label{thm:optimal_monitors_disjhmld}
For all $\vfvar \in \disjhmld$, one can effectively construct a monitor that is satisfaction-complete and violation-optimal for $\vfvar$.
\end{restatable}
Yet, as soon as we add conjunctions to get \chmld{}, this becomes impossible. Indeed, from a violation-optimal monitor one can build a semi-algorithm to decide unsatisfiability of \chmld{}. Since we have a semi-algorithm for satisfiability of \chmld{} (\Cref{cor:sat-chml-re}), this would yield an algorithm to decide satisfiability of \chmld{}, contradicting \Cref{thm:satisfiability_chmld}.
\begin{restatable}{theorem}{optimalMonitorscHMLdUndecidable}
\label{thm:optimal_monitors_chmld_undecidable}
No effective procedure can construct violation-optimal monitors for \chmld{}.
\end{restatable}

\section{Satisfaction-Completeness: Beyond \chmld{}}
\label{sec:beyond_chmld}
We just established that \chmld{} is a fragment of \rechmld{} that can be monitored in a sound and satisfaction-complete way with the synthesis procedure of \Cref{fig:monitors_guessing} (\Cref{thm:mon_satisfactions_cHMLd}), which generalises the finite alphabet case~\cite[Proposition~4.15]{DBLP:journals/pacmpl/AcetoAFIL19}.
Moreover, our model of monitors is expressively equivalent to register automata (\Cref{thm:monitors_register_automata}), which generalises~\cite[Section 4.2]{DBLP:journals/jlap/AcetoAFIK20}, with the major difference that our monitors cannot be made deterministic.

We now show that, in contrast to the finite alphabet case~\cite[Proposition~4.18]{DBLP:journals/pacmpl/AcetoAFIL19}, that fragment is however not maximal: there are properties that admit sound and satisfaction-complete monitors that cannot be expressed in \chmld{}. \Cref{prop:langfsefblocks_not_chmld} presents such a property, which can be expressed in the larger fragment \minhmldg{} that we introduce. The latter is ``almost maximal'': it is maximal \emph{within} \minhmld{}, \ie{} \rechmld{} without greatest fixed-points (\Cref{cor:maximality-of-guarded-in-min}), but not in the full \rechmld{}. This is for a good reason: there cannot exist a maximally monitorable fragment of \rechmld{} that is effective (\Cref{cor:undec-monit}).
\subsection{A Candidate Maximal Fragment\dots}
\label{sec:candidate_maximal_fragment}
In general, universally quantified formulae are not monitorable for satifactions, as they require checking infinitely many instantiations of the quantified variable. Consider, e.g., the formula
$
  \uni{\binder{\vdvar}}{\recmin{\vpvar}{\pos{\indata = x}{\tru} \lor \pos{\indata \neq x}{\vpvar}}}
$
which states that all data values appear in the input. It is satisfiable, since we assumed that the data domain is countable. Yet, it is not monitorable for satisfactions: any finite prefix only contains finitely many data values and can be continued by, e.g., $\#^\omega$, yielding an input which violates the formula.

Nevertheless, some formulae containing universal quantifiers \emph{are} monitorable. Consider the property which states that the input is divided into blocks separated by dollar and sharp symbols, and that all data values that appear in the second block appear in the first block (formalised in Equation~\ref{eq:def_langfsefblocks}). It is monitorable for satisfactions: the monitor reads the first two blocks by waiting to see the $\$$ and then the $\#$; if this never happens it means that the input violates the property. Otherwise, the monitor can check that all data values in the second block appear in the first one by processing them one by one, going back and forth.
\begin{align}
  \label{eq:def_langfsefblocks}
  \langfsefblocks &= \left\{ d_1 \dots d_k \$ e_1 \dots e_l \# \dots \; \middle| \; \forall 1 \leq j \leq l,  \exists 1 \leq i \leq k, d_i = e_j \right\}, \text{ expressed as:} \\
  \formguardedfsefblocks &= \exi{\binder{\vdvar}}{\vgvar(\vdvar)} \land \uni{\binder{\vdvar}}{\left( \vgvar(\vdvar) \lor \vvfvar(\vdvar) \right)}, \text{ where:} \notag \\
  \vgvar(\vdvar) &= \recmin{\vpvar}{\pos{\indata \neq \$}{\vpvar} \lor
  \pos{\indata = \$}{\recmin{\vvpvar}{\pos{\indata = \#}{\tru} \lor \pos{\indata \neq \vdvar}{\vvpvar}}}} \notag \\
  \vvfvar(\vdvar) &= \recmin{\vvvpvar}{\pos{\indata = \vdvar}{\tru} \lor \pos{\indata \neq \$}{\vvvpvar}} \notag
\end{align}
The formula $\vgvar(\vdvar)$ is called the \emph{guard}, and sets a monitorable bound on the maximal position where a candidate $x$ violating the formula $\vvfvar$ can be found. This way, once the bound is found, the monitor knows that the subsequent data values that appear need not be checked.
Here, it expresses that the trace starts with two blocks---ended by \$ and \#, respectively---and that $\vdvar$ does not appear in the second block:
first, look for a $\$$; once it is found, look for a $\#$, and if in the meantime $\vdvar$ is encountered, the formula cannot recurse and therefore rejects.

The formula $\vvfvar(\vdvar)$ is the universally quantified property, and since we are looking for satisfactions, its universal quantification has to be limited to finitely many values to ensure that it has a finite witness, hence the disjunction with the guard. Here, it expresses that $\vdvar$ appears in the first block. Summing up, the conjunct $\exi{\binder{\vdvar}}{\vgvar(\vdvar)}$ ensures that a trace has the form $w_1 \$ w_2\# u$ for some $w_1, w_2 \in \dat^*$ and $u \in \trc$, while the conjunct $\uni{\binder{\vdvar}}{\left( \vgvar(\vdvar) \lor \vvfvar(\vdvar) \right)}$ yields that every $d \in \dat$ occurring in $w_2$ also occurs in $w_1$.

The above property cannot however be expressed in \chmld{}. Indeed, the length of $w_2$ is unbounded, and its elements have to be compared to elements that appear \emph{before} in the input, so they cannot be manipulated only using existential quantifiers, even with fixpoints. This is made formal by going through monitors, thanks to \Cref{prop:correctness_synthesis_chmld}, which can only carry boundedly many data values accross the $\$$ sign. Thus, \chmld{} is not a maximally monitorable fragment (the details are in \Cref{app:beyond_chml}).
\begin{restatable}{proposition}{langfsefblocksNotcHMLd}
  \label{prop:langfsefblocks_not_chmld}
 There does not exist any formula $\vfvar \in \chmld$ such that $\sem{\vfvar} = \langfsefblocks$.
\end{restatable}

We now proceed to characterise the collection of formulae without greatest fixed points that can be monitored in a sound and satisfaction-complete fashion.
Given a formula $\vguard$, a data variable $\vdvar$, and a finite set $F \subset \dvar
$ of data variables,
we use the following notations:
\begin{align*}
  &Fx \defeq F \cup \{x\};
  \quad
  F\bar{x} \defeq F \setminus \{x\};
  \quad
  x \neq F \defeq \bigwedge_{\vvdvar \in F\bar{x}} \pos{x \neq y}{\tru};
  \quad
  x \sim F \defeq \bigvee_{\vvdvar \in F\bar{x}} \pos{x = y}{\tru};
  ~~\text{ and} \\
  &\gforall{\vguard}{F}{\vdvar}.~\vfvar \defeq
  \exi{\binder{\vdvar}}{ (x \neq F \land \vguard)} \land \uni{\binder{\vdvar}}{\left(\left(x \neq F \land \vguard \right) \lor \vfvar \right)  }.
\end{align*}
The formula $x \neq F$ describes that the value of $x$ is different from every value assigned to any element of $F$ (except for $x$ if $x \in F$), and $x \sim F$ conversely describes that the value of $x$ coincides with the value assigned to some other element of $F$.

The quantifier in $\gforall{\vguard}{F}{\vdvar}.~\vfvar$ intuitively bounds the quantification of $\vdvar$, as we only need to verify $\vfvar$ for all data values that
are assigned to variables in $F\bar{\vdvar}$ and the ones for which $\vguard$ is not true. As such, we say that $\vguard$ is a guard or bound for $\vdvar$, or that $\vdvar$ is guarded.
We need to keep track of the free and guarded variables.
Hence, we parameterize the definition of our fragment with respect to two finite sets of data variables.
For all finite $F \subset \dvar
$ and $V \subseteq  F
$,
we define $\minhmldg[V,F]$ as the set of formulae
that are produced from
$\vfvar_{V,F}$ in the following grammar whose grammar variables are parameterized with respect to $V$ and $F$:
\begin{align*}
  \hspace{-1.5em} \vfvar_{V,F}, \vguard_{V,F}
  &\bnfdef~
   \tru ~\mid~ \fls ~\mid~ X ~\mid~ \recmin{\vpvar}{\vfvar_{V,F}}
  ~\mid~ \pos{\vbexp(F) \cand
  \star \neq V
  }{\vfvar_{V,F}}
  ~\mid~ \vfvar_{V,F} \land \vfvar_{V,F} ~\mid~ \vfvar_{V,F} \lor \vfvar_{V,F}
  \\
  &
  ~\mid~ \gforall{\vguard_{Vx,Fx}}{F} \vdvar.\vfvar_{V\bar{x},Fx}
  ~\mid~ \exi{\binder{\vdvar}}{(\vdvar \neq V
  \land \vfvar_{V\bar{x},Fx}) \lor (
  x \sim V
  \land \vfvar_{Vx,Fx})}
\end{align*}
We then define $\minhmldg = \bigcup_{V \subseteq F \subset \dvar}\minhmldg[V,F]$.
If $\vfvar \in \minhmldg$ has no free data variables, then
$\vfvar \in \minhmldg[\varnothing,\varnothing]$.
In the above grammar, the set $F$ keeps track of the \emph{free variables} in $ \vfvar$, and $V$ of the ``\emph{guarded}'' \emph{free variables}.
Here, ``$x$ is guarded'' means that the value of $x$ is not encountered in the trace while we evaluate the formula (but this value is still assigned to some variable).
This is ensured by the ``$\star \neq V$'' conjunct in the diamonds and by guaranteeing that, during the existential quantification of $y$, if the value of $y$ matches that of some $x$ in $V$, then $y$ is added to $V$.
Hence $ \vfvar_{Vx,F}(x) $ can only be true for values of $x$ that do not appear in some finite annotation of $\vfvar$.

The main characteristic of this fragment is that every universal quantification is \emph{guarded} by a bound on the positions where a candidate $x$ violating the formula can be found.
This is achieved by partitioning the potential values of $x$ into those that appear during the (finite) evaluation of the guard (and must be checked against $ \vfvar$) and those that do not (and therefore satisfy the guard).
Thus, when monitoring or evaluating $\gforall{\vguard_{Vx,Fx}}{F} \vdvar.\vfvar_{V\bar{x},Fx} $, we only need to consider a fixed number of cases for the value of $x$ when checking the subformula $\vguard_{Vx,Fx}$, and therefore, $\vguard_{Vx,Fx}$ is, in a sense, easier to monitor for, or evaluate, than $\vfvar_{V\bar{x},Fx} $. Then, the number of cases that we need to consider for the value of $x$ when checking the subformula $\vfvar_{V\bar{x},Fx} $ is finite and depends on how we evaluated $\vguard_{Vx,Fx}$.

In $\formguardedfsefblocks$,
the evaluation of the guard $\gamma$ is complete at the end of the second block. Therefore, to evaluate $\forall x . (\gamma(x) \lor \psi(x)) $, it suffices to check values of $x$ for $ \psi $ that appear during the evaluation of $ \gamma $---specifically in the second block.
More generally, the grammar of $\minhmldg$ induces a recursive strategy to evaluate a formula while only remembering finitely many cases for the values assigned to its variables.
As we see below, this allows us to find a finite witness for the satisfaction of a formula, using a guarded version of annotations, and, subsequently, to monitor for the satisfaction of all formulae in $\minhmldg$.

\subparagraph*{Guarded-branching Annotations}
We can extend the definitions for annotations for $\chmld$ from \cref{sec:annotations} to guarded-branching annotations for $\minhmld$.
For annotation $(A, \dep{})$, we
replace the quantifier conditions with:
\begin{description}
  \item[if
  $a = (\gforall{\vguard}{F} \vdvar.\vfvar, \vdenv, \vtrc
  ) \in A,$]
  there is some finite $D \cup \{ \vdat_* \} \subseteq \dat$ such that $\vdat_* \notin D$, and:
  \begin{enumerate}
      \item $(\vguard, \vdenv[x\mapsto \vdat_*], \vtrc ) \in A$ and $a \dep{} (
        x \neq F \land \vguard, \vdenv[x\mapsto \vdat_*], \vtrc )$;
      \item for every $\vdat \in D
      $,
      $(
        \vfvar, \vdenv[x\mapsto \vdat], \vtrc ) \in A$ and $a \dep{} (
        \vfvar
        , \vdenv[x\mapsto \vdat], \vtrc )$, or
        $(
        \vguard, \vdenv[x\mapsto \vdat], \vtrc ) \in A$ and $a \dep{} (
        \vguard
        , \vdenv[x\mapsto \vdat], \vtrc )$; 
      \item for every $\vdat \in \dat$, having transition $(
        {\vguard},\vdenv[x \mapsto \vdat_*],\vtrc) \dep{}^* (\vvfvar,\vvdenv,\vdat\vvtrc)$ implies $\vdat \in D$;
      and
      \item $\{ \vdenv(x) \mid x \in \dvar\} \cup
      (F \cap \dat)
      \subseteq D$.
  \end{enumerate}
  \item[if $a = (\exi{\binder{\vdvar}}{(\vdvar \neq V
  \land \vfvar_1) \lor (
  x \sim V
  \land \vfvar_2)},\vdenv,\vtrc) $,] then there is some $\vdat \in \dat$, such that
  either
  \begin{itemize}
    \item $\vdat \neq \vdenv(\vvdvar)$
    for every $\vvdvar \in V\bar{\vdvar}$,
    and $(
        \vfvar_1, \vdenv[x\mapsto \vdat], \vtrc ) \in A$ and $a \dep{} (
        \vfvar_1
        , \vdenv[x\mapsto \vdat], \vtrc )$; or
      \item $\vdat = \vdenv(\vvdvar)$
      for some
      $\vvdvar \in V\bar{\vdvar}$,
      and $(
        \vfvar_2, \vdenv[x\mapsto \vdat], \vtrc ) \in A$ and $a \dep{} (
        \vfvar_2
        , \vdenv[x\mapsto \vdat], \vtrc )$.
  \end{itemize}
\end{description}

The condition for the existential quantifier is used to delineate the existential quantifier as it appears in the grammar and the one hidden inside the guarded universal quantifier.

\begin{theorem}\label{thm:guarded-annotation-semantics}
    For every closed $\vfvar \in \minhmldg$, $\vdenv \in \denv$, and $\vtrc \in \trc$,
        $\vtrc \in \sem{\vfvar,\vdenv}$ \emph{if and only if} $(\vfvar, \vdenv,\vtrc)$ has a finite guarded-branching annotation.
\end{theorem}

\begin{proof}[Proof Sketch]
  For guarded variables (the ones in $V$) and for variables whose value does not appear in the annotation, the specific value does not affect the evaluation of the formula, which allows us to show the equivalence of annotations with (finite) guarded-branching ones.
For the universal quantifier, $D$ represents the values that we must explicitly consider and $d_\star$ is a ``dummy'' value that represents all other values.
  See \Cref{app:proof:thm:guarded-annotation-semantics} for the full proof.
\end{proof}

\subparagraph*{The Monitorable Least-fixed-point formulae}

The guarded-branching annotation semantics for $\minhmldg$ yields that the fragment is effectively monitorable for satisfactions, in the sense that satisfactions can be monitored by a Turing machine.
The monitors of \Cref{sec:satisfaction_complete_monitorability} are equivalent to alternating register automata (\Cref{thm:monitors_register_automata}),
which are computable, so:
\begin{theorem}\label{thm:chml-is-effective}
    Every formula in $\chmld$ is effectively monitorable for satisfactions.
\end{theorem}
Now, as a consequence of \Cref{thm:guarded-annotation-semantics}:
\begin{corollary}
  Let $\vfvar \in \minhmldg$. If $\vtrc \in \sem{\vfvar}$, then $\vtrc$ has a good prefix for $\vfvar$.
\end{corollary}
\begin{corollary}\label{cor:min-effective-mon}
  Every $\vfvar \in \minhmldg$ is effectively monitorable for satisfactions.
\end{corollary}

Moreover, the fragment $\minhmldg$ \emph{characterizes} the monitorable properties in $\minhmld$.
There, not all formulae are monitorable, but they are optimally effectively monitorable for satisfactions, in the sense that there exists a satisfaction-optimal monitor for them:


\begin{theorem}\label{cor:maximality-of-guarded-in-min}
  Every formula $\vfvar \in \minhmld$ is optimally effectively monitorable for satisfactions.
  A formula $\vfvar \in \minhmld$ is monitorable if and only if
  $\vfvar \equiv \vvfvar$ (\ie{} $\sem{\vfvar} = \sem{\vvfvar}$)
  for some $\vvfvar \in \minhmldg$.
\end{theorem}

To prove this theorem, we introduce $\guard$, that turns every $\minhmld$ formula into a guarded form in $\minhmldg$.
Let $\vfvar$ be a closed formula without $\max$ operators and let $Vr(\vfvar) \subseteq \dvar$ be the set of data variables that appear in $\vfvar$.
%
  For every subformula $\vvfvar$ of $\vfvar$,  finite $V \subseteq F \subset Vr(\vfvar)$,
  let $\vrvar_{V,F}$ be a new recursion variable associated with $\vrvar$ and $V,F$.
  For each finite
  $\Pi \subseteq (2^{Vr(\vfvar)})^2$,
  we define
  $\guard(\vvfvar,V,F,\Pi)$ by double recursion on $(2^{Vr(\vfvar)})^2 \setminus \Pi$ and $\vvfvar$:




  \begin{itemize}
    \item $\guard(\vvfvar,V,F,\Pi) = \vvfvar$, when $\vvfvar = \tru$ or $\vvfvar = \fls$;
    \item
  $\guard(\vrvar,V,F,\Pi)=\vrvar_{V,F}$,
  when $(V,F)\in \Pi$;
  \item
  $\guard(\vrvar,V,F,\Pi)=\guard(\fx{\vrvar},V,F,\Pi)$, when $(V,F)\notin \Pi$;
  \item
  $\guard(\min \vrvar.\vvfvar,V,F,\Pi) = \min \vrvar_{V,F}.\guard(\vvfvar,V,F,\Pi\cup\{(V,F)\})$;
  \item
  $\guard(\forall \vdvar.\vvfvar,V,F,\Pi) =
  \gforall{\guard(\vvfvar,V x,F x,\Pi)}{F}{\vdvar}.
  \guard(\vvfvar,V\bar{x},Fx,\Pi)$;
  \item
    $\guard(\exists x.\vvfvar, V, F,\Pi) = \exists x.(\vdvar \neq V \land  \guard(\vvfvar,V\bar{\vdvar}, F x,\Pi)) \lor (
    x \sim V
    \land \guard(\vvfvar,V\vdvar, F x,\Pi))$;
  \item
  \(
    \guard(\pos{\vbexp}{\vvfvar},V, F,\Pi) =
  \pos{\vbexp \land \bigwedge_{x \in V}(x \neq *
  )}{\guard(\vvfvar,
    V, F,\Pi
  )}
  \);
  \end{itemize}
  and $\guard(-,V, F,\Pi)$ commutes with
  $\land$ and $\lor$.
  Observe that for all $\vvfvar$ and $V \subseteq Var$, $\guard(\vvfvar, V, F,\Pi) \in \minhmldg[V,F]$.
  We then define $\guard(\vvfvar, V, F) = \guard(\vvfvar, V, F,\varnothing)$ and
  $\guard(\vvfvar) = \guard(\vvfvar, \varnothing, \varnothing)$, where $\vvfvar$ has no free recursion variables, and, respectively, no free data variables.


The idea behind $\guard$ is to leverage the existence of \emph{good prefixes} for a formula to construct a formula in the guarded fragment. To do so, $\guard$ guards the universal quantification in $ \forall x . \psi(x) $ by a ``\emph{more monitorable}'' formula $ \gamma(x) $ that is constructed from $ \psi(x) $ by guarding $x$. Intuitively, a \emph{good prefix} $p$ for $ \gamma(x) $ (which exists if the trace is a satisfying one and the formula/guard is monitorable) provides a bound on the part of the trace to consider when looking for candidate values violating $ \psi(x) $. Data values outside $p$ are irrelevant: they satisfy $ \gamma(x) $ and do not need to be verified against $ \psi(x) $.

The operation $\guard$ produces formulae with good monitorability properties when applied to formulae in $\minhmld$.
In fact, for each $\varphi \in \minhmld{}$, $\guard(\varphi) \in \minhmldg$ and therefore is monitorable for satisfactions. Furthermore, the sound and complete monitor for $\guard(\varphi)$ is \emph{optimal} for $\varphi$, in that it can detect all good prefixes for $\varphi$; finally, if $\varphi$ is monitorable for satisfactions, then $\varphi$ and $\guard(\varphi)$ are equivalent.
The full arguments are in \Cref{app:beyond_chml}.

\subsection{\dots That Is Not Maximal in General}

In the finite alphabet case, one can turn greatest fixed points into least fixed points while preserving monitorable consequences by a procedure analogous to determinisation of word automata~\cite[Section~5]{DBLP:conf/csl/AcetoAFIL21}. Over data domains, this is not the case anymore~\cite[Section~4]{DBLP:journals/tcs/KaminskiF94}.
In this section, we show that the addition of $\max$ strictly increases the monitorable fragment, and establish that it is undecidable to check if a formula is (effectively) monitorable.

\begin{lemma}\label{lem:2formulae-for-halting}
  For each deterministic Turing machine $M$, we can construct a formula:
  \begin{enumerate}
    \item $\vvfvar_M^e \in \shmld$, such that $\sem{\vvfvar_M^e}$ is the set of traces that encode the run of $M$ on
    0;
    and
    \item $\vvfvar_M^{\neg{H}}$, such that $\sem{\vvfvar_M^{\neg{H}}}$ is the set of traces that encode a
    non-empty prefix of a run
    of $M$, but do not encode a \emph{terminating} run of $M$. 
  \end{enumerate}
\end{lemma}

\begin{corollary}
  $\vvfvar_T^{\neg{H}} \in \rechmld$ is monitorable for satisfactions, but not effectively monitorable for satisfactions for every $T$.
\end{corollary}
\begin{proof}
  The formula
  $\vvfvar_T^{\neg{H}} \in \rechmld$ is monitorable for satisfactions, because
  every satisfying trace $t$ extends a finite trace $p$ that encodes the starting configuration of $T$ on input $x$. Indeed, if $T$ on $x$ terminates, then every satisfying trace
    is not a correct encoding of a run of $T$,
    and therefore has a good prefix. If $T$ on $x$ does not terminate, then every extension of $p$ satisfies the formula, and therefore $p$ is a good prefix.
  Therefore, every satisfying $t$ has a good prefix that extends $p$, yielding that $\vvfvar_T^{\neg{H}} \in \rechmld$ is monitorable for satisfactions.

  If $\vvfvar_T^{\neg{H}}$ were effectively monitorable, then there would exist a Turing machine $M$ that would recognize the good prefixes of $\vvfvar_T^{\neg{H}}$. $M$ would accept $x$ whenever $T$ does not terminate on $x$, yielding that the Halting problem is co-recursively enumerable, which is a contradiction.
\end{proof}

\begin{corollary}\label{cor:undec-monit}
  Monitorability and effective monitorability for satisfactions for $\shmld$ and $\rechmld{}$ are undecidable.
\end{corollary}
\begin{proof}
  Observe that $\vvfvar_T^e$ is (effectively) monitorable for satisfactions if and only if $T$ terminates on 0: if $T$ on 0 terminates, then every satisfying trace has a good prefix where an error in the encoding has occurred, or where the full encoding of a run has appeared. Conversely, if $T$ on 0 does not terminate, then the trace that encodes the run of $T$ on 0 has no good prefix, as every prefix can be extended in a way that does not encode the run.
\end{proof}


The above result yields the impossibility of a decidable, maximal monitorable fragment of $\rechmld$, and similarly for an effectively monitorable fragment.

\label{references}
\bibliography{bibliography}

\newpage

\appendix
\section{Appendix}

\subsection{Formulae Examples}
\label{app:formulae_examples}

\begin{example}
  \label{ex:rechmld_properties}
  To give an intuition of the logic and its expressiveness, here are a few elementary \rechmld{} properties, along with their respective fragments:
  \begin{itemize}
  \item The first and second data values are equal (\hmld{}):
    \begin{equation}
      \label{eq:first_second_equal}
      \formfsequal \defeq \exi{\binder{\vdvar}}{\pos{\vdvar = \star}{\pos{\vdvar = \star}{\tru}}}
    \end{equation}
    Indeed, the only way for the first modality $\pos{\vdvar = \star}$ to be satisfied is if $\vdvar$ takes the value of the first data value. Then, the second modality $\pos{\vdvar = \star}$ is satisfied iff the second value is equal to $\vdvar$, hence to the first value.
  \item The first data value appears again (\disjhmld{}):
    \begin{equation}
      \label{eq:first_repeats}
      \formfrepeats \defeq \exi{\binder{\vdvar}}{\pos{\vdvar = \star}{\recmin{\vpvar}{\pos{\vdvar = \star}{\tru} \lor \pos{\vdvar \neq \star}{\vpvar}}}}
    \end{equation}
    where we use $\vdvar \neq \star$ to abbreviate $\neg (\vdvar = \star)$. As above, $\vdvar$ stores the first data value. Then, we use recursion to look for the second occurrence. Intuitively, on encoutering a fixed-point variable $\vpvar$ the formula recurses, i.e. we can replace $\vpvar$ with the whole $\recmin{\vpvar}{\vfvar}$ that encloses it, as expressed by \Cref{prop:fixed_point_equation}. Here, the formula recurses while it encounters values satisfying $\vdvar \neq \star$, and is satisfied (reaching $\tru$) if it encounters a value satisfying $\vdvar = \star$, viz. the first value in the trace. Since this is a least fixed point ($\min$), the formula is violated if it recurses ad infinitum, i.e. if the first value never appears again.
  \item \emph{Some} data value appears at least twice (\disjhmld{}):
    \begin{align}
      \label{eq:repeats}
      \formrepeats \defeq &\exi{\binder{\vdvar}}{\recmin{\vpvar}{\pos{\vdvar = \star}{\recmin{\vvpvar}{\pos{\vdvar \neq \star}{\vpvar} \lor
       \pos{\vdvar = \star}{\tru} \lor \pos{\vdvar \neq \star}{\vvpvar} }}}}
    \end{align}
    For a given value of $\vdvar$, the formula accepts only if this value is found once (first disjunct of the first $\min$) and then again (first disjunct of the second, nested $\min$). Overall, the formula accepts whenever there exists such a value, which thus appears twice.
  \item All data values are pairwise distinct (negation by dualisation of a \disjhmld{} formula):
    \begin{align}
      \label{eq:pairwise_distinct}
      \formpairwisedistinct \defeq &\uni{\binder{\vdvar}}{\recmax{\vpvar}{\nec{\vdvar = \star}{\recmax{\vvpvar}{\nec{\vdvar \neq \star}{\vpvar} \land
      \nec{\vdvar = \star}{\fls} \land \nec{\vdvar \neq \star}{\vvpvar} }}}}
    \end{align}
    Dually to the above one, this formula \emph{rejects} whenever some value appears twice.
    \item The first data value eventually repeats, and in between all data values are pairwise distinct (\chmld{}):
    \begin{align}
      \label{eq:first_repeats_pairwise_distinct}
      \formfirstrepeatspairwisedistinct \defeq~& \exi{\binder{\vdvar}}{\pos{\star = \vdvar}{\recmin{\vpvar}{\pos{\vdvar = \star}{\tru} \lor \big(\exi{\binder{\vvdvar}}{\pos{\star = \vvdvar}{
                                               \recmin{\vvpvar}{\pos{\star = \vdvar}{\tru} \lor \pos{\star \neq \vdvar \land \star \neq \vvdvar}{\vvpvar}}}}\big) \land \notag \\
      &\pos{\star \neq \vdvar}{\vpvar}}}}
    \end{align}
  \item There exists a data value that never appears (\rechmld{}):
    \begin{equation}
      \label{eq:never_appears}
      \formneverappears \defeq \exi{\binder{\vdvar}}{\recmax{\vpvar}{\nec{\vdvar = \star}\fls \wedge \nec{\vdvar \neq \star}\vpvar}}
    \end{equation}
    As for $\formpairwisedistinct$, the $\max$ allows one to forbid a data value (existentially guessed using the $\exists$ quantifier) from appearing in a trace.
  \end{itemize}
\end{example}

\subsection{Expressiveness Comparisons}
\label{def:expressiveness}
In many parts of the paper, we compare the expressiveness of various fragments of the logic, so we make the notion precise. Given two classes of formulae $\vclass$ and $\vvclass$, we say that $\vclass$ \emph{can be expressed in} $\vvclass$, written $\vclass \sqsubseteq \vvclass$, whenever for every formula in $\vclass$, there is some formula in $\vvclass$ that is equivalent to it. Moreover, $\vclass \equiv \vvclass$ means $\vclass \sqsubseteq \vvclass$ and $\vvclass \sqsubseteq \vclass$. Finally, we say that $\vclass$ is \emph{strictly less expressive than} $\vvclass$, written $\vclass \sqsubsetneq \vvclass$, whenever $\vclass \sqsubseteq \vvclass$ and $\vclass \not \equiv \vvclass$, i.e. there is some formula in $\vvclass$ that cannot be expressed in $\vclass$.

\subsection{Extensions of \rechmld{}}
\label{sec:rechmld_extensions}

\subsubsection{Data Labels}
\label{sec:extension_data_labels}

In the usual definition, each data-word letter consists of a pair $(\vlabel, \vdat) \in \alphabet \times \dat$ of a data value $\vdat$ labelled by some letter from a finite alphabet $\Sigma$. This models the actions of the system, where data is often equipped with additional information. To reduce the technicality level, in this paper we omit those labels, since they play no role in our constructions and results. It should be clear that they can be straightforwadly adapted, but if it is not the case let us just mention that they can also be simulated as follows: to simulate a formula $\vfvar \in \rechmld$ manipulating data values labelled by an alphabet $\Sigma = \{a_1, \dots, a_{n-1}\}$, define a formula $\vfvar' = \exi{\vdvar_{\vlabel_1}}{\dots \exi{\vdvar_{\vlabel_n}}{\pos{\bigwedge_{1 \leq i < j} x_{\vlabel_i} \neq x_{\vlabel_j}}{\tt}}} \land \vfvar''$, where $\vfvar''$ alternately reads a data value representing the label (this corresponds to some $\pos{\star = \vdvar_{\vlabel_i}}$ for some $1 \leq i \leq n$) and an actual data value. Such a formula describes data words of the form $\vdat_{\vlabel_0} \vdat_0 \vdat_{\vlabel_1} \vdat_1 \dots$ (where $\vdat_{\vlabel_i}$ is the data value encoding $\vlabel_i$) which encode corresponding labelled data words $(\vlabel_0, \vdat_0) (\vlabel_1, \vdat_1) \dots$ that satisfy the original formula. This encoding preserves the notions that we consider in this paper. Note that the presence of $\exists$ and $\land$ means that the encoding produces formulae that are outside some fragments of \rechmld{} but it is easy to adapt it for those cases.

\subsubsection{Constants}
\label{sec:extension_constants}

Another common feature of data word formalisms is to handle constants, so that formulae can mention specific data values of the domain, e.g. $\pos{\star = 1}{}$. For the same reasons as above, we omit them: they would increase the technicality level without increasing the depth of our contributions. For completeness, let us point out that they can be simulated (one can also extend our proofs in a straightforward way) in a similar way as finite labels\footnote{They could also be directly simulated directly by finite labels, by encoding the pair $(\vlabel, \vconst)$ by a pair $((\vlabel, \vconst), \vdat)$, where the choice of $\vdat \in \dat$ does not matter, but we prefer not to compose encodings.}: to simulate a set of constants $\vconst = \{\vconst_1, \dots,  \vconst_k\}$, turn $\vfvar$ into $\vfvar' = \exi{\vdvar_{\vconst_1}}{\dots \exi{\vdvar_{\vconst_k}}{\pos{\bigwedge_{1 \leq i < j} x_{\vlabel_i} \neq x_{\vlabel_j}}{\tt}}} \land \vfvar'''$, where in $\vfvar'''$ each $\vdvar = \vconst_i$ is replaced with $\vdvar = \vdvar_{\vconst_i}$. This encoding again preserves the relevant notions.

\subsection{Missing Proofs of \Cref{sec:rechmld}}
\label{app:proofs_rechmld}
\subsubsection{\rechmld{} is Strictly More Expressive than LTL with Freeze}
\label{app:proof:prop:ltl_freeze_rechmld}
By extending the encoding for the finite alphabet case (see, e.g., \cite[Section~12]{esslli1994-Dam} or~\cite{DBLP:journals/tcs/CranenGR11}), we can show that \rechmld{} can express LTL with the freeze quantifier~\cite{DBLP:journals/iandc/DemriLN07}, since quantifiers can simulate the freeze quantifier. The inclusion is moreover strict, for the same reasons as in the finite alphabet case:
\begin{restatable}{proposition}{LTLFreezecHMLd}
  \label{prop:ltl_freeze_rechmld}
  LTL with freeze is strictly less expressive than $\rechmld{}$.
\end{restatable}
\begin{proof}
  The main LTL constructs can be encoded as in the regular setting~\cite[Section 12]{esslli1994-Dam}:
  \begin{itemize}
  \item $\ltlNext{\vfvar} \defeq \pos{\ltru}{\vfvar}$
  \item $\ltlUntil{\vfvar}{\vvfvar} \defeq \recmin{\vpvar}{\vvfvar \lor (\vfvar \land \pos{\ltru}{\vpvar})}$
  \end{itemize}
  Then, the freeze quantifier (storing the current data value in $r$) is encoded as $ \downarrow_r \phi(r) \defeq \exi{r}{\pos{\star = r}{\phi(r)}}$. Finally, its dual (checking that the current data value is equal to the content of $r$) is encoded as $\uparrow_r \phi(r) \defeq \pos{\star = r}{\phi(r)}$.

  Now, over finite alphabets, LTL and LTL with freeze have the same expressiveness, since the different values that a frozen variable may take can be simulated by copies of a formula. Moreover, LTL with freeze and \rechmld{} can both restrict to a finite alphabet using constants. Then, since $\rechmld{}$ can express $\rechml{}$ formulae (up to adding constants) and LTL is strictly less expressive than $\rechml{}$, we get the expected result. A typical separating language is $L = (aa)^*b^\omega$, which can be expressed in $\rechmld{}$ as $\recmin{\vpvar}{\pos{\vdvar = a}{\pos{\vdvar = a}{\vpvar}} \lor \pos{\vdvar = b}{\recmax{\vvpvar}{\pos{\vdvar = b}{\vvpvar}}}}$, but cannot be expressed in LTL~\cite[Chapter~5]{BaiKat:MC:2008} (see also~\cite[Example~1.2 along with Theorem~4.5]{DBLP:conf/lata/Pin20}).

  The reader may grow the feeling that this example is not separating in an interesting way, since it is about the finite alphabet behaviour. However, using similar techniques, one can show that $L' = \left\{w \in \trc \mid \text{there exists $x, y \in \dat$ such that } w \in (xx)^*y^\omega\right\}$ (awaiting for an analogue of Kamp's theorem for data languages).
\end{proof}

\subsubsection{Proof of \Cref{thm:validity_disjhmld}}
\label{app:proof_thm_validity_disjhmld}
\validityDisjHMLd*
\begin{proof}
  We prove the undecidability of the satisfiability problem of \conjhmld{}, the dual fragment of \disjhmld, which is the complement of the validity problem for \disjhmld.
  The proof relies on ideas similar to that of~\cite[Theorem~5.1]{DBLP:journals/tocl/NevenSV04}. However, to stay within \conjhmld{}, we must instead reduce from the non-halting problem and allow the available tape to grow, similarly to what is done in the last paragraph of the proof of~\cite[Theorem~4.50]{ExibardThesis}. Note that the original encoding suffices if one is only concerned with the satisfiability problem for the full logic \rechmld.

  Thus, we reduce from the non-halting problem for deterministic Turing machines, which is undecidable. Let $M$ be a deterministic Turing machine with states $Q = \{p_0, \dots, p_k\}$ and transition function $\delta : Q \times \{0,1\} \rightarrow \{0,1\} \times \{\leftarrow, \rightarrow\} \times Q$, where $\delta(p,c) = (c',m,p')$ means that in state $p$, if the cell under the head contains $c$, then the machine writes $c'$ instead, the control state changes to $p'$ and if $m = \rightarrow$, the head moves one step to the right, otherwise one step to the left. We assume that $M$ halts by entering a distinguished final sink state $p_f \in Q$.
Configurations of $M$ are of the form $(q,t,h)$, where $q \in Q$ is the current state, $t \in \{0,1\}^n$ for some $n \in \N$ is the content of the tape and $h \in \{0, \dots, n-1\}$ is the position of the head. We encode such a configuration as follows: let $d^{p_0}, \dots, d^{p_k}, d^0, d^1, d^{\headOn{0}}, d^{\headOn{1}}, d^{\#}, i_0, i_1, \dots$ be pairwise distinct data values, where each $d^s$ encodes the corresponding symbol $s$ ($\#$ will be used as a separator) and each $i_j$ will be used as a unique identifier for each cell of the tape (note that there are unboundedly many of them, possibly even infinitely many). Then, $(q,t,h)$ is encoded as $\encoding(q,t,h) = d^q \cdot i_0 \cdot w_0 \cdot i_1 \cdot w_1 \cdot \cdots \cdot i_{n-1} w_{n-1}$, where $w_i = d^{\headOn{t_i}}$ if $h = i$, and $w_i = d^{t_i}$ otherwise (recall that $t_i$ is either $0$ or $1$).

  Then, the run of $M$ on input $0$ (recall $M$ is deterministic) $(q_0, 0, 0) (q_1, t_1, h_1) (q_2, t_2, h_2) \dots$ is encoded as follows:
  \begin{equation*}
    d^\# d^{p_0} d^{p_1} \dots d^{p_k} d^0 d^1 d^{\headOn{0}} d^{\headOn{1}}
    d^\# \encoding(q_0,0,0) d^\# \encoding(q_1, t_1, h_1) d^\# \encoding(q_2, t_2, h_2) d^\# \dots
  \end{equation*}
  Thus, it consists of an initial block of pairwise distinct data values that correspond to the $d^s$, followed by the encoding of the successive configurations, separated by $d^\#$. Then, $M$ does not halt on input $0$ if and only there is no occurrence of $p_f$ in the run.\footnote{We run the Turning machine $M$ on $0$ instead of the usual choice of the empty input, as that allows us to maintain a non-empty tape content, and therefore always have a position for the tape head in our encodings.}
  We now describe how the above encoding can be described using an \conjhmld{} formula.

  When $x$ is a data variable, we will use $[x]\psi$ as a shorthand for $[\star = x] \psi$;  $[\_]\psi$ as a shorthand for $[\ltru]\psi$;  $\rhd  x.\psi$ as a shorthand for $\recmax{X}{[\star \neq x]X \land [x]\psi}$; and  $\unrhd  x.\psi$ as a shorthand for $\recmax{X}{[\_]X \land [x]\psi}$.
  That is, $\rhd  x.\psi$ asserts that $\psi$ is true at the next position where the value of $x$ occurs, if there is such an occurence, and
  $\unrhd  x.\psi$ asserts that $\psi$ is true at every position after the current one where the value of $x$ occurs.
  In the following, a \emph{block} is a part of the encoding trace that starts with $d^\#$ and ends at the position right before the next occurence of $d^\#$ (or never ends if $d^\#$ does not occur again, but our formulae ensure this does not happen).

  The length of the first block is fixed, equal to
  $k+5$,
  so
  we can
  design an \hmld{} formula that
  we will use
  as a context to name these data values with
  $k+5$
  data variables and refer to them later.
    Specifically, let $S = \{\#, {p_0}, {p_1}, \ldots ,{p_k}, 0, 1, {\headOn{0}}, {\headOn{1}}
    \}$ and let
    \begin{align*}
    \varphi_{M}
    = &
      \forall  x^\# x^{p_0} x^{p_1} \dots x^{p_k} x^0 x^1 x^{\headOn{0}} x^{\headOn{1}}
      x^\# .
      ~ [ x^\#] [ x^{p_0}] [  x^{p_1}] \dots [ x^{p_k}] [x^0] [ x^1] [ x^{\headOn{0}}] [ x^{\headOn{1}}]
      \psi ,
    \end{align*}
    where $\psi$ is described in the following:
  \begin{itemize}
    \item
    It is now straightforward to
    express that
    these
    $k+5$
    values of the first block are pairwise distinct and the
    second block encodes an initial configuration, $(p_0,t,0)$, where $t \neq \varepsilon$ (we specify $t$ to be $0$ later):
    \[
    \psi_1 =
      \left[\star \neq x^\# \lor \bigvee_{\substack{s_1,s_2 \in S \\ s_1 \neq s_2}} x^{s_1} = x^{s_2} \right] \fls
      \land [\_] [\star \neq x^{p_0}] \fls
      \land [\_] [\_] [\_] [\star \neq x^{\headOn{0}}] \fls
    \]
    and they do not appear on any odd position of an encoding block $\# \encoding(q,t,h)$ ---~except for $\#$:
    \[
      \psi_2 =
        \recmax{X}{[\_][\_]X \land \left[ \bigvee_{s \in S \setminus \{\#
        \}} \star = x^s \right]\fls}
    \]
  \item The difficulty lies in checking the unique identifiers. To do so, one checks the following:
    \begin{itemize}
    \item After the initial block, $d^\#$ is always followed by $d \cdot i_0$, where $d$ does not matter (we later check that it actually encodes some state and that the transition relation is satisfied, but let's not get ahead of ourselves)
    \[
      \psi_3 =
      \forall x^{i_0}.
      [\_][\_][x^{i_0}]
      \unrhd x^\#.
      [\_][\star \neq x^{i_0}]\fls
    \]
    \item Within a block (except for the initial one), all data values encoding identifiers (i.e. those situated at
    even positions from the symbol $d^\#$ that begins the block)
    are pairwise distinct.
    \[
      \psi_4 =
      \forall
      x. [x^\# ][ \_ ] \recmax{X}{[\star \neq x^\#][\_] X \land[\star = x^\#][\_] X \land [x][\_]\recmax{Y}{[\star \neq x^\#][\_]Y \land [x]\fls}}
    \]
    \item After the initial block, any identifier $i_j$ that does not encode the rightmost cell (i.e. that is more than two steps from the ending $d^\#$) is always followed by $d i_{j+1}$, where $d$ again does not matter (we later check that it encodes $0, 1, \headOn{0}$ or $\headOn{1}$ and that the transition relation is satisfied) but $i_{j+1}$ is always the same data value
    \[
      \psi_5 =
      \forall x. \forall y. [x^\# ][ \_ ] \recmax{X}{[\_][\_] X \land [x][\_][\star = y \neq x^\#][\_]\recmax{Y}{[\_][\_]Y \land [x][\_][\star \neq y]\fls}}
    \]
    \item Finally, we need to handle the case where a new identifier is introduced, i.e. when the tape extends: for each data value that does not encode some symbol $s$ nor $i_0$, on its first occurrence it is followed by $dd^\#$, where $d$ does not matter. This ensures that at most one cell is added at a time, and that the identifier is distinct from all the others.
    \[
      \psi_6 =
      \forall x. [x^\# ]
      [ \_ ]
      \recmax{X}{[\star \neq x ][\_] X \land [x][\_][\star \neq x^\#]\fls}
    \]
    We note that a consequence of $\psi_6$ is that after every occurence of $\#$, there is eventually another occurence of $\#$.
    \end{itemize}
  \item Once we have unique identifiers, it is fairly straightforward to check that the transition relation is satisfied:
    \begin{itemize}
    \item Check that in each block, $d^q$ does encode some $q \in Q$, that the $w_i$ indeed encode some cell possibly decorated with the head position, i.e. $0, 1, \headOn{0}$ or $\headOn{1}$:
    \begin{align*}
      \psi_7 =
        \unrhd x^\#.
        \left[\bigwedge_{i=1}^k \star \neq x^{p_k} \right] \fls  \land [\star \neq x^\#] \left[\bigwedge_{s = 0, 1, {\headOn{0}}, {\headOn{1}}} \star \neq x^s \right] \fls
        ,
    \end{align*}
    that the head is in at least one cell (i.e. there is an occurrence of either $\headOn{0}$ or $\headOn{1}$):
    \begin{align*}
      \psi_8 =
          \unrhd x^\#.
          \recmax{Y}{
            [\star \neq x^{\headOn{0}} \land \star \neq x^{\headOn{1}} \land \star \neq x^\#] Y \land [x^\#]\fls
          },
    \end{align*}
    and
    that the head is in at most one cell (i.e. there is at most one occurrence of either $\headOn{0}$ or $\headOn{1}$):
    \begin{align*}
      \psi_9 =
        \recmax{X}{
          [\_]X \land [\mathsf{head}]
          \recmax{Y}{
            [\neg \mathsf{head} \land \star \neq x^\#] Y \land [\mathsf{head}]\fls
            }
          } ,
    \end{align*}
    where $\mathsf{head}$ stands for $\star = x^{\headOn{0}} \lor \star = x^{\headOn{1}}$.
    \item We can think of a transition $\delta(p, c) = (c',m,p')$ as a mapping of state $p$ and a triple of symbols $s_1,s_2,s_3$ to state $q'$ and triple $s'_1,s'_2,s'_3$, where $s_2$ is the symbol at the position of the tape head --- either ${\headOn{0}}$ or ${\headOn{0}}$; $s_1$ is either the symbol ($0$ ir $1$) on the left of the tape head, or $p$, if the head is on the leftmost position; and, similarly, $s_3$ is either the symbol on the right of the tape head, or $p'$, if the head is on the leftmost position.
    Then, $s_2 = c'$ and $s'_1$ and $s'_3$ may change from $s_1$ and $s_3$ respectively, depending on whether the tape head moves to the left or the right.

    For each transition $\delta(p, c) = (c',m,p')$, which we think of as an element $\tau = p,s_1,s_2,s_3 \mapsto p',s'_1,s'_2,s'_3$ of $\delta$, and for each block encoding a configuration $(q,t,h)$, check that if $p = q$ (i.e., $d^p = d^q$) and $t_h = c$ (i.e. the cell with the head is $d^{\headOn{c}}$), then the next block encodes some configuration $(q',t',h')$ with certain required properties:
    \[
      \psi_{10} =
      \bigwedge_{
        \tau =
        p,s_1,s_2,s_3 \mapsto p',s'_1,s'_2,s'_3
        \in \delta
        } \forall x.
      \unrhd x^\#.
      [x^p]
      \recmax{Y}{[\star \neq x^\#]Y \land [x][s_1][\_][s_2][\_][s_3]
      \psi_{stt}^\tau  \land
      \psi_{blc}^\tau
      } ,
    \]
    where for each
    $\tau =
    p,s_1,s_2,s_3 \mapsto p',s'_1,s'_2,s'_3
    \in \delta$,
    $\psi^\tau_{stt} $ and $ \psi^\tau_{blc}$ assert the following:
      \begin{itemize}
      \item $q' = p'$ (i.e. $d^{q'} = d^{p'}$):
      \[
        \psi_{stt}^\tau =
        \rhd x^\#.
        [\star \neq x^{p'}]\fls
      \]
    \item the cell where the tape head is at and the ones right next to it change according to the right transition in $\delta$:
    \[
      \psi_{blc}^\tau =
      \recmax{Y}{[\star \neq x]Y \land [x]([\star \neq s'_1] \fls \land  [s'_1][\_]([\star \neq s'_2]\fls \land [s'_2][\_][\star \neq s'_3]\fls ))
      }
    \]
    \item finally, we must ensure that all cells but the one with the head or the ones next to it are unchanged. This is done with the help of the unique identifiers: when at index $j$, bind the corresponding identifier $i_j$ to a data variable, and check that in the next block, when $i_j$ occurs, the content of the cell (which is encoded by the next data value) is the same:
      \[
        \psi_{11} = \forall x. \forall x^i. [x^\#]
        \left[
          \bigwedge_{s \in S
          } x^i \neq x^s \land x^{\headOn{1}} \neq x \neq x^{\headOn{0}}\right] \recmax{X}{[\_]X \land
        [\neg \mathsf{head}][x^i][x][\_][\neg \mathsf{head}]
        \rhd
        x^i
        .
        [\star \neq x] \fls
        }
      \]
  \end{itemize}
  \end{itemize}
  \item Finally, one checks that the machine never halts, i.e.\ that $d^{p_f}$ never occurs:
  \[
    \psi_{\bar{h}} =
    \unrhd x^{p_f}.\fls
  \]
  \end{itemize}
  Then, we define
  \[ \psi = \psi_{1} \land \psi_{2} \land \psi_{3} \land \psi_{4} \land \psi_{5} \land \psi_{6} \land \psi_{7} \land \psi_{8} \land \psi_{9} \land \psi_{10} \land \psi_{11} \land \psi_{\bar{h}} .\]
  By construction, each $\psi$ belongs to \conjhmld. Since the overall formula consists of a conjunction of all of them, it also belongs to \conjhmld.

  Then, it follows from the construction that the said formula is satisfiable if and only if $M$ does not halt.
\end{proof}
\subsubsection{Proof of \Cref{thm:satisfiability_chmld}}
\label{app:proof_thm_satisfiability_chmld}
\satisfiabilitycHMLd*
\begin{proof}
  Similarly to the proof of \Cref{thm:validity_disjhmld}, we reduce the Halting problem to the satisfiability problem of \chmld{}.
  For our convenience, we now assume that every sequence of data that starts with $d^{p_f}$ encodes a halting configuration. That is, we only check the encoding of a run until we encounter $d^{p_f}$, at which point we can declare that the run of $M$ halts.
  Thus, we can describe that the halting configuration is reached with $\psi_{h} = \neg \psi_{\bar{h}} = \recmin{X}{\pos{\star = x^{p_f}}{\tru} \lor \pos{\ltru}{X} }$.
  Then, we can bound fixpoint formulas to the occurence of $d^{p_f}$.
  For example, we would replace $\unrhd x.\chi$ by $\recmax{X}{[\star \neq x^{p_f}]X \land [x]\chi}$, which, on a trace where $d^{p_f}$ appears, is equivalent to $\recmin{X}{[\star \neq x^{p_f}]X \land [x]\chi}$.
  This way, we can replace greatest-fixed-points with least-fixed-points in $\psi_1$ through $\psi_{11}$.
  Observe that all the formulas in  the proof of \Cref{thm:validity_disjhmld} can be rewritten so that every universal quantifier $\forall x$ appears immediately before a box of the form $[x]$. Furthermore, $\forall x.[x]\chi $ is equivalent to $\exists x.\pos{x}{\chi}$.
  Finally, since $[\vbexp]\chi$ is equivalent to $\pos{\neg \vbexp}{\tru} \lor \pos{\vbexp}{\chi}$, we can use these  replacements on the  formulas in  the proof of \Cref{thm:validity_disjhmld} to construct a \chmld{}-formula that is satisfiable if and only if $M$ halts.
\end{proof}

\subsection{Missing Proofs of \Cref{sec:annotations}}
\label{sec:app:annotations}

\subsubsection{An Iterative Characterisation of Least Fixed Points}
\label{sec:iterative_fixed_points}
In our proofs, we use the iterative construction of the fixed points (see, e.g., \cite[Subsection~3, in particular 3.2 and 3.4]{DBLP:books/el/07/BradfieldS07}, although this dates back to the Knaster-Tarski theorem), which provides a more computational view of the $\min$ (and dually, of the $\max$) operator. Let $\vfvar \in \rechmld$, and $\vdenv \in \denv$. For an ordinal $\zeta$, we define the semantics of $\recmin{\vpvar^{\zeta}}{\vfvar}$ as:
\begin{itemize}
\item $\sem{\recmin{\vpvar^{0}}{\vfvar}, \vdenv, \rho} =
\varnothing
$;
\item $\sem{\recmin{\vpvar^{\zeta + 1}}{\vfvar}, \vdenv, \rho} = \sem{\recmin{\vpvar^{\zeta}}{\vfvar}, \vdenv,\rho[\vpvar \mapsto \lambda \vvdenv.\sem{\recmin{\vpvar^{\zeta }}{\vfvar}, \vvdenv, \rho}]}$; and
\item $\sem{\recmin{\vpvar^{\zeta }}{\vfvar}, \vdenv, \rho} = \bigcup_{\eta < \zeta} \sem{\recmin{\vpvar^{\eta}}{\vfvar}, \vdenv,\rho}$, if $\zeta$ is a limit ordinal.
\end{itemize}

\begin{lemma}[Iterative Characterization of the Least Fixed Point, \cite{DBLP:books/el/07/BradfieldS07}, see also {\cite[Lemma~2.12]{aceto2024complexity}}]\label{lem:iterativeTarski}
  For every environment $\rho$, $\vdenv \in \denv$, and $\recmin{\vpvar}{\vfvar}$, there exists some ordinal $\xi$ such that
  \begin{equation*}
    \sem{\recmin{\vpvar}{\vfvar},\vdenv,\rho} = \sem{\vfvar,\vdenv,\rho[\vpvar \mapsto \sem{\recmin{\vpvar^{\xi}}{\vfvar},\vdenv,\rho}]} = \sem{\recmin{\vpvar^{\xi}}{\vfvar},\vdenv,\rho} .
    \end{equation*}
\end{lemma}

\subsubsection{An Example of Annotation}
\label{app:example_annotation}

\begin{example}
  Consider the formula $\formfrepeats$ in \Cref{eq:first_repeats} on trace $\vvtrc = 0 2 0 1^\omega$ starting from the empty data valuation. A minimal annotation witnessing that $\vvtrc \in \sem{\formfrepeats}$ is:
  \begin{align*}
    & (\exi{\binder{\vdvar}}{\pos{\vdvar = \star}{\recmin{\vpvar}{\left(\pos{\vdvar = \star}{\tru} \lor \pos{\vdvar \neq \star}{\vpvar}}  \right)}}, \varnothing, 0 2 0 1^\omega) \\
    & \dep{} ({\pos{\vdvar = \star}{\recmin{\vpvar}{\left(\pos{\vdvar = \star}{\tru} \lor \pos{\vdvar \neq \star}{\vpvar}}  \right)}}, \vdvar \mapsto 0, 0 2 0 1^\omega) \\
    & \dep{} ({\recmin{\vpvar}{\left(\pos{\vdvar = \star}{\tru} \lor \pos{\vdvar \neq \star}{\vpvar}}  \right)}, \vdvar \mapsto 0, 2 0 1^\omega) \\
    & \dep{} (\pos{\vdvar = \star}{\tru} \lor \pos{\vdvar \neq \star}{\vpvar}, \vdvar \mapsto 0, 2 0 1^\omega) \\
    & \dep{} (\pos{\vdvar \neq \star}{\vpvar}, \vdvar \mapsto 0, 2 0 1^\omega) \dep{} ({\vpvar}, \vdvar \mapsto 0, 0 1^\omega) \\
    & \dep{} ({\recmin{\vpvar}{\left(\pos{\vdvar = \star}{\tru} \lor \pos{\vdvar \neq \star}{\vpvar}}  \right)}, \vdvar \mapsto 0, 0 1^\omega) \\
    & \dep{} (\pos{\vdvar = \star}{\tru} \lor \pos{\vdvar \neq \star}{\vpvar}, \vdvar \mapsto 0, 0 1^\omega) \\
    & \dep{} (\pos{\vdvar = \star}{\tru}, \vdvar \mapsto 0, 0 1^\omega) \dep{} ({\tru}, \vdvar \mapsto 0, 1^\omega)
    \end{align*}
\end{example}

\subsubsection{Proof of \Cref{prop:annotation-semantics}}
\label{app:proof:prop:annotation-semantics}
\annotationSemantics*
  \begin{proof} We prove each implication.
    \subparagraph{Left-to-right: annotation implies satisfaction}
      Let $(A,\dep{})$ be an annotation.
      For each $a \in A$, we define $\vpenv_a$ as follows: for every recursion variable $\vpvar \in \subform{\vfvar}$ and every $\vdenv \in \denv$, $\vpenv_a(\vpvar)(\vdenv) = \{ \vvtrc \mid \vanv \dep{X}^* (\vfvar_\vpvar, \vdenv, \vvtrc) \}$.
      We prove that for every $\vvfvar \in \subform{\vfvar}$ and all\footnote{Note that there may not exist such an $a$.}
      $a = (\vvfvar, \vdenv, \vtrc) \in A$,
      $ \vtrc \in \sem{\vvfvar,\vdenv, \vpenv_a}$.
      We proceed by induction on $\vvfvar$.
      The interesting cases are:
      \begin{itemize}
      \item $\vvfvar = \exi{\binder{\vdvar}}{\vvvfvar}$. Then, since $(\exi{\binder{\vdvar}}{\vvvfvar}, \vdenv, \vtrc) \in A$, there exists some $\vdat \in \dat$ such that $(\vvvfvar, \vdenv[\vdvar \mapsto \vdat], \vtrc) \in A$ and $(\exi{\binder{\vdvar}}{\vvvfvar}, \vdenv, \vtrc) \dep{} (\vvvfvar, \vdenv[\vdvar \mapsto \vdat], \vtrc)$. By induction, this means that $\vtrc \in \sem{\vvvfvar, \vdenv[\vdvar \mapsto \vdat], \vpenv_a}$. As a consequence, $\vtrc \in \sem{\exi{\binder{\vdvar}}{\vvvfvar}, \vdenv, \vpenv_a} = \bigcup_{\vdat \in \dat} \sem{\vvvfvar,\vpenv_a,\vdenv[\vdvar\mapsto\vdat]}$.
      \item $\vvfvar = \uni{\binder{\vdvar}}{\vvvfvar}$. This case is dual: since $(\uni{\binder{\vdvar}}{\vvvfvar}, \vdenv, \vtrc) \in A$, we know that for all $\vdat \in \dat$, $(\vvvfvar, \vdenv[\vdvar \mapsto \vdat], \vtrc) \in A$ and $(\uni{\binder{\vdvar}}{\vvvfvar}, \vdenv, \vtrc) \dep{} (\vvvfvar, \vdenv[\vdvar \mapsto \vdat], \vtrc)$. By induction, this means that for all $\vdat \in \dat$, $\vtrc \in \sem{\vvvfvar, \vdenv[\vdvar \mapsto \vdat], \vpenv_a}$. As a consequence, $\vtrc \in \sem{\uni{\binder{\vdvar}}{\vvvfvar}, \vdenv, \vpenv_a} = \bigcap_{\vdat \in \dat} \sem{\vvvfvar,\vpenv_a,\vdenv[\vdvar\mapsto\vdat]}$.
      \item $\vvfvar = \recmax{\vpvar}{\vfvar_\vpvar}$. Let $a = (\recmax{\vpvar}{\vfvar_\vpvar}, \vdenv, \vtrc) \in A$ (if such an $a$ exists). Recall that $\sem{\recmax{\vpvar}{\vfvar_\vpvar}, \vpenv_a, \vdenv} = \big(\bigsqcup\big\{F \mid F \sqsubseteq \lambda\vdenv'.\sem{\vfvar_\vpvar, \vpenv_a[\vpvar \mapsto F], \vdenv'} \big\}\big)(\vdenv)$. Thus, we need to show that there exists some $F$ such that $\vtrc \in F(\vdenv)$ and
        \begin{equation}
          \label{eq:greatest_fixed_point_union}
          F \sqsubseteq \lambda \vdenv'.\sem{\vfvar_\vpvar, \vpenv_a[\vpvar \mapsto F], \vdenv'}.
        \end{equation}
        Take $F = \vpenv_a(\vpvar) = \lambda \vvdenv . \{ \vvtrc \mid \vanv \dep{X}^* (\vpvar, \vvdenv, \vvtrc) \}$.
        First, since $a = (\recmax{\vpvar}{\vfvar_\vpvar}, \vdenv, \vtrc) \in A$, by definition of an annotation this implies $(\vfvar_\vpvar, \vdenv, \vtrc) \in A$ and $a \dep{X} (\vfvar_\vpvar, \vdenv, \vtrc)$. Thus, by definition of $F$, $\vtrc \in \F(\vdenv)$. Now, it remains to show that $F$ satisfies equation~\eqref{eq:greatest_fixed_point_union}. Let $\vdenv' \in \denv$ and $\vvtrc \in F(\vdenv')$. We need to show that $\vvtrc \in \sem{\vfvar_\vpvar, \vpenv_a[\vpvar \mapsto F], \vdenv'}$. Since $\vvtrc \in F(\vdenv')$,  $\vanv \dep{X}^* (\vfvar_\vpvar, \vvdenv, \vvtrc) = b$. By the induction hypothesis, this means that $\vvtrc \in \sem{\vfvar_\vpvar, \vvdenv, \vpenv_b}$. Since $a \dep{X}^* b$, $\vpenv_b \sqsubseteq \vpenv_a = \vpenv_a[\vpvar \mapsto F]$, so $\sem{\vfvar_\vpvar, \vvdenv, \vpenv_b} \subseteq \sem{\vfvar_\vpvar, \vvdenv, \vpenv_a[X \mapsto F]}$, which concludes the proof.
      \item $\vvfvar = \recmin{\vpvar}{\vfvar_\vpvar}$. Let $A_\vvfvar = \{(\vvvfvar, \vdenv, \vtrc) \in A \mid \vvvfvar = \vvfvar\}$.
        For every $a,b \in A_\vvfvar$, we define $a \leq b$ if $b \dep{\vpvar}^* a$.
        Since $(A,\dep{})$ is lfp-consistent and $\vpvar$ is a lfp variable, there is no infinite $\dep{\vpvar}$-sequence in $A$ where $\vpvar$ appears infinitely often.
        Therefore, $(A_\vvfvar,\leq)$ is well-founded and we can use well-founded induction on $(A_\vvfvar,\leq)$
        to prove that for every $a = (\vvfvar, \vdenv, \vtrc) \in A_\vvfvar$, there is some ordinal $\zeta(a)$, such that $\vtrc \in \sem{\recmin{\vpvar^{\zeta(a)}}{\vfvar_\vpvar},\vdenv,\rho_a}$.

        The base case is that $a = (\vvfvar,\vdenv,\vtrc) \in A_\vvfvar$ is $\leq$-minimal, and we observe that then,
        $\rho_a = \rho_a[\vpvar \mapsto \lambda \vvvdvar.\varnothing]$.
        Therefore, for $b =  (\vfvar_\vpvar,\vdenv,\vtrc) \in A$, we have:
        \[\sem{\recmin{\vpvar^0}{\vfvar_\vpvar},\vdenv,\rho_a} = \sem{\vfvar_\vpvar,\vdenv,\rho_a[\vpvar \mapsto \lambda \vvvdvar.\varnothing]} = \sem{\vfvar_\vpvar,\vdenv,\rho_a}
          = \sem{\vfvar_\vpvar,\vdenv,\rho_b}
          ,
        \]
        Indeed, $a \dep{} b$ and $a \dep{\vvpvar} b$ for any $\vvpvar$ so $\rho_a = \rho_b$.
        By the inductive hypothesis for $\vfvar_\vpvar$ from the formula induction, $\vtrc \in \sem{\vfvar_\vpvar,\vdenv,\rho_b}$, and therefore $\vtrc \in \sem{\recmin{\vpvar^0}{\vfvar_\vpvar},\vdenv,\rho_a}$, which completes the base case for the well-founded induction.

        For the inductive case, let $a = (\vvfvar,\vdenv,\vtrc) \in A_\vvfvar$ that is not $\leq$-minimal.
        By the inductive hypothesis, for each $b = (\vvfvar, \vvdenv, \vtrc) \in A_\vvfvar$, if  $b < a$, then there exists an ordinal $\zeta(b)$ such that
        $\vtrc \in \sem{\recmin{\vpvar^{\zeta(b)}}{\vfvar_\vpvar},\vvdenv,\rho_b}$.
        For $b<a$, by the monotonicity of $\vfvar_\vpvar$ and the observation that $\rho_b(\vpvar) \sqsubseteq \rho_a(\vpvar)$, we see that
        $\vtrc \in \sem{\recmin{\vpvar^{\zeta(b)}}{\vfvar_\vpvar},\vvdenv,\rho_b} \subseteq \sem{\recmin{\vpvar^{\zeta(b)}}{\vfvar_\vpvar},\vvdenv,\rho_a}$, and therefore
        $\vtrc \in \sem{\recmin{\vpvar^{\zeta(b)}}{\vfvar_\vpvar},\vvdenv,\rho_a}$.
        Let $\eta = \bigcup_{b<a} \zeta(b)$.
        Then,
        $\vtrc \in \sem{\recmin{\vpvar^{\eta}}{\vfvar_\vpvar},\vvdenv,\rho_a}$
        for every
        $b = (\vvfvar, \vvdenv, \vvtrc) \in A_\vvfvar$ such that $\vvtrc = \vtrc$ and $b < a$.
        Therefore, $\rho_a(X)(\vvdenv) \subseteq \sem{\recmin{\vpvar^{\eta}}{\vfvar_\vpvar},\vvdenv,\rho_a}$ for every $\vvdenv$, which yields
        $\rho_a(X) \sqsubseteq \lambda \vvdenv.\sem{\recmin{\vpvar^{\eta}}{\vfvar_\vpvar},\vvdenv,\rho_a}$.
        By the monotonicity of $\vfvar_\vpvar$,
        \[\sem{\vfvar_\vpvar,\vdenv,\rho_a}
          \subseteq
          \sem{\vfvar_\vpvar,\vdenv,\rho_a[\vpvar \mapsto \lambda \vvdenv.\sem{\recmin{\vpvar^{\eta}}{\vfvar_\vpvar},\vvdenv,\rho_a}]}
          =
          \sem{\recmin{\vpvar^{\eta+1}}{\vfvar_\vpvar},\vdenv,\rho_a}.\]
        By the inductive hypothesis for $\vfvar_\vpvar$ from the formula induction, $\vtrc \in \sem{\vfvar_\vpvar,\vvdenv,\rho_a}$, and therefore $\vtrc \in \sem{\recmin{\vpvar^{\eta + 1}}{\vfvar_\vpvar},\vvdenv,\rho_a}$, which completes  the well-founded induction.

        Finally,
        from the iterative characterization of the least fixed points (\Cref{lem:iterativeTarski}),
        we have that for every $\vdenv, \vpenv$,
        and ordinal $\zeta$,
        $\sem{\recmin{\vpvar^\zeta}{\vfvar_\vpvar},\vdenv,\rho} \subseteq
        \sem{\vvfvar,\vdenv,\rho}$, and therefore
        for every $a = (\vvfvar, \vdenv, \vtrc) \in A_\vvfvar$, $\vtrc \in \sem{\recmin{\vpvar}{\vfvar_\vpvar},\vdenv,\rho_a}$.
      \end{itemize}
      This completes the induction on $\vvfvar$.
    \subparagraph{Right-to-left: satisfaction implies annotation}
      For this case, we use a similar argument as \cite[Theorem 2.13]{aceto2024complexity}, which has a similar statement and proof.

      To handle the case of free recursion variables,
      we extend the definition of an annotation $A$ for $\vfvar$ to open formulae in the context of an environment $\rho$, such that the condition
      \textquote{if $(\vpvar,\vdenv,\vtrc) \in A$,
        then $(\vpvar, \vdenv, \vtrc) \dep{} (\vfvar_\vpvar, \vdenv, \vtrc)$}
      is omitted when $X$ is free in $\varphi$ and $\vtrc \in \rho(X)(\vdenv)$.
      Thus, for the remainder of this proof, we maintain that if $\vtrc \in \sem{\vfvar,\vdenv,\rho}$, then $(\vfvar,\vdenv,\vtrc)$ appears in an annotation. We proceed by induction on $\vfvar$.

      The interesting cases are:
      \begin{itemize}
      \item $\vfvar = \exi{\binder{\vdvar}}{\vvfvar}$. By definition of $\sem{\vfvar, \vdenv, \vpenv}$, if $\vtrc \in \sem{\vfvar, \vdenv, \vpenv}$, then there exists some $\vdat \in \dat$ such that $\vtrc \in \sem{\vfvar, \vdenv[\vdvar \mapsto \vdat], \vpenv}$. By the induction hypothesis, this implies that $\vvfvar, \vdenv[\vdvar \mapsto \vdat]$ have an annotation $A$ on $\vtrc$ in the context of $\vpenv$. Then, $B = A \cup \{(\exi{\binder{\vdvar}}{\vvfvar}, \vdenv, \vtrc)\}$ with the dependency relation $\dep{} \cup \{((\exi{\binder{\vdvar}}{\vvfvar}, \vdenv, \vtrc), (\vvfvar, \vdenv[\vdvar \mapsto \vdat], \vtrc))\}$ is an annotation of $\vfvar, \vdenv$ on $\vtrc$ in the context of $\vpenv$.
      \item $\vfvar = \uni{\binder{\vdvar}}{\vvfvar}$. This case is again dual. By definition of $\sem{\vfvar, \vdenv, \vpenv}$, if $\vtrc \in \sem{\vfvar, \vdenv, \vpenv}$, then for all $\vdat \in \dat$, $\vtrc \in \sem{\vfvar, \vdenv[\vdvar \mapsto \vdat], \vpenv}$. By the induction hypothesis, this implies that for all $\vdat \in \dat$, $\vvfvar, \vdenv[\vdvar \mapsto \vdat]$ have an annotation $A_{\vdat}$ on $\vtrc$ in the context of $\vpenv$. Then, $B = \bigcup_{\vdat \in \dat} A_{\vdat} \cup \{(\uni{\binder{\vdvar}}{\vvfvar}, \vdenv, \vtrc)\}$ with the dependency relation $\bigcup_{\vdat \in \dat} \dep{}_{\vdat} \cup \left\{((\uni{\binder{\vdvar}}{\vvfvar}, \vdenv, \vtrc), (\vvfvar, \vdenv[\vdvar \mapsto \vdat], \vtrc)) \mid \vdat \in \dat\right\}$ is an annotation of $\vfvar, \vdenv$ on $\vtrc$ in the context of $\vpenv$.
      \item $\varphi = \max X.\psi$. Then
      by the induction hypothesis there exists
      an annotation $(A,\dep{})$ for $\psi, \vdenv$ in the context of $\rho[X \mapsto \lambda \vvdenv.\sem{\varphi,\vvdenv,\rho}]$.
      Then, let $B = A \cup \{(\varphi,\vdenv,\vtrc)\}$ and $\dep{}' = \dep{} \cup \{((\varphi,\vdenv,\vtrc),(\psi,\vdenv,\vtrc)),((\psi,\vdenv,\vtrc),(\varphi,\vdenv,\vtrc))\}$.
      If $Y$ is a least-fixed-point variable, then any new $\dep{}'$-path where $Y$ appears infinitely often visit $X$ infinitely often, and therefore $(B,\dep{})$ is still lfp-consistent, so it is an annotation of $\vfvar, \vdenv$ in te context of $\vpenv$.
      \item $\varphi = \min X.\psi$. Then by Lemma \ref{lem:iterativeTarski}, there exists some ordinal $\xi$, with $\sem{\varphi,\vdenv,\rho} = \sem{\psi,\vdenv,\rho[X \mapsto \sem{\min X^\xi.\psi,\vdenv,\rho}]}$.
      By the induction hypothesis, there exists
      an annotation $(A,\dep{})$ for $\psi,\vvdenv$ on each $\vvtrc \in \sem{\psi,\vvdenv,\rho[X \mapsto \lambda \vvdenv.\varnothing]}$ in the context of $\rho[X \mapsto \lambda \vvdenv.\varnothing]$ --- and therefore also in the context of $\rho$. It is not hard to extend $(A,\dep{})$ to an annotation for $\psi,\vdenv$ on each $\vvtrc \in \sem{\psi,\vvdenv,\rho[X \mapsto \lambda \vvdenv.\varnothing]}$.
      For each $\zeta \leq \xi$, we define $(A_\zeta,\dep{}_\zeta)$ and prove that it is an lfp-consistent
      annotation for $\varphi,\delta$ on every $\vvtrc \in \sem{\psi,\vvdenv,\rho[X \mapsto \sem{\min X^\xi.\psi,\vdenv,\rho}]}$ and for $\psi,\delta$ on every $\vvtrc \in \sem{\psi,\vvdenv,\rho[X \mapsto \sem{\min X^\xi.\psi,\vdenv,\rho}]}$ in the context of
      $\rho$, where $X$ does not appear infinitely often on any $\dep{X}_\zeta$ path:
      \begin{description}
      \item[Case $\zeta =0$:]
        let $(A_0,\dep{}_0) = (A,\dep{})$.

      \item[Case $\zeta = \eta+ 1$ for some $\eta$:] then by the inductive hypothesis, there exists an annotation
        $(B_\zeta,\dep{}^B_\zeta)$ for $\psi$ in the context of $\rho[X \mapsto \lambda \vvdenv.\sem{\min X^\eta.\psi,\vdenv,\rho}]$.
        Let
        \[A_{\eta+1} = A_\eta \cup B_{\eta+1} \cup \{ (\varphi,\vvdenv,\vvtrc) \mid (\psi,\vvdenv,\vvtrc) \in  B_{\eta + 1}\}, \text{ and} \]
        \[\dep{}_{\eta + 1} = \dep{}_\eta \cup \dep{}_{\eta+1}^B \cup \{ ((X,\vvdenv,\vvtrc),(\varphi,\vvdenv,\vvtrc)), ((\varphi,\vvdenv,\vvtrc),(\psi,\vvdenv,\vvtrc)) \in A_{\eta + 1}^2
          \},\]
          that is, $\dep{}_{\eta + 1}$ includes all pairs from $\dep{}_{\eta}$ and $\dep{}_{\eta + 1}^B$, and adds some pairs for the recursion cases that may be missing.
      \item[if $\zeta$ is a limit ordinal] we define
        $
        A_\zeta = \bigcup_{\eta < \zeta} A_\eta
        $
        and
        $
        \dep{}_{\zeta} = \bigcup_{\eta < \zeta} \dep{}_{\eta}
        $.
      \end{description}
      It is straightforward to verify that $(A_\zeta,\dep{}_{\zeta})$ is an annotation for
      $\varphi$ in the context of $\rho$.
      Notice that if $(X,\vvdenv,\vvtrc) \dep{}_{\zeta} (\varphi,\vvdenv,\vvtrc)$, then
      either both $(X,\vvdenv,\vvtrc), (\varphi,\vvdenv,\vvtrc) \in A_\eta$
      for some $\eta < \zeta$, or
      $(X,\vvdenv,\vvtrc) \in A_{\zeta}$, in which case either $(\varphi,\vvdenv,\vvtrc) \in A_\eta$
      or $\vvtrc \in
      \sem{\min X^\eta.\psi,\vvdenv,\rho}
      $
      for some $\eta < \zeta$.
      In all three cases we conclude that
      $(\varphi,\vvdenv,\vvtrc) \in A_\eta$, and
      therefore, no $\dep{X}_\zeta$-path can have infinitely many occurrences of $X$, and $(A_\zeta,\dep{}_{\zeta})$ is an lfp-consistent annotation for $\varphi,\vdenv$ in the context of $\rho$, thus completing the inductive proof.
      \qedhere
      \end{itemize}
  \end{proof}
\subsubsection{Proof of \Cref{prop:annotation-chmld}}
\label{app:proof:prop:annotation-chmld}
\begin{restatable}{proposition}{annotationcHMLd}
  \label{prop:annotation-chmld}
  Let $\vfvar \in \chmld$ and $\vtrc \in \trc$. If $\vfvar$ has an annotation on $\vtrc$, then it has a finite one.
\end{restatable}
\begin{proof}
  Let $\vfvar \in \chmld$ and $\vtrc \in \trc$. Consider a minimal annotation $(A_f,\dep{f})$, such that $(\vfvar, \vdenv, \vtrc) \in A_f$ for some $\vdenv \in \denv$. In \cref{def:annotation}, all rules except that for $\uni{\binder{\vdvar}}{\vfvar}$ only induce finite branching, and \chmld\ does not contain the $\forall$ operator, so the tree unfolding of $(A_f,\dep{f})$ is finitely branching. Moreover, since $(A_f,\dep{f})$ is lfp-consistent and $\chmld$ does not allow $\max$, every path is finite, otherwise we would find an lfp variable that appears infinitely often. Therefore, $(A_f,\dep{f})$ is a finite annotation for $\vfvar, \vdenv$ on $\vtrc$.
  \end{proof}

\subsection{Renamings and Types}
\label{sec:app:renamings}

A key property of data words is that they can only be manipulated through predicates of the domain. Thus, when there are no unary predicates (in particular, no constants), data values do not matter: only the \emph{relations} between them do, i.e. how they compare with regards to the predicates.

When the only predicate is `=', this is elegantly captured by the notion of \emph{nominal set}~\cite{Pitts2013}.
Here, we borrow a few notions from this framework to ease the manipulation of data words. For a more comprehensive introduction, see~\cite{Bojanczyk19, DBLP:journals/corr/BojanczykKL14}. First, note that in~\cite{Bojanczyk19}, sets that are stable under renamings are called \emph{equivariant}; we choose a different terminology to avoid jargon.
\begin{definition}
  \label{def:renaming}
  A \emph{renaming} of data words is a bijection $\vren: \dat \rightarrow \dat$. A set $\vset$ built over elements of $\dat$ using union, intersection, concatenation etc. is \emph{stable under renaming} if for all renamings $\vren: \dat \rightarrow \dat$, $\vren(\vset) = \vset$ (where $\vren$ is extended to elements of $\vset$ in a natural way).
\end{definition}
\begin{remark}
  To handle constants, one restricts to renamings that preserve constants. Formally, given a finite set of constants $\const \subset \dat$, one asks that $\vren : \dat \rightarrow \dat$ is a bijection and additionally satisfies that elements of $\const$ are fixed points of $\vren$, that is $\vren(\vconst) = \vconst$ for every $\vconst \in \const$.
\end{remark}
The only subsets of $\dat$ that are stable under renaming are $\varnothing$ and $\dat$. An example of a non-trivial set that is stable under renaming is the set of data words containing pairwise distinct data values, $\{d_0 d_1 \dots \in \dat^\omega \mid \forall i \neq j, d_i \neq d_j\}$, which is described by the formula $\formpairwisedistinct$ in \Cref{ex:rechmld_properties}.

Observe that all equations of \Cref{fig:rechml_LT} are stable under renaming when the formula is closed. As a consequence, so are the sets of traces that satisfy closed formulae:
\begin{proposition}
  \label{prop:invariance_renaming_rechmld}
  For each closed \rechmld{} formula $\vfvar$, the set $\sem{\vfvar}$ is stable under renaming.
\end{proposition}

This is also the case for our model of monitors. Since all rules in \Cref{fig:monitors_guessing} (\pageref{fig:monitors_guessing}) are invariant under renaming, we get:
\begin{proposition}
  \label{prop:invariance_renaming_monitors}
Let $\sigma : \mathbb{D} \rightarrow \mathbb{D}$ be a bijection. For all configurations $\vcon$ and $\vvcon$, and all words $w \in \mathbb{D}^*$, we have $\vcon \xrightarrow{w} \vvcon$ if and only if $\sigma(\vcon) \xrightarrow{\sigma(w)} \sigma(\vvcon)$.
\end{proposition}

A second useful notion is that of type, which appears in many guises in the literature. We choose the following definition, which is the most convenient for our purpose.
\begin{definition}
  \label{def:type}
  A function $f: X \rightarrow \dat$ naturally induces an equivalence relation $\eqrel[X]{f}$ over $X$ defined, for all $x, y \in X$, as $x \eqrel[X]{f} y$ whenever $f(x) = f(y)$, sometimes called the \emph{kernel} of $f$.
  We define the \emph{type} $\type(\vdenv)$ of a valuation $\vdenv: \DVar \rightarrow \dat$ as its kernel, i.e. the set of equivalence classes of the relation $\vdvar \sim \vvdvar$ whenever $\vdenv(\vdvar) = \vdenv(\vvdvar)$.
  We then extend this definition to finite data words $w \in \dat^*$ by letting $\type(\vword) = \type(\vdenv_\vword)$, where $\vdenv_\vword: \{0, \dots, n-1\} \rightarrow \dat$ is defined for $n = \length{w}$ as, for all $0 \leq i < n$, $\vdenv_\vword(i) = \vword_i$.

Note that the type of a word $w \in
\dat^*$ can be expressed a boolean expression $\bform{w}(x_0, \dots, x_{n-1}) = \bigwedge_{0 \leq i < n} \bigwedge_{0 \leq j < n} x_i \eqorneq_{i,j} x_j$, where $\eqorneq_{i, j}$ stands for $=$ if $w_i = w_j$, and for $\neq$ otherwise.
\end{definition}
The type of a word is uniquely determined by its length and by the equality relations between its different letters, e.g. $\type(121) = \type(232)$ but $\type(121) \neq \type(1211)$ and $\type(121) \neq \type(123)$. Observe that, more generally, for any two data words $\vword$ and $\vvword$, we have that $\type(\vword) = \type(\vvword)$ if and only if there exists a renaming $\vren: \dat \rightarrow \dat$ such that $\vren(\vword) = \vren(\vvword)$ (see~\cite[Claim~4.12]{Bojanczyk19}).
A consequence of this observation is the following:
\begin{proposition}
  \label{prop:log_form_finite_words}
  Let $n \geq 0$, and let $S \subseteq \dat^n$ be stable under renaming. Then, there exists $\bform{S}(x_0, \dots, x_{n-1})$ such that $S = \left\{ w \in \dat^n \mid \bform{S}(w_0, \dots, w_{n-1}) \right\}$.
\end{proposition}
\begin{proof}
  We redo the proof for completeness. Let $n \geq 0$ and $S \subseteq \dat^n$ that is stable under renaming. Consider the set $B = \{\bform{w} \mid w \in S\}$. There are only $2^{n^2}$ possibilities for $\bform{w}$, so $B$ is finite. Then, define $\bform{S}(x_0, \dots, x_{n-1}) = \bigvee_{b \in B} b(x_0, \dots, x_{n-1})$.

  Let $w \in \dat^n$. If $w \models \bform{S}$, then $w \models b$ for some $b \in B$, so by definition $w \models \bform{w'}$ for some $w' \in S$. By~\cite[Claim~4.12]{Bojanczyk19}, this means there exists a renaming $\vren : \dat \rightarrow \dat$ such that $\vren(w') = w$, so $w \in S$ since $S$ is stable under renaming.

  Conversely, if $w \in S$, then $\bform{w} \in B$ and $w \models \bform{S}$ since $w \models \bform{w}$.
\end{proof}

\subsection{Missing Proofs from \Cref{sec:complete_monitorability}}
\label{sec:proofs:complete_monitorability}

Before we start, let us establish the following (stability under renaming is formally defined in \Cref{def:renaming} on page~\pageref{def:renaming}):
  \begin{lemma}
    \label{lem:good_bad_prefixes_stable_renamings}
    Let $\vstrc$ be a set of traces that is stable under renamings, and $\goodprefixes, \badprefixes \subseteq \ftrc$ respectively be the set of its good and bad prefixes. Then, $\goodprefixes$ and $\badprefixes$ are both closed under renamings.
  \end{lemma}
  \begin{proof}
    Let $\vstrc$ be a set of traces that is stable under renamings. Let $\vren$ be a renaming and $g \in \goodprefixes$ be a good prefix for $\vstrc$. We show that $\vren(g)$ is a good prefix for $\vstrc$ as well (the proof for bad prefixes is dual). Let $\vtrc \in \trc$. $g$ is a good prefix for $\vstrc$, so in particular $g \vren^{-1}(\vtrc) \in \vstrc$. Since $\vstrc$ is stable under renamings, $\vren(g \vren^{-1}(\vtrc)) = \vren(g) \vtrc \in \vstrc$. This holds for all $\vtrc \in \trc$, so $\vren(g)$ is a good prefix for $\vstrc$.
  \end{proof}
Over finite alphabets, complete monitorability is characterised as having bounded discriminating prefixes. As we show, this generalises to our setting.
  \begin{definition}
    $\vstrc$ has \emph{bounded discriminating prefixes} when there exists $n \in \N$ such that for all finite traces $\vftrc \in \ftrc$, if $\length{\vftrc} \geq n$, then $\vftrc$ is either a good or a bad prefix for $\vstrc$.
  \end{definition}
  For instance, the property $\vstrc = \{d^n d' \vtrc \mid n \geq 0, d \neq d', \vtrc \in \trc\}$ does not have bounded discriminating prefixes (since $d'$ can appear arbitrarily far in the input), but for all $k \geq 0$, $\vstrc_k = \{d^n d' \vtrc \mid k \geq n \geq 0, d \neq d', \vtrc \in \trc\}$ does, since $d'$ has to appear within the first $k+1$ elements.

\begin{restatable}{proposition}{completeMonitorabilityHMLd}
  \label{prop:CM_HML}
  Let $\vstrc \subseteq \trc$ be a set of traces that is stable under renaming. The following are equivalent:
  \begin{enumerate}[(i)]
  \item \label{prop:CM_HML:itm:complete_mon} $\vstrc$ is completely monitorable;
  \item \label{prop:CM_HML:itm:bounded_pref} $\vstrc$ has bounded discriminating prefixes;
  \item \label{prop:CM_HML:itm:good_pref} There exists $n \in \N$ such that $\vstrc = G \dat^\omega$ for some $G \subseteq \dat^n$ that is stable under renaming;
  \item \label{prop:CM_HML:itm:bool_expr} There exist $n \in \N$ and a boolean expression $\vbexp(\vdvar_0, \dots, \vdvar_{n-1})$ such that $\vstrc = \sem{\exi{\vdvar_0}{\pos{\star = \vdvar_0}{\exi{\vdvar_1}{\pos{\star = \vdvar_1}{\dots \exi{\vdvar_{n-1}}{\pos{\star = \vdvar_{n-1} \cand \vbexp(\vdvar_0, \dots, \vdvar_{n-1})}{\tru}}}}}}}$;
  \item \label{prop:CM_HML:itm:HML} $\vstrc$ can be expressed in \hmld{}.
  \end{enumerate}
\end{restatable}
  \begin{proof}
    Let $\vstrc \subseteq \trc$ be a set of traces that is stable under renamings.

    \ref{prop:CM_HML:itm:complete_mon} $\Rightarrow$ \ref{prop:CM_HML:itm:bounded_pref}: Assume that $\vstrc$ is completely monitorable and let $\goodprefixes, \badprefixes \subseteq \dat^*$ respectively be the set of its good and bad prefixes.

Consider the (possibly infinite) directed graph $\typeGraph$ whose vertices consist of $V = \{\type(\vftrc) \mid \vftrc \in \ftrc \backslash (\goodprefixes \cup \badprefixes)\}$. Since $\vstrc$ is stable under renaming, so are $\goodprefixes$ and $\badprefixes$ (\Cref{lem:good_bad_prefixes_stable_renamings}), and we get that \emph{for all} finite traces $\vftrc \in \ftrc$ such that $\type(\vftrc) \in V$, $\vftrc \notin \goodprefixes \dat^*$ and $\vftrc \notin \badprefixes \dat^*$.

We now define the set of edges $E$ of $\typeGraph$, writing $\vtype \rightarrow \vvtype$ instead of $(\vtype, \vvtype) \in E$: for all $\vtype, \vvtype \in V$, we let $\vtype \rightarrow \vvtype$ whenever there exists $\vftrc \in \ftrc$ and $\vdat \in \dat$ such that $\type(\vftrc) = \vtype$ and $\type(\vftrc \cdot \vdat) = \vvtype$. Note that if $\vtype \rightarrow \vvtype$, then $\length{\vvtype} = \length{\vtype} + 1$, and since for a fixed length, there are only finitely many different types, we get that that $\typeGraph$ is finitely branching. Moreover, since the property of not being a discriminating prefix is stable under taking prefixes, $\typeGraph$ is connected. Besides, if $\vtype \rightarrow \vvvtype$ and $\vvtype \rightarrow \vvvtype$, then $\vtype = \vvtype$. Finally, there is a single type of length $0$, so $\typeGraph$ is actually a tree.

  Towards a contradiction, assume that $\typeGraph$ is infinite. Then, by König's lemma, it has an infinite path, i.e. there exists $\vtype_0, \vtype_1, \dots \in V$ such that for all $i \geq 0$, $\vtype_i \rightarrow \vtype_{i+1}$. We build by induction a trace $\vtrc \in \trc$ such that for all $i \geq 0$, $\type(\vtrc[:i]) = \vtype_i$, where $\vtrc[:i] \defeq \vtrc[0] \dots \vtrc[i]$.

  Since $\typeGraph$ is a tree, $\vtype_0 = \type(\varepsilon)$, so the property holds for $i = 0$. Now, assume we have built $\vtrc$ up to index $i \geq 0$. Since $\vtype_i \rightarrow \vtype_{i+1}$, we know that  there exists $\vftrc \in \ftrc$ and $\vdat \in \dat$ such that $\type(\vftrc) = \vtype_i$ and $\type(\vftrc \cdot \vdat) = \vtype_{i+1}$. Since $\type(\vtrc[:i]) = \vtype_i = \type(\vftrc)$, we know that there exists a renaming $\vren: \dat \rightarrow \dat$ such that $\vren(\vftrc) = \vtrc[:i]$. Then, since renaming preserves types, $\type(\vren(\vftrc \cdot \vdat)) = \vtype_{i+1}$, so by letting $\vtrc[i+1] = \vren(\vdat)$, we get that $\type(\vtrc[:i+1]) = \vtype_{i+1}$.

  Now, since for all $i \geq 0$, $\type(\vtrc[:i]) = \vtype_i$, we get that $\vtrc[:i] \in \ftrc \backslash (\goodprefixes \cup \badprefixes)$. Thus, no prefix of $\vtrc$ is a good prefix, nor a bad prefix, which contradicts the assumption that $\vstrc$ is completely monitorable. Thus, $\typeGraph$ is finite. In particular, there is a bound $n \in \N$ such that for all $\vtype \in V$, $\length{\vtype} \leq n$, thus for all $\vftrc \in \vstrc \backslash (\goodprefixes \cup \badprefixes)$, $\length{\vftrc} \leq n$. In other words, $\vstrc$ has bounded discriminating prefixes.

  \ref{prop:CM_HML:itm:bounded_pref} $\Rightarrow$ \ref{prop:CM_HML:itm:good_pref}: Assume that $\vstrc$ has bounded discriminating prefixes, and let $n \in \N$ be the associated bound. Then, let $\goodprefixes = \{\vftrc \in \dat^n \mid \vftrc \text{ is a good prefix for }\vstrc\}$. Clearly, $\goodprefixes \cdot \dat^\omega \subseteq \vstrc$. Conversely, let $\vtrc \in \vstrc$, and let $\vftrc = \vtrc[:n]$. Since $\length{\vftrc} \geq n$, $\vftrc$ is either a good or a bad prefix, so it is a good prefix since $\vtrc \in \vstrc$, hence $\vftrc \in \goodprefixes$. Thus, $\vstrc \subseteq \goodprefixes \cdot \dat^\omega$, and we have that $\vstrc = \goodprefixes \cdot \dat^\omega$. Finally, since $\vstrc$ is stable under renaming, so is $\goodprefixes$ (\Cref{lem:good_bad_prefixes_stable_renamings}).

  Let us close the first cycle of implications here, by establishing \ref{prop:CM_HML:itm:good_pref} $\Rightarrow$ \ref{prop:CM_HML:itm:complete_mon}: if there exists $\goodprefixes$ as above, then $\vstrc$ is completely monitorable, as witnessed by the set of good prefixes $\goodprefixes \dat^*$ and the set of bad prefixes $(\dat^n \backslash \goodprefixes) \dat^*$.

  \ref{prop:CM_HML:itm:good_pref} $\Rightarrow$ \ref{prop:CM_HML:itm:bool_expr}: let $\goodprefixes$ be as above.
  Since $\goodprefixes$ is stable under renaming, by~\Cref{prop:log_form_finite_words}, there exists a logical formula $\bform{\goodprefixes}$ such that $\goodprefixes = \{w \in \dat^n \mid w \models \bform{\goodprefixes}\}$.
  Then, by letting $\vfvar = \exi{\vdvar_0}{\pos{\star = \vdvar_0}{\exi{\vdvar_1}{\pos{\star = \vdvar_1}{\dots \exi{\vdvar_{n-1}}{\pos{\star = \vdvar_{n-1} \cand \bform{\goodprefixes}(\vdvar_0, \dots, \vdvar_{n-1})}{\tru}}}}}}$, we get $\vstrc = \sem{\vfvar}$.



  Now, by definition of \hmld{}, \ref{prop:CM_HML:itm:bool_expr} $\Rightarrow$ \ref{prop:CM_HML:itm:HML}.

  Then, a routine induction on the height of the syntactic tree of a formula in \hmld{} establishes that for all formulae $\vfvar$ of height $n \geq 0$, the discriminating prefixes of $\sem{\vfvar}$ are bounded by $n$, so \ref{prop:CM_HML:itm:HML} $\Rightarrow$ \ref{prop:CM_HML:itm:bounded_pref}. Thus, \ref{prop:CM_HML:itm:HML} $\Rightarrow$ \ref{prop:CM_HML:itm:bounded_pref} $\Rightarrow$ \ref{prop:CM_HML:itm:good_pref} $\Rightarrow$ \ref{prop:CM_HML:itm:complete_mon}, and the full cycle is closed.
\end{proof}
\begin{remark}
The above proposition yields a procedure to synthesise a monitor from a property, as soon as one is able to compute the maximum length of the discriminating prefixes and the corresponding $n$ along with the formula $\vbexp$ in item~\ref{prop:CM_HML:itm:bool_expr}. Indeed, the monitor reads the first $n$ values, stores them and evaluates the formula $\vbexp$ on them. This necessitates a very weak computing capacity: $n$ (immutable) variables, no loops and no non-determinism. And indeed, having bounded discriminating prefixes is a very strong property, which drastically limits expressiveness; in particular, properties cannot be recursive.
\end{remark}

\subsection{Parallel Monitors}
\label{app:missing_proofs_parallel_monitors}

In our proofs, we need a few technical lemmata that summarise the behaviour of parallel monitors.
First, as expected, sums and products of monitors can be decomposed and recomposed:
\begin{restatable}{lemma}{parallelMonitorsIff}
  \label{lem:parallel_monitors_iff}
  Let $\vcon_1, \vcon_2, \vvcon_1, \vvcon_2$ be configurations, $\vftrc$ be a finite trace and $\parany \in \{\parsum, \parprod\}$. There exists a derivation $\vcon_1 \parany \vcon_2 \xRightarrow{\vftrc} \vvcon_1 \parany \vvcon_2$ if and only if there exists derivations $\vcon_1 \xRightarrow{\vftrc} \vvcon_1$ and $\vcon_2 \xRightarrow{\vftrc} \vvcon_2$.
\end{restatable}
\begin{proof}
  \begin{description}
  \item[$\Rightarrow$:] We proceed by induction on the length $n$ of the derivation. We prove a slightly stronger result, which will prove useful for the following lemmata. Namely, if there exists a derivation $\vcon_1 \parany \vcon_2 \xRightarrow{\vftrc} \vvcon_1 \parany \vvcon_2$ \emph{of length $n$}, then there exists derivations $\vcon_1 \xRightarrow{\vftrc} \vvcon_1$ and $\vcon_2 \xRightarrow{\vftrc} \vvcon_2$ \emph{of length at most $n$}.
    \begin{itemize}
    \item For $n = 1$, three rules apply: $\msyn$, $\masyl$, $\masyr$. All cases are immediate.

    \item Now, let $n \geq 1$, and assume the result holds for all derivations of length up to $n$. Consider a derivation $\vcon_1 \parany \vcon_2 \xRightarrow{\vftrc} \vvcon_1 \parany \vvcon_2$ (with the notations of the lemma statement) of length $n+1$. Let $\vvvcon \in \monConfs$, $\mu \in \dat \cup \{\tau\}$, $\vvftrc \in \ftrc$ and rule $\vrul$ be such that $\vcon_1 \parany \vcon_2 \xrightarrow[\vrul]{\mu} \vvvcon \xRightarrow{\vvftrc} \vvcon_1 \parany \vvcon_2$, with $\mu \vvftrc = \vftrc$. The derivation associated to $\vvftrc$ is of length $n$. Besides, necessarily, $\vrul \in \{\msyn, \masyl, \masyr\}$. We distinguish cases:
    \begin{itemize}
    \item $\vrul = \msyn$. Then, $\mu = \vdat$ and $\vvvcon = \vvvcon_1 \parany \vvvcon_2$, with $\vcon_1 \xrightarrow{\vdat} \vvvcon_1$ and $\vcon_2 \xrightarrow{\vdat} \vvvcon_2$. By the induction hypothesis, we have two derivations $\vvvcon_1 \xRightarrow{\vvftrc} \vvcon_1$ and $\vvvcon_2 \xRightarrow{\vvftrc} \vvcon_2$, each of length at most $n$. Hence, we get that there exists two derivations $\vcon_1 \xrightarrow{\vdat} \vvvcon_1 \xRightarrow{\vvftrc} \vvcon_1$ and $\vcon_2 \xrightarrow{\vdat} \vvvcon_2 \xRightarrow{\vvftrc} \vvcon_2$, each length at most $n+1$. This is the required result since $\vftrc = \vdat \vvftrc$.
    \item $\vrul = \masyl$. Then, $\mu = \tau$ and $\vvvcon = \vvvcon_1 \parany \vcon_2$ with $\vcon_1 \xrightarrow{\tau} \vvvcon_1$. By the induction hypothesis, we have two derivations $\vvvcon_1 \xRightarrow{\vvftrc} \vvcon_1$ and $\vcon_2 \xRightarrow{\vvftrc} \vvcon_2$, each of length at most $n$. As a consequence, there exists a derivation $\vcon_1 \xrightarrow{\tau} \vvvvcon_1 \xRightarrow{\vftrc} \vvcon_1$ of length at most $n+1$ and a derivation $\vcon_2 \xRightarrow{\vftrc} \vvcon_2$ of length at most $n \leq n+1$, as expected.
    \item $\vrul = \masyr$. This case is symmetric to the above one.
    \end{itemize}
    \end{itemize}
  \item[$\Leftarrow$:] We now proceed by induction on the sum $s$ of the lengths of the two derivations.

    \begin{itemize}
    \item When $s=0$, the result trivially holds.
    \item Now, assume there exists a derivation $\vcon_1 \xRightarrow{\vftrc} \vvcon_1$ of length $l \geq 0$ and a derivation $\vcon_2 \xRightarrow{\vftrc} \vvcon_2$ of length $m \geq 0$ such that $s = l + m > 0$. At least one of them is of length at least one. We assume this is the case for the first one, the other case is symmetric. Let $\vvvcon_1 \in \monConfs$, $\mu \in \dat \cup \{\tau\}$ and $\vvftrc \in \ftrc$ be such that $\vcon_1 \xrightarrow{\mu} \vvvcon_1 \xRightarrow{\vvftrc} \vvcon_1$, with $\vftrc = \mu \vvftrc$. There are two cases:
      \begin{itemize}
      \item $\mu = \tau$. Then, $\vvftrc = \vftrc$, $\vcon_1 \xrightarrow{\tau} \vvvcon_1$ and the derivation $\vvvcon_1 \xRightarrow{\vftrc} \vvcon_1$ is of length $l - 1$.  By applying rule $\masyl$, we get that $\vcon_1 \parany \vcon_2 \xrightarrow{\tau} \vvvcon_1 \parany \vcon_2$. Besides, by the induction hypothesis, since $l-1 + m < s$, we have that there exists a derivation $\vvvcon_1 \parany \vcon_2 \xRightarrow{\vftrc} \vvcon_1 \parany \vvcon_2$. Overall, $\vcon_1 \parany \vcon_2 \xrightarrow{\tau} \vvvcon_1 \parany \vcon_2 \xRightarrow{\vftrc} \vvcon_1 \parany \vvcon_2$, which is the required result.
      \item $\mu = \vdat \in \dat$. Then, $\vftrc = \vdat \vvftrc$, $\vcon_1 \xrightarrow{\vdat} \vvvcon_1$ and the derivation $\vvvcon_1 \xRightarrow{\vvftrc} \vvcon_1$ is of length $l - 1$. Now, the derivation $\vcon_2 \xRightarrow{\vftrc} \vvcon_2$ cannot be empty since $\vftrc \neq \varepsilon$. There are two cases:
        \begin{itemize}
        \item $\vcon_2 \xrightarrow{\vdat} \vvvcon_2 \xRightarrow{\vvftrc} \vvcon_2$. Following rule $\msyn$, we get that $\vcon_1 \parany \vcon_2 \xrightarrow{\vdat} \vvvcon_1 \parany \vvvcon_2$. The sum of the length of the derivations from $\vvvcon_1$ and $\vvvcon_2$ is $l-1 + m-1 < s$, so by the induction hypothesis, we get that $\vvvcon_1 \parany \vvvcon_2 \xRightarrow{\vvftrc} \vvcon_1 \parany \vvcon_2$. As a consequence, $\vcon_1 \parany \vcon_2 \xrightarrow{\vdat} \vvvcon_1 \parany \vvvcon_2 \xRightarrow{\vvftrc} \vvcon_1 \parany \vvcon_2$, as expected.
        \item $\vcon_2 \xrightarrow{\tau} \vvvcon_2 \xRightarrow{\vftrc} \vvcon_2$. By rule $\masyr$, $\vcon_1 \parany \vcon_2 \xrightarrow{\tau} \vcon_1 \parany \vvvcon_2$. The sum of the length of derivations $\vcon_1 \xRightarrow{\vftrc} \vvcon_1$ and $\vvvcon_2 \xRightarrow{\vtrc} \vvcon_2$ is $l + m-1 < s$, so by the induction hypothesis, $\vcon_1 \parany \vvvcon_2 \xRightarrow{\vtrc} \vvcon_1 \parany \vvcon_2$. Overall, $\vcon_1 \parany \vcon_2 \xrightarrow{\tau} \vcon_1 \parany \vvvcon_2 \xRightarrow{\vftrc} \vvcon_1 \parany \vvcon_2$, the expected result.
  \qedhere
        \end{itemize}
      \end{itemize}
    \end{itemize}
  \end{description}
\end{proof}

As a consequence, we have:
\begin{restatable}{lemma}{parallelMonitorsParsumYesDecompose}
  \label{lem:parallel_monitors_parsum_yes_decompose}
  Let $\vcon_1, \vcon_2, \vvcon_1, \vvcon_2$ be configurations and $\vftrc$ be a finite trace. If there exists a derivation $\vcon_1 \parsum \vcon_2 \xRightarrow{\vftrc} \under{\yes}{\vdenv}$ (for some $\vdenv \in \denv$) of length $n$, then either there exists a derivation $\vcon_1 \xRightarrow{\vftrc} \under{\yes}{\vvdenv}$ or a derivation $\vcon_2 \xRightarrow{\vftrc} \under{\yes}{\vvvdenv}$ (for some $\vvdenv, \vvvdenv \in \denv$), in both cases of length smaller than $n$.
  \end{restatable}
  \begin{proof}
    Necessarily, the derivation is of the form $\vcon_1 \parsum \vcon_2 \xRightarrow{\vvftrc} \under{\yes}{\vdenv} \parsum \vvcon_2 \xrightarrow[\mvrdb]{\tau} \under{\yes}{\vdenv} \xRightarrow[\mvrd]{\vvvftrc} \under{\yes}{\vvdenv}$, with $\vvftrc \cdot \vvvftrc = \vftrc$ (note that $\vvvftrc$ can be the empty word), or symmetrically if we instead reach $\vvcon_1 \parsum \under{\yes}{\vvvdenv}$ after reading $\vvftrc$, and apply the symmetric rule of $\mvrdb$. We treat the first case, the other one is symmetric.

    By applying \Cref{lem:parallel_monitors_iff} to the derivation $\vcon_1 \parsum \vcon_2 \xRightarrow{\vvftrc} \under{\yes}{\vdenv} \parsum \vvcon_2$ we obtain that there exists a derivation $\vcon_1 \xRightarrow{\vvftrc} \under{\yes}{\vdenv}$, so $\vcon_1 \xRightarrow{\vvftrc} \under{\yes}{\vdenv} \xRightarrow[\mvrd]{\vvvftrc} \under{\yes}{\vvdenv}$ by applying $\mvrd$ as above. Since the derivation contains at most the same rules as the initial one, minus the application of $\mvrdb$, it is of length $<n$.
  \end{proof}
  Besides, we can show that:
\begin{restatable}{lemma}{parallelMonitorsParprodYesDecompose}
  \label{lem:parallel_monitors_parprod_yes_decompose}
  Let $\vcon_1, \vcon_2, \vvcon_1, \vvcon_2$ be configurations and $\vftrc$ be a finite trace. If there exists a derivation $\vcon_1 \parprod \vcon_2 \xRightarrow{\vftrc} \under{\yes}{\vdenv}$ (for some $\vdenv \in \denv$) of length $n$, then there exists derivations $\vcon_1 \xRightarrow{\vftrc} \under{\yes}{\vvdenv}$ and $\vcon_2 \xRightarrow{\vftrc} \under{\yes}{\vvvdenv}$ (for some $\vvdenv, \vvvdenv \in \denv$), each of length $< n$.
  \end{restatable}
  \begin{proof}
    Necessarily, the derivation is of the form $\vcon_1 \parprod \vcon_2 \xRightarrow{\vvftrc} \under{\yes}{\vdenv} \parprod \vvcon_2 \xrightarrow[\mvrca]{\tau} \vvcon_2 \xRightarrow[\mvrd]{\vvvftrc} \under{\yes}{\vvdenv}$, with $\vvftrc \cdot \vvvftrc = \vftrc$ (note that $\vvvftrc$ can be the empty word), or symmetrically if we instead reach $\vvcon_1 \parprod \under{\yes}{\vvvdenv}$ after reading $\vvftrc$, and apply the symmetric rule of $\mvrca$. We treat the first case, the other one is symmetric.

    By applying \Cref{lem:parallel_monitors_iff} to the derivation $\vcon_1 \parprod \vcon_2 \xRightarrow{\vvftrc} \under{\yes}{\vdenv} \parprod \vvcon_2$ we obtain that there exists a derivation $\vcon_1 \xRightarrow{\vvftrc} \under{\yes}{\vdenv}$ and a derivation $\vcon_2 \xRightarrow{\vvftrc} \vvcon_2$, each of length at most $n$. So, on the one hand, $\vcon_1 \xRightarrow{\vvftrc} \under{\yes}{\vdenv} \xRightarrow[\mvrca]{\vvvftrc} \under{\yes}{\vvdenv}$ by applying $\mvrca$ as above.  On the other hand, $\vcon_2 \xRightarrow{\vvftrc} \vvcon_2 \xRightarrow{\vvvftrc} \under{\yes}{\vvdenv}$. Since both derivation contains at most the same rules as the initial one, minus the application of $\mvrca$, it is of length $<n$.
  \end{proof}
\begin{restatable}{lemma}{parallelMonitorsParprodYesCompose}
  \label{lem:parallel_monitors_parprod_yes_compose}
  Let $\vcon_1, \vcon_2, \vvcon_1, \vvcon_2$ be configurations and $\vftrc$ be a finite trace. If there exists derivations $\vcon_1 \xRightarrow{\vftrc} \under{\yes}{\vvdenv}$ and $\vcon_2 \xRightarrow{\vftrc} \under{\yes}{\vvvdenv}$ (for some $\vvdenv, \vvvdenv \in \denv$), then there exists a derivation $\vcon_1 \parprod \vcon_2 \xRightarrow{\vftrc} \under{\yes}{\vdenv}$ (for some $\vdenv \in \denv$).
  \end{restatable}
  \begin{proof}
    By \Cref{lem:parallel_monitors_iff}, with the hypotheses and notations of the statement, we have that $\vcon_1 \parprod \vcon_2 \xRightarrow{\vftrc} \under{\yes}{\vdenv} \parprod \under{\yes}{\vvdenv}$ (for some $\vdenv, \vvdenv \in \denv$). By applying rule $\mvrca$, $\under{\yes}{\vdenv} \parprod \under{\yes}{\vvdenv} \xrightarrow[\mvrca]{\tau} \under{\yes}{\vdenv}$.
  \end{proof}

  Finally, we can show that the converse of \Cref{lem:parallel_monitors_parsum_yes_decompose} also holds. To that end, we first show that our monitors are reactive (in the sense of~\cite[Definition~3.4]{DBLP:journals/pacmpl/AcetoAFIL19}). The proof follows the same lines as~\cite[Proposition~4.14]{DBLP:journals/pacmpl/AcetoAFIL19}.
  \begin{lemma}
    \label{lem:monitors_reactive}
    Let $\vcon \in \monConfs$ be a monitor such that each occurrence of a recursion variable $\vpvar$ is preceded by $(\vbexp).$ and $\vdat \in \dat$. There exists $\vvcon \in \monConfs$ such that $\vcon \xRightarrow{\vdat} \vvcon$.
    \end{lemma}
    As a consequence:
\begin{restatable}{lemma}{parallelMonitorsParsumYesCompose}
  \label{lem:parallel_monitors_parsum_yes_compose}
  Let $\vcon_1, \vcon_2, \vvcon_1, \vvcon_2$ be configurations and $\vftrc$ be a finite trace. If there exists a derivation $\vcon_1 \xRightarrow{\vftrc} \under{\yes}{\vvdenv}$ or a derivation $\vcon_2 \xRightarrow{\vftrc} \under{\yes}{\vvvdenv}$ (for some $\vvdenv, \vvvdenv \in \denv$), then there exists a derivation $\vcon_1 \parsum \vcon_2 \xRightarrow{\vftrc} \under{\yes}{\vdenv}$ (for some $\vdenv \in \denv$).
  \end{restatable}
\begin{proof}
  Assume that there exists a derivation $\vcon_1 \xRightarrow{\vftrc} \under{\yes}{\vdenv}$ for some $\vcon_1, \vcon_2, \vvcon_1, \vvcon_2 \in \monConfs$ and $\vftrc \in \ftrc$, the other case is symmetric. Let us show by induction on the length $n$ of the derivation then there exists a derivation $\vcon_1 \parsum \vcon_2 \xRightarrow{\vftrc} \under{\yes}{\vdenv}$ (for some $\vdenv \in \denv$).

  For $n = 0$, we get $\vcon_1 = \under{\yes}{\vdenv}$ and $\vftrc = \varepsilon$. By applying rule $\mvrdb$, we get that $\under{\yes}{\vdenv} \parsum \vcon_2 \xrightarrow{\tau} \under{\yes}{\vdenv}$, i.e. $\vcon_1 \parsum \vcon_2 \xRightarrow{\vftrc} \under{\yes}{\vdenv}$.

  Now, let $n \geq 0$, and assume the result holds for all derivations of length $n$. Consider a derivation $\vcon_1 \xrightarrow{\mu} \vvvcon_1 \xRightarrow{\vvftrc} \under{\yes}{\vdenv}$. There are two cases:
  \begin{itemize}
  \item $\mu = \tau$. Then, $\vcon_1 \xrightarrow{\tau} \vvvcon_1$ and $\vvvcon_1 \xRightarrow{\vftrc} \under{\yes}{\vdenv}$. By the induction hypothesis $\vvvcon_1 \parsum \vcon_2 \xRightarrow{\vftrc} \under{\yes}{\vdenv}$. Overall, $\vcon_1 \parsum \vcon_2 \xrightarrow[\masyl]{\tau} \vvvcon_1 \parsum \vcon_2 \xRightarrow{\vftrc} \under{\yes}{\vdenv}$, which is the expected result.
  \item $\mu = \vdat$. Then, $\vcon_1 \xrightarrow{\vdat} \vvvcon_1$ and $\vvvcon_1 \xRightarrow{\vvftrc} \under{\yes}{\vdenv}$. By \Cref{lem:monitors_reactive}, we know that there exists $\vvvcon_2$ such that $\vcon_2 \xRightarrow{\vdat} \vvvcon_2$. Applying rule $\msyn$, we get $\vcon_1 \parsum \vcon_2 \xrightarrow{\vdat} \vvvcon_1 \parsum \vvvcon_2$. By the induction hypothesis, since $\vvvcon_1 \xRightarrow{\vvftrc} \under{\yes}{\vdenv}$, we have that $\vvvcon_1 \parsum \vvvcon_2 \xRightarrow{\vvftrc} \under{\yes}{\vdenv}$. Overall, $\vcon_1 \parsum \vcon_2 \xrightarrow{\vdat} \vvvcon_1 \parsum \vvvcon_2 \xRightarrow{\vvftrc} \under{\yes}{\vdenv}$, which is the sought result.
    \qedhere
  \end{itemize}
\end{proof}

Summing up, we have:
\begin{restatable}{proposition}{parallelSumAcc}
  \label{prop:parallel_sum_acc}
  For all monitors $\vmon, \vvmon$, all $\vdenv \in \denv$ and all traces $\vtrc \in \trc$: $\acc(\vmon \parsum \vvmon, \vdenv, \vtrc)$ if and only if $\acc(\vmon, \vdenv, \vtrc)$ or $\acc(\vvmon, \vdenv, \vtrc)$.
\end{restatable}

\begin{restatable}{proposition}{parallelProdAcc}
  \label{prop:parallel_prod_acc}
  For all monitors $\vmon, \vvmon$, all $\vdenv \in \denv$ and all traces $\vtrc \in \trc$: $\acc(\vmon \parprod \vvmon, \vdenv, \vtrc)$ if and only if $\acc(\vmon, \vdenv, \vtrc)$ and $\acc(\vvmon, \vdenv, \vtrc)$.
\end{restatable}

\subsection{Detailed Execution of \Cref{ex:token_server}}
\label{app:ex:token_server}
Now, the system emits $1$, and following rule $\mact$ we get:
\begin{equation*}
  \xrightarrow[\mact]{1}{}
    \under{
        \rec{\vpvar}{\left(
          \prf{(\star=\vdvar)}{\yes}
          \parsum
          \prf{(\star\neq\vdvar)}{\vpvar}
      \right)
      }
  }{\vdvar\mapsto 1}
\end{equation*}
We apply rule $\mrec$, then the monitor forks into two parallel components using $\mfork$:
\begin{align*}
  \xrightarrow[\mrec, \mfork]{\tau}{} &
    \big(\under{
          \prf{(\star=\vdvar)}{\yes}
  }{\vdvar \mapsto 1})
  \parsum (\under{
          \prf{(\star\neq\vdvar)}{ \\
        & \rec{\vpvar}{\left(
          \prf{(\star=\vdvar)}{\yes}
          \parsum
          \prf{(\star\neq\vdvar)}{\vpvar}
      \big)
      }
          }
  }{\vdvar\mapsto 1}\right)
\end{align*}
Now, there are two submonitors that evolve in parallel, each with a local copy of $\vdenv = \vdvar \mapsto 1$ and the system emits $0$. Following rule $\mnact$ (since the guard $\star = \vvdvar$ is violated), the monitor on the left reaches an inconclusive verdict $\inc$. The one on the right follows rule $\mact$ and we get:
\begin{align*}
  \xrightarrow[\mnact, \mact, \msyn]{0}{}&
  \left(
    \under{
      \inc
      }{\vdvar\mapsto 1}
    \right)
          \parsum
          \big(
          \under{
        \rec{\vpvar}{\big(
          \prf{(\star=\vdvar)}{\yes}
          \parsum \\
          & \prf{(\star\neq\vdvar)}{\vpvar}
      \big)
      }
  }{\vdvar\mapsto 1}
  \big)
\end{align*}
We again use rule $\mrec$ on the right of the $\parsum$, while the left monitor makes no progress (rule $\masyr$):
\begin{align*}
  \xrightarrow[\mrec, \masyr]{\tau}{}&
    \left(\under{
          \inc
        }{\vdvar\mapsto 1}\right)
          \parsum
          \big(
          \big(
          \prf{(\star=\vdvar)}{\yes}
          \parsum
          \prf{(\star\neq\vdvar)}{ \\
        &\rec{\vpvar}{\left(
          \prf{(\star=\vdvar)}{\yes}
          \parsum
          \prf{(\star\neq\vdvar)}{\vpvar}
      \right)
      }
    }
  \big)
  ,\vdvar\mapsto 1
  \big)
\end{align*}
Now, the system emits $1$. Following rule $\mvrd$ on the left monitor, $\mact$ on the middle one and $\mnact$ on the right one, combining them using $\msyn$ twice, we get:
\begin{align*}
  \xrightarrow[\mvrd, \mact, \mnact, \msyn, \msyn]{1}{}
  \left(
    \under{
      \inc
      }{\vdvar \mapsto 1}
  \right)
          \parsum
          \big(
            \left(
          \under{
          \yes
          }{\vdvar \mapsto 1}
        \right)
          \parsum \\
          \left(
          \under{
          \inc
          }{\vdvar \mapsto 1}
        \right)
        \big)
\end{align*}
Finally, by applying $\mvrdb$ and its symmetric, we get:
\begin{equation*}
  \xrightarrow[\mvrdb, \mvrdb']{\tau}{}
    \under{
      \yes
  }{\vvdvar\mapsto 1}
\end{equation*}

\subsection{Missing proofs of \Cref{prop:correctness_synthesis_chmld}}
In this section, we prove the correctness of the synthesis procedure of \Cref{fig:monitors_guessing} on page~\pageref{fig:monitors_guessing}, i.e.:
\begin{restatable}{proposition}{correctnessSynthesiscHMLd}
  \label{prop:correctness_synthesis_chmld}
  For all $\vfvar \in \chmld$, $\syn{\vfvar}$ is a sound and satisfaction-complete monitor for $\vfvar$.
\end{restatable}
\subsection{Soundness}
\label{sec:proof:sec:soundness}
Recall that \chmld{} is defined in \Cref{fig:rechml_LT} on page~\pageref{fig:rechml_LT}. We show the following:
\begin{restatable}{lemma}{synSoundness}
  \label{lem:syn_soundness}
  For all $\vfvar \in \chmld$ and all $\vdenv \in \denv$, if $\acc(\syn{\vfvar}, \vdenv, \vtrc)$, then $\vtrc \in \sem{\vfvar, \vdenv}$.
\end{restatable}
\begin{proof}
By unpacking the definition, we have that $\acc(\syn{\vfvar}, \vdenv, \vtrc)$ whenever there exists a finite trace $\vftrc \prec \vtrc$ such that $\syn{\vfvar}, \vdenv \xRightarrow{\vftrc} \yes, \vvdenv$ for some $\vvdenv \in \denv$.

We prove the result by complete induction on the length $n$ of the derivation $\syn{\vfvar}, \vdenv \xRightarrow{\vftrc} \yes, \vvdenv$.
\begin{itemize}
\item For $n = 0$, necessarily $\syn{\vfvar} = \yes$, which is the case only when $\vfvar = \tru$ (see \Cref{fig:monitors_guessing}). Then, $\vtrc \in \sem{\tru, \vdenv} = \trc$.
\item Now, assume the result holds for all derivations of length $n \geq 0$, and let $\vfvar \in \chmld$, $\vdenv, \vvdenv, \vvvdenv \in \denv$ such that $\syn{\vfvar}, \vdenv \xrightarrow{\mu} \vcon \xRightarrow{\vvftrc} \yes, \vvdenv$, where $\mu \in \dat \cup \{\tau\}$ is such that $\mu \vvftrc = \vftrc$, $\vcon \in \monConfs$ and the derivation $\vcon \xRightarrow{\vftrc} \yes, \vvdenv$ is of length $n$. We distinguish cases based on the shape of $\vfvar \in \chmld{}$:
  \begin{itemize}
    \item $\vfvar = \tru$: this case has been treated above.
    \item $\vfvar = \pos{\vbexp}{\vvfvar}$. Then, $\syn{\vfvar} = \prf{(\vbexp)}{\syn{\vvfvar}}$. The first derivation $\syn{\vfvar}, \vdenv \xrightarrow{\mu} \vcon$ can be of two shapes, depending on the rule used:
      \begin{itemize}
      \item Rule $\mact$. Then, there is some $\vdat \in \dat$ such that $\mu = \vdat$, $\evalrel{\vbexp\vdenv[\star\mapsto\vdat]}{\ltru}$ and we have $\syn{\vfvar}, \vdenv \xrightarrow[\mact]{\vdat} \syn{\vvfvar}, \vdenv \xRightarrow{\vvftrc} \yes, \vvdenv$, with $\vdat \vvftrc = \vftrc$. As $\vftrc \prec \vtrc$, we can write $\vtrc = \vftrc \vvtrc = \vdat \vvftrc \vvtrc$. By the induction hypothesis, since $\acc(\syn{\vvfvar}, \vdenv, \vvftrc \vvtrc)$, we know that $\vvftrc \vvtrc \in \sem{\vvfvar, \vdenv}$. As a consequence, since $\evalrel{\vbexp\vdenv[\star\mapsto\vdat]}{\ltru}$, $\vdat \vvftrc \vvtrc \in \sem{\pos{\vbexp}{\vvfvar}, \vdenv}$,
        i.e. $\vtrc \in \sem{\vfvar, \vdenv}$.
      \item $\mnact$. Then, $\mu = \vdat$, $\evalrel{\vbexp\vdenv[\star\mapsto\vdat]}{\lfls}$ and we have $\syn{\vfvar}, \vdenv \xrightarrow[\mnact]{\vdat} \inc, \vdenv$. From this point, the only rule that applies is $\mvrd$ so we cannot have $\inc, \vdenv \xRightarrow{\vvftrc} \yes, \vvdenv$ for any $\vvdenv \in \denv$, and the result vacuously holds.
      \end{itemize}
      \item $\vfvar = \exi{\binder{\vdvar}}{\vvfvar}$. Then, $\syn{\vfvar} = \prf{\guess{\vdvar}}{\syn{\vvfvar}}$. Only rule $\mgs$ can be used, and we have $\syn{\vfvar}, \vdenv \xrightarrow[\mgs]{\tau} \syn{\vvfvar}, \vdenv[\vdvar \mapsto \vdat] \xRightarrow{\vvftrc} \yes, \vvdenv$. This means that $\acc(\syn{\vvfvar}, \vdenv\map{\vdvar \mapsto \vdat}, \vtrc)$. By the induction hypothesis, we get that $\vtrc \in \sem{\vvfvar, \vdenv\map{\vdvar \mapsto \vdat}} \subseteq \sem{\exi{\binder{\vdvar}}{\vvfvar}, \vdenv}$, so $\vtrc \in \sem{\vfvar, \vdenv}$.
      \item $\vfvar = \vvfvar \lor \vvvfvar$. Then, $\syn{\vfvar} = \syn{\vvfvar} \parsum \syn{\vvvfvar}$ and we have a derivation $\under{\syn{\vvfvar} \parsum \syn{\vvvfvar}}{\vdenv} \xRightarrow{\vftrc} \under{\yes}{\vvdenv}$. Necessarily, the first rule that applies is $\mfork$ and we get $\under{\syn{\vvfvar} \parsum \syn{\vvvfvar}}{\vdenv} \xRightarrow[\mfork]{\tau} \under{\syn{\vvfvar}}{\vdenv} \parsum \under{\syn{\vvvfvar}}{\vdenv} \xRightarrow{\vftrc} \under{\yes}{\vvdenv}$.
        By \Cref{lem:parallel_monitors_parsum_yes_decompose}, this implies that either $\under{\syn{\vvfvar}}{\vdenv} \xRightarrow{\vftrc} \under{\yes}{\vvdenv}$ or $\under{\syn{\vvvfvar}}{\vdenv} \xRightarrow{\vftrc} \under{\yes}{\vvdenv}$, in both cases with a derivation of length at most $n$. We assume the former, the latter is symmetric. Then, by the induction hypothesis, this means that $\vtrc \in \sem{\vvfvar, \vdenv}$, so $\vtrc \in \sem{\vfvar, \vdenv} = \sem{\vvfvar, \vdenv} \cup \sem{\vvvfvar, \vdenv}$.
      \item $\vfvar = \vvfvar \land \vvvfvar$. Then, $\syn{\vfvar} = \syn{\vvfvar} \parprod \syn{\vvvfvar}$ and we have a derivation $\under{\syn{\vvfvar} \parprod \syn{\vvvfvar}}{\vdenv} \xRightarrow{\vftrc} \under{\yes}{\vvdenv}$. Again, the first rule that applies is necessarily $\mfork$ and we get $\under{\syn{\vvfvar} \parprod \syn{\vvvfvar}}{\vdenv} \xRightarrow[\mfork]{\tau} \under{\syn{\vvfvar}}{\vdenv} \parprod \under{\syn{\vvvfvar}}{\vdenv} \xRightarrow{\vftrc} \under{\yes}{\vvdenv}$.
        By \Cref{lem:parallel_monitors_parprod_yes_decompose}, this implies that $\under{\syn{\vvfvar}}{\vdenv} \xRightarrow{\vftrc} \under{\yes}{\vvdenv}$ and $\under{\syn{\vvvfvar}}{\vdenv} \xRightarrow{\vftrc} \under{\yes}{\vvdenv}$, in both cases with a derivation of length at most $n$. By the induction hypothesis, this means that $\vtrc \in \sem{\vvfvar, \vdenv}$ and $\vtrc \in \sem{\vvvfvar, \vdenv}$,  so $\vtrc \in \sem{\vfvar, \vdenv} = \sem{\vvfvar, \vdenv} \cap \sem{\vvvfvar, \vdenv}$.
      \item $\vfvar = \recmin{\vpvar}{\vvfvar}$ Then, $\syn{\vfvar} = \rec{\vpvar}{\syn{\vvfvar}}$. Thus, the transition sequence is necessarily $\under{\rec{\vpvar}{\syn{\vvfvar}}}{\vdenv} \xrightarrow{\tau} \under{\syn{\vvfvar}\map{\frac{\rec{\vrvar}{\syn{\vvfvar}}}{\vrvar}}}{\vdenv} \xRightarrow{\vftrc} \under{\yes}{\vvdenv}$. Now, observe that, since the synthesis procedure is compositional, $\syn{\vvfvar}\map{\frac{\rec{\vrvar}{\syn{\vvfvar}}}{\vrvar}} = \syn{\vvfvar\map{\frac{\recmin{\vrvar}{\vvfvar}}{\vrvar}}}$. By the induction hypothesis, we get $\vtrc \in \sem{\vvfvar\map{\frac{\recmin{\vrvar}{\syn{\vvfvar}}}{\vrvar}}, \vdenv}$ and, since $\recmin{\vpvar}{\vvfvar}$ satisfies the fixed-point equation $\sem{\recmin{\vpvar}{\vvfvar}, \vdenv} = \sem{\vvfvar\map{\frac{\recmin{\vpvar}{\vvfvar}}{\vrvar}}, \vdenv}$, the required result follows.
        \qedhere
  \end{itemize}
\end{itemize}
\end{proof}
As a consequence:
\begin{proposition}
  \label{prop:soundness_cHMLd}
  For all $\vfvar \in \chmld$, $\syn{\vfvar}$ is a monitor for $\sem{\vfvar}$ that is sound for satisfactions.
\end{proposition}
The proof that we can dually generate monitors that are sound for violations for $\shmld$ is dual.

\subsection{Partial Completeness}
\label{sec:proof:sec:partial_completeness}
\begin{definition}[Closure]
Given a closed formula $\vfvar$, we define for each of its subformulae $\vvfvar \in \subform{\vfvar}$ the \emph{closure} of $\vvfvar$ \emph{within} $\vfvar$, denoted $\cl[\vfvar]{\vvfvar}$ (we omit to mention $\vfvar$ when it is clear from the context) as follows: if $\vvfvar$ is closed, we let $\cl[\vfvar]{\vvfvar} = \vvfvar$, otherwise we pick (according to some fixed arbitrary order on recursion variables) an $\vpvar$ among the $\leq$-minimal variables and we let $\cl[\vfvar]{\vvfvar} = \cl[\vfvar]{\vvfvar\map{\frac{\fx[\vfvar]{\vpvar}}{\vpvar}}}$, where $\vvfvar\map{\frac{\vvvfvar}{\vpvar}}$ is the usual substitution operation.
\end{definition}
\begin{restatable}{proposition}{partialCompletenesscHMLd}
  \label{prop:partial_compl_cHMLd}
  For all $\vfvar \in \chmld$, all $\vdenv \in \denv$ and all $\vtrc \in \trc$, if $\vtrc \in \sem{\vfvar, \vdenv}$ then $\acc(\syn{\vfvar}, \vdenv, \vtrc)$.
\end{restatable}
\begin{proof}
Let $\vfvar$ be a \chmld{} formula, $\vdenv \in \denv$ be a data valuation and $\vtrc \in \trc$ be a trace. Assume that $\vfvar, \vdenv$ have an annotation on $\vtrc$, that we call $A$. By \Cref{prop:annotation-chmld}, we can assume that $A$ is a finite annotation, i.e. it is finite and acyclic (see \Cref{def:annotation} on page~\pageref{def:annotation}). Moreover, up to restricting to the connected component containing the vertex $(\vfvar, \vdenv, \vtrc)$, we assume that $A$ is connected.

Let us show, by induction on the height\footnote{We recall that the height $h(v)$ of a vertex $v$ is $h(v) = 0$ if $v$ has no outgoing edges; otherwise, $h(v) = 1 + \max \{h(v') \mid v \dep{} v'\}$. This is well-defined in any directed acyclic graph.} of a vertex, that for each vertex $(\vvfvar, \vvdenv, \vvtrc)$ of $A$, we have $\acc(\syn{\cl[\vfvar]{\vvfvar}}, \vvdenv, \vvtrc)$.
\begin{itemize}
\item 
  If $v = (\vvfvar, \vvdenv, \vvtrc)$ is of height $0$, then by definition of an annotation, $\vvfvar = \tru$ otherwise $v$ has at least one outgoing edge. Thus, $\syn{\cl{\vvfvar}} = \yes$ and $\acc(\syn{\cl{\vvfvar}}, \vvdenv, \vvtrc)$.
\item
  Now, let $v = (\vvfvar, \vvdenv, \vvtrc)$ be a vertex of $A$ of height $h \geq 1$. We distinguish cases based on the shape of $\vvfvar$:
  \begin{itemize}
  \item If $\vvfvar = \tru$, then, by the same reasoning as above, we have $\acc(\syn{\cl{\vvfvar}}, \vvdenv, \vvtrc)$.

  \item The case $\vvfvar = \fls$ is vacuous by definition of an annotation.

  \item If $\vvfvar = \pos{\vbexp}{\vvvfvar}$, then $v = (\pos{\vbexp}{\vvvfvar}, \vvdenv, \vdat \vvvtrc)$ for some $\vdat$ such that $\evalrel{\vbexp\vvdenv[\star\mapsto\vdat]}{\ltru}$, $(\vvvfvar, \vvdenv, \vvvtrc) \in A$ and $v \dep{} (\vvvfvar, \vvdenv, \vvvtrc) \in A$. Necessarily, $(\vvvfvar, \vvdenv, \vvvtrc)$ is of height at most $h-1$, so by the induction hypothesis we get that $\acc(\syn{\cl{\vvvfvar}}, \vvdenv, \vvvtrc)$. Since $\syn{\cl{\vvfvar}} = \prf{(\vbexp)}{\syn{\cl{\vvvfvar}}}$, by rule $\mact$, we have $\under{\syn{\cl{\vvfvar}}}{\vvdenv} \xrightarrow[\mact]{\vdat} \under{\syn{\cl{\vvvfvar}}}{\vvdenv}$, so, since $\acc(\syn{\cl{\vvvfvar}}, \vvdenv, \vvvtrc)$, we get that $\acc(\syn{\cl{\vvfvar}}, \vvdenv, \vdat \vvvtrc)$.

  \item If $\vvfvar = \exi{\binder{\vdvar}}{\vvvfvar}$, then, since $v = (\exi{\binder{\vdvar}}{\vvvfvar}, \vvdenv, \vvtrc) \in A$, we know that there exists some $\vdat \in \dat$ such that $(\vvvfvar, \vvdenv[\vdvar\mapsto\vdat], \vvtrc) \in A$ and $(\exi{\binder{\vdvar}}{\vvvfvar}, \vvdenv, \vvtrc) \dep{} (\vvvfvar, \vvdenv[\vdvar\mapsto\vdat], \vvtrc)$. Again, the target vertex $(\vvvfvar, \vvdenv[\vdvar\mapsto\vdat], \vvtrc)$ is of height at most $h-1$, so the induction hypothesis yields $\acc(\syn{\cl{\vvvfvar}}, \vvdenv[\vdvar\mapsto\vdat], \vvtrc)$.
    By the definition of $\syn{-}$, $\syn{\cl{\vvfvar}} = \prf{\guess{\vdvar}}{\syn{\cl{\vvvfvar}}}$. By rule $\mgs$, we have that $\under{\syn{\cl{\vvfvar}}}{\vvdenv} \xrightarrow[\mgs]{\tau} \under{\syn{\cl{\vvvfvar}}}{\vvdenv[\star\mapsto\vdat]}$, which means that $\acc(\syn{\cl{\vvfvar}}, \vvdenv, \vvtrc)$ since $\acc(\syn{\cl{\vvvfvar}}, \vvdenv[\vdvar\mapsto\vdat], \vvtrc)$.

  \item If $\vvfvar = \vvvfvar \lor \vvvvfvar$, we know that either $(\vvvfvar, \vvdenv, \vvtrc) \in A$ and $(\vvvfvar \lor \vvvvfvar, \vvdenv, \vvtrc) \dep{} (\vvvfvar, \vvdenv, \vvtrc)$, or symmetrically $(\vvvvfvar, \vvdenv, \vvtrc) \in A$ and $(\vvvfvar \lor \vvvvfvar, \vvdenv, \vvtrc) \dep{} (\vvvvfvar, \vvdenv, \vvtrc)$. We treat the first case; the second is symmetric. By definition, $\syn{\cl{\vvfvar}} = \syn{\cl{\vvvfvar}} \parsum \syn{\cl{\vvvvfvar}}$. Now, by the induction hypothesis, we know that $\acc(\syn{\cl{\vvvfvar}}, \vvdenv, \vvtrc)$. Then, by \Cref{prop:parallel_sum_acc}, we get that $\acc(\syn{\cl{\vvfvar}}, \vvdenv, \vvtrc)$.

  \item If $\vvfvar = \vvvfvar \land \vvvvfvar$, we know that both $(\vvvfvar, \vvdenv, \vvtrc), (\vvvvfvar, \vvdenv, \vvtrc) \in A$ and that $(\vvvfvar \land \vvvvfvar, \vvdenv, \vvtrc) \dep{} (\vvvfvar, \vvdenv, \vvtrc)$ and $(\vvvfvar \land \vvvvfvar, \vvdenv, \vvtrc) \dep{} (\vvvvfvar, \vvdenv, \vvtrc)$.
By the induction hypothesis, we have that $\acc(\syn{\cl{\vvvfvar}}, \vvdenv, \vvtrc)$ and $\acc(\syn{\cl{\vvvvfvar}}, \vvdenv, \vvtrc)$. Since, $\syn{\cl{\vvfvar}} = \syn{\cl{\vvvfvar}} \parprod \syn{\cl{\vvvvfvar}}$, by \Cref{prop:parallel_prod_acc}, we get that $\acc(\syn{\cl{\vvfvar}}, \vvdenv, \vvtrc)$.

\item  If $\vvfvar = \vpvar$, then $(\recform{\vpvar}, \vvdenv, \vvtrc) \in A$ and $v \dep{} (\recform{\vpvar}, \vvdenv, \vvtrc)$.
  By definition, $\cl{\vpvar} = \cl{\fx{\vpvar}} = \cl{\recmin{\vpvar}{\recform{\vpvar}}}$. Thus, it is of the form $\cl{\vpvar} = \recmin{\vpvar}{\varphi'_{\vpvar}}$, where $\varphi'_{\vpvar}$ is the closure of $\recform{\vpvar}$ considering $\vpvar$ is not free (i.e., we close all the other recursion variables). Thus, $\syn{\cl{\vpvar}} = \rec{\vpvar}{\syn{{\varphi'_{\vpvar}}}}$. By rule $\mrec$, $\under{\rec{\vpvar}{\syn{{\varphi'_{\vpvar}}}}}{\vvdenv} \xRightarrow[\mrec]{\tau} \under{\syn{{\varphi'_{\vpvar}}}\map{\frac{\rec{\vpvar}{\syn{{\varphi'_{\vpvar}}}}}{\vpvar}}}{\vvdenv}$.

  Let us have a closer look at this latter monitor:
  \begin{align*}
      & \syn{{\varphi'_{\vpvar}}}\map{\frac{\rec{\vpvar}{\syn{{\varphi'_{\vpvar}}}}}{\vpvar}} \\
    =~& \syn{{\varphi'_{\vpvar}}}\map{\frac{\syn{\cl{\vpvar}}}{\vpvar}} & &  \text{ from } \syn{\cl{\vpvar}} = \rec{\vpvar}{\syn{{\varphi'_{\vpvar}}}} \\
    =~& \syn{\varphi'_{\vpvar}\map{\frac{\cl{\vpvar}}{\vpvar}}} & & \text{ compositionality of synthesis}\\
    =~& \syn{\cl{\recform{\vpvar}\map{\frac{\fx{\vpvar}}{\vpvar}}}} & & \varphi'_{\vpvar} \text{ is the closure of }\recform{\vpvar} \text{ considering }\vpvar\text{ not free}\\
    =~& \syn{\cl{\recform{\vpvar}}} & & \text{ definition of closure}
    \end{align*}
 By the induction hypothesis, since $(\recform{\vpvar}, \vvdenv, \vvtrc) \in A$, we have that  $\acc(\syn{\cl{\recform{\vpvar}}}, \vvdenv, \vvtrc)$, so we finally get $\acc(\syn{\cl{\vpvar}}, \vvdenv, \vvtrc)$.

\item If $\vvfvar = \recmin{\vpvar}{\recform{\vpvar}}$, then $(\recform{\vpvar}, \vvdenv, \vvtrc) \in A$ and $(\recmin{\vpvar}{\recform{\vpvar}}, \vvdenv, \vvtrc) \dep{} (\recform{\vpvar}, \vvdenv, \vvtrc)$. By definition, $\cl{\vvfvar} = \cl{\recmin{\vpvar}{\recform{\vpvar}}} = \cl{\vpvar}$. Besides, by the induction hypothesis, we have that $\acc(\syn{\cl{\recform{\vpvar}}}, \vvdenv, \vvtrc)$. Thus, we are back to the above case, and $\acc(\syn{\cl{\vvfvar}}, \vvdenv, \vvtrc)$.
  \end{itemize}
\end{itemize}
Overall, for all vertices $v = (\vvfvar, \vvdenv, \vvtrc)$ of $A$, we have $\acc(\syn{\cl{\vvfvar}}, \vvdenv, \vvtrc)$. This is the case in particular for $v_0 = (\vfvar, \vdenv, \vtrc)$. Besides, since $\vfvar$ is closed, $\cl{\vfvar} = \vfvar$. As a consequence, we obtain that for all $\vfvar \in \chmld$, all $\vdenv \in \denv$ and all $\vtrc \in \trc$, if $\vfvar, \vdenv$ have an annotation on $\vtrc$, then $\acc(\syn{\vfvar}, \vdenv, \vtrc)$. Thus, if $\vtrc \in \sem{\vfvar, \vdenv}$, then $\acc(\syn{\vfvar}, \vdenv, \vtrc)$.
\end{proof}

\subsection{Proof of \Cref{thm:monitors_register_automata}}
\label{sec:app:register_automata_monitors}
\begin{definition}[\cite{Bojanczyk19}]
  \label{def:register_automaton}
  An \emph{alternating register automaton with existential guessing} (or \emph{register automaton} for short) $\vregatm$ consists of:
  \begin{itemize}
  \item a finite non-empty set $\locations = \exilocs \disjunion \unilocs$ of \emph{locations}, partitioned into existential ($\exilocs$) and universal ($\unilocs$) locations;
  \item a finite set $\registers$ of \emph{registers};
  \item an \emph{initial location} $\iloc \in \locations$, an initial valuation $\idval : \registers \rightarrow \dat$ and a set of \emph{accepting locations} $\flocs \subseteq \locations$;
  \item a \emph{transition relation} $\vtransrel$ whose elements are of the form:
    \begin{itemize}
    \item $\vloc \xrightarrow{\vbexp} \vvloc$ for some quantifier-free formula $\vbexp$ with free variables in $\registers \cup \{\star\}$ and predicate $=$ (i.e. an expression, as defined in \cref{fig:rechml_LT}).
    \item $\vloc \xrightarrow{\raguess{\vreg}} \vvloc$ where $q \in \exilocs$ is an existential location and $\vreg \in \registers$ is a register\footnote{Note that we do not allow guessing from universal states, hence the name of ``existential guessing''. This is in line with the fact that our model of monitors only has existential guessing. As we will see in \Cref{sec:candidate_maximal_fragment}, this design decision strictly restricts expressiveness, as witnessed by $\langfsefblocks$ in \cref{eq:def_langfsefblocks}, see the proof of \cref{prop:langfsefblocks_not_chmld}. That proof is phrased in the context of monitors but carries through to automata, since those models are closely related as we demonstrate in this section.}.
    \end{itemize}
  \end{itemize}
  A \emph{data valuation} is a function\footnote{Note that it is isomorphic with a data environment.} $\vdval : \registers \rightarrow \dat$. A \emph{state} of a register automaton is a pair $\regcfg{\vloc}{\vdenv}$. A state $\regcfg{\vvloc}{\vvdenv}$ is a \emph{successor} of a state $\regcfg{\vloc}{\vdenv}$ on reading $\mu \in \dat \cup \{\tau\}$ following transition $\vtrs$, written $\regcfg{\vloc}{\vdenv} \xrightarrow{\mu} \regcfg{\vvloc}{\vvdenv}$, whenever:
  \begin{itemize}
    \item $\vtrs = \vloc \xrightarrow{\vbexp} \vvloc$, $\mu = \vdat \in \dat$, $\evalrel{\vbexp \vdval[\star \leftarrow \vdat]}{\ltru}$ and $\vvdenv = \vdenv$, or
    \item $\vtrs = \vloc \xrightarrow{\raguess{\vreg}} \vvloc$, $\mu = \tau$, $\vvdenv = \vdenv[\vreg\leftarrow \vdat]$ for some data value $\vdat \in \dat$.
  \end{itemize}

  Then, we say that $\vregatm$ \emph{has a final run from state $\regcfg{\vloc}{\vdval}$ on a data word $\vdwrd \in \dat^*$} if one of the following conditions applies\footnote{This slightly departs from the usual definition, but it is straightforward to prove that both are equivalent.}:
  \begin{itemize}
  \item $\vdwrd = \varepsilon$ is the empty word and $\vloc \in \flocs$ is a final location;
  \item $\vdwrd = \mu \vvdwrd$, $\vloc \in \exilocs$ is an existential location and there exists a successor state $\regcfg{\vloc}{\vdenv} \xrightarrow{\mu} \regcfg{\vvloc}{\vvdenv}$ such that $\vregatm$ has a final run from $\regcfg{\vvloc}{\vvdenv}$ on $\vvdwrd$;
  \item $\vdwrd = \mu \vvdwrd$, $\vloc \in \unilocs$ is a universal location and for all successor states $\regcfg{\vloc}{\vdenv} \xrightarrow{\mu} \regcfg{\vvloc}{\vvdenv}$, $\vregatm$ has a final run from $\regcfg{\vvloc}{\vvdenv}$ on $\vvdwrd$\footnote{Note that we could do away with the notion of final location by encoding them as universal locations with no successors.}.
  \end{itemize}
  The run is moreover accepting if $\regcfg{\vloc}{\vdval} = \regcfg{\iloc}{\idval}$ is the initial state.
  Finally, the \emph{language} of a register automaton $\vregatm$, denoted $\lang(\vregatm)$, is the set of data words $\vdwrd$ such that $\vregatm$ has an accepting run on $\vdwrd$. Since we sometimes change the initial state, we also write $\lang(\vregatm, \vloc, \vdval)$ to denote the language of the register automaton $\vregatm_{\regcfg{\vloc}{\vdval}}$ which is identical to $\vregatm$ except its initial state is $\regcfg{\vloc}{\vdval}$.
\end{definition}

In this section, we show that monitors can be converted to alternating register automata with existential guessing and the other way around, by adapting the construction of~\cite[Section~4.2]{DBLP:journals/jlap/AcetoAFIK20}. In our setting, the $\guess{}$ construct corresponds to non-deterministic reassignment~\cite{DBLP:journals/ijfcs/KaminskiZ10} (a.k.a. existential ``guessing''~\cite{Bojanczyk19}), and parallel sum and parallel product respectively correspond to non-deterministic and universal choice.

\begin{proposition}
  \label{prop:monitor_to_RA}
  Let $\vmon \in \mon$ and $\vdenv \in \denv$. There exists a register automaton $\regatm{\vmon}$ such that for all finite traces $\vftrc \in \ftrc$, $\vftrc \in \lang(\regatm{\vmon}, \vdenv)$ if and only if $\vmon, \vdenv \xRightarrow{\vftrc} \yes, \vdenv''$ for some $\vdenv'$.
\end{proposition}
\begin{proof}
  Given a monitor $\vmon$, we show how to construct a register automaton $\regatm{\vmon}$ that accepts the same finite traces as $\vmon$. To ease the construction, the register automaton $\regatm{\vmon}$ we build has $\varepsilon$-transitions, but they can be simulated by a dummy $\guess{}$ transition or directly removed (see~\cite[Exercise~2]{Bojanczyk19}). Moreover, up to renaming recursion variables, we assume that each recursion variable $\vpvar$ appears in a unique submonitor $\recmon{\vpvar} = \rec{\vpvar}{\recedmon{\vpvar}}$.

  As in~\cite{DBLP:journals/jlap/AcetoAFIK20}, we use a slightly different semantics and replace rule $\mrec$ with rules $\mrecf$ and $\mrecb$:
  \begin{mathpar}
      \inferrule*[left=\mrecf]
      {~}
      {\under{\rec{\vrvar}{\recedmon{\vrvar}}}{\vdenv} \trans[\tau] \under{\recedmon{\vrvar}}{\vdenv}}
      \qquad
      \inferrule*[left=\mrecb]
      {~}
      {\under{\vrvar}{\vdenv} \trans[\tau] \under{\recmon{\vrvar}}{\vdenv}}
    \end{mathpar}
    The proof that the use of the rules $\mrecf$ and $\mrecb$ in lieu of $\mrec$ does not modify the semantics of monitors regarding acceptance is the same as in the original paper: just note that those rules do not modify the data environment.

  Then, we define $\regatm{\vmon} = (\locations, \registers, \iloc, \flocs, \transrel)$ as:
  \begin{itemize}
  \item $\locations = \submonitors(\vmon)$, where $\submonitors(\vmon)$ denotes the set of submonitors of $\vmon$. All locations are existential except monitors that consist in parallel products, formally $\unilocs = \{\vvmon \in \submonitors(\vmon) \mid \vvmon = \vvvmon \parprod \vvvvmon \text{ for some }\vvvmon, \vvvvmon \in \mon\}$ and $\exilocs = \locations \setminus \unilocs$;
  \item $\registers = \vars(\vmon)$, where $\vars(\vmon)$ denotes the set of data variables occurring in $\vmon$;
  \item $\iloc = \vmon$;
  \item $\flocs = \{\yes\}$ if $\yes \in \locations$ and $\flocs = \varnothing$ otherwise;
  \item $\transrel$ is defined as follows:
    \begin{itemize}
    \item If $\vloc = \yes$, then 
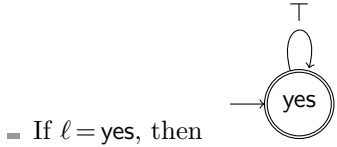
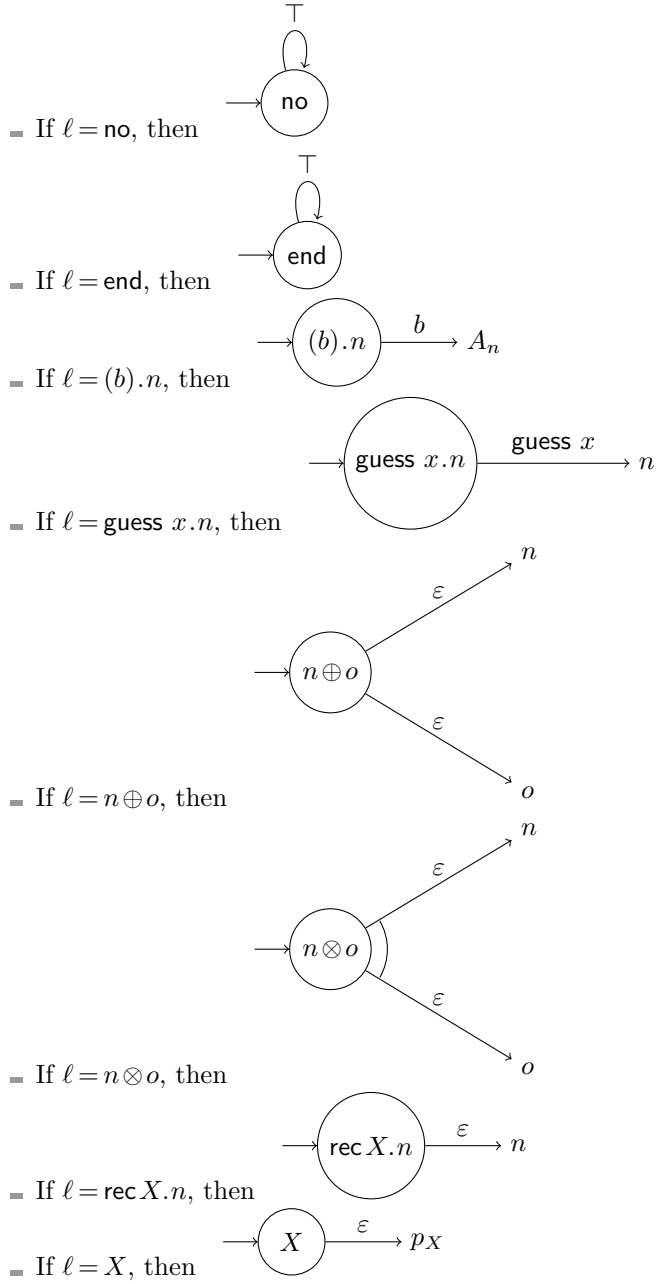
\begin{tikzpicture}
   \node[state, initial, initial text = {}, accepting] (q_0)   {$\yes$};
    \path[->]
    (q_0) edge [loop above]  node {$\top$} ();
\end{tikzpicture}

    \item If $\vloc = \no$, then \begin{tikzpicture}
   \node[state, initial, initial text = {}] (q_0)   {$\no$};
    \path[->]
    (q_0) edge [loop above]  node {$\top$} ();
\end{tikzpicture}

    \item If $\vloc = \inc$, then \begin{tikzpicture}
   \node[state, initial, initial text = {}] (q_0)   {$\inc$};
    \path[->]
    (q_0) edge [loop above]  node {$\top$} ();
\end{tikzpicture}

    \item If $\vloc = \prf{(\vbexp)}{\vvmon}$, then \begin{tikzpicture}
   \node[state, initial, initial text = {}] (q_0)   {$\prf{(\vbexp)}{\vvmon}$};
   \node[right = of q_0] (Am)   {$\regatm{\vvmon}$};
    \path[->]
    (q_0) edge node[above]{$\vbexp$} (Am);
\end{tikzpicture}

    \item If $\vloc = \prf{\guess{\vdvar}}{\vvmon}$, then \begin{tikzpicture}
   \node[state, initial, initial text = {}] (q_0)   {$\prf{\guess{\vdvar}}{\vvmon}$};
   \node[right = 2cm of q_0] (Am)   {$\vvmon$};
    \path[->]
    (q_0) edge node[above]{$\raguess{\vdvar}$} (Am);
\end{tikzpicture}

    \item If $\vloc = \vvmon\parsum\vvvmon$, then \begin{tikzpicture}
   \node[state, initial, initial text = {}] (q_0)   {$\vvmon\parsum\vvvmon$};
   \node[above right = 1cm and 2cm of q_0] (Am)   {$\vvmon$};
   \node[below right = 1cm and 2cm of q_0] (An)   {$\vvvmon$};
    \path[->]
    (q_0) edge node[above]{$\varepsilon$} (Am);
    \path[->]
    (q_0) edge node[above]{$\varepsilon$} (An);
\end{tikzpicture}

    \item If $\vloc = \vvmon\parprod\vvvmon$, then \begin{tikzpicture}
   \node[state, initial, initial text = {}] (q_0)   {$\vvmon\parprod\vvvmon$};
   \node[above right = 1cm and 2cm of q_0] (Am)   {$\vvmon$};
   \node[below right = 1cm and 2cm of q_0] (An)   {$\vvvmon$};
    \path[->]
    (q_0) edge node[above]{$\varepsilon$} (Am);
    \path[->]
    (q_0) edge node[above]{$\varepsilon$} (An);
    \draw ([shift = (-30:0.75cm)]q_0.center) arc (-30:30:0.75cm);
\end{tikzpicture}

    \item If $\vloc = \rec{\vrvar}{\vvmon}$, then \begin{tikzpicture}
   \node[state, initial, initial text = {}] (q_0)   {$\rec{\vrvar}{\vvmon}$};
   \node[right = of q_0] (Am)   {$\vvmon$};
    \path[->]
    (q_0) edge node[above]{$\varepsilon$} (Am);
\end{tikzpicture}

    \item If $\vloc = \vpvar$, then \begin{tikzpicture}
   \node[state, initial, initial text = {}] (q_0)   {$\vpvar$};
   \node[right = of q_0] (pX)   {$\recmon{\vpvar}$};
    \path[->]
    (q_0) edge node[above]{$\varepsilon$} (pX);
\end{tikzpicture}

    \end{itemize}
    where the initial $\rightarrow$ marks the entry point of the associated subautomaton.
  \end{itemize}
  Now, let us show that for all $\vftrc \in \ftrc$ and all $\vdenv \in \denv$, $\vftrc \in \lang(\regatm{\vmon}, \vdenv)$ if and only if $\under{\vmon}{\vdenv} \xRightarrow{\vftrc} \under{\yes, \vdenv'}$ for some $\vdenv'$.
  We show a stronger statement, namely that for all locations $\vloc \in \locations$, all valuations $\vdenv \in \denv$ and all $\vftrc \in \ftrc$, $\vftrc \in \lang(\regatm{\vmon}, {\vloc}, \vdenv)$ if and only if $\under{\vloc}{\vdenv} \xRightarrow{\vftrc} \under{\yes}{\vvdenv}$ for some $\vvdenv$.

  We first show the left-to-right implication by induction on the proof of existence of a final run of $\lang(\regatm{\vmon}, {\vloc}, \vdenv)$:
  \begin{itemize}
  \item If $\vftrc = \varepsilon$, then necessarily the run is in the only final location $\vloc = \yes$, and we indeed have that $\under{\yes}{\vdenv} \xRightarrow{\vftrc} \under{\yes}{\vdenv}$.
  \item Otherwise, we distinguish cases based on $\vloc$:
    \begin{itemize}
    \item If $\vloc = \yes$, then the run is of the form $\under{\yes}{\vdenv} \xrightarrow{\vdat} \vrun$, where $\vrun$ is a run starting in $\under{\yes}{\vdenv}$, since the only available transition is the self-loop labelled with $\top$ (which thus accepts every data value). Moreover, $\vftrc = \vdat \vvftrc$ where $\vvftrc$ is the finite trace accepted along the final run $\vrun$. Then, by rule $\mvrd$ and the induction hypothesis, we indeed have that $\under{\yes}{\vdenv} \xRightarrow{\vdat} \under{\yes}{\vdenv} \xRightarrow{\vvftrc} \under{\yes}{\vdenv}$.
    \item If $\vloc = \inc, \no$, then one cannot have an accepting run, since the corresponding state is a sink non-accepting state.
    \item If $\vloc = \prf{(\vbexp)}{\vvmon}$, then there is a single available transition and the run is of the form $\under{\prf{(\vbexp)}{\vvmon}}{\vdenv} \xrightarrow{\vdat} \vrun$, where $\evalrel{\vbexp \vdenv\map{\star \leftarrow \vdat}}{\ltru}$ and $\vrun$ is an accepting run from state $\vvmon$ on some trace $\vftrc \in \ftrc$. By induction, we know that $\under{\vvmon}{\vdenv} \xRightarrow{\vftrc} \under{\yes}{\vvdenv}$, so $\under{\prf{(\vbexp)}{\vvmon}}{\vdenv} \xrightarrow[\mact]{\vdat} \under{\vvmon}{\vdenv} \xRightarrow{\vftrc} \under{\yes}{\vvdenv}$.
    \item If $\vloc = \prf{\guess{\vdvar}}{\vmon}$, then the run is of the form $\under{\vloc = \prf{\guess{\vdvar}}{\vmon}}{\vdenv} \xrightarrow{\varepsilon} \vrun$, where $\vrun$ is an accepting run from state $\vvmon, \vdenv[\vdvar \leftarrow \vdat]$ on some trace $\vftrc \in \ftrc$. Then, $\under{\prf{\guess{\vdvar}}{\vmon}}{\vdenv} \xrightarrow[\mgs]{\tau} \under{\vvmon}{\vdenv\map{\vdvar \leftarrow \vdat}} \xRightarrow{\vftrc} \under{\yes}{\vvdenv}$.
    \item If $\vloc = \vvmon \parsum \vvvmon$. Then, the first transition of the run is either of the form $\under{\vvmon \parsum \vvvmon}{\vdenv} \xrightarrow{\varepsilon} \under{\vvmon}{\vdenv}$ or $\under{\vvmon \parsum \vvvmon}{\vdenv} \xrightarrow{\varepsilon} \under{\vvvmon}{\vdenv}$. We treat the former case, the latter is symmetric. By induction, we know that $\under{\vvmon}{\vdenv} \xRightarrow{\vftrc} \under{\yes}{\vvdenv}$ for some $\vftrc \in \ftrc$. Then, by \Cref{prop:parallel_sum_acc}, we know that $\under{\vvmon \parsum \vvvmon}{\vdenv} \xRightarrow{\vftrc} \under{\yes}{\vvdenv}$.
    \item If $\vloc = \vvmon \parprod \vvvmon$. Then, the run is of the form
      \begin{tikzpicture}
        \node (parprod) {$\under{\vvmon \parprod \vvvmon}{\vdenv}$};
        \node[below left = of parprod] (runm) {$\vrun$};
        \node[below right = of parprod] (runn) {$\vvrun$};

        \path[->]
        (parprod) edge node[above left] {$\varepsilon$} (runm);
        \path[->]
        (parprod) edge node[above right] {$\varepsilon$} (runn);
        \end{tikzpicture}
        where $\vrun$ (respectively, $\vvrun$) is an accepting run from $\under{\vvmon}{\vdenv}$ (respectively, from $\under{\vvvmon}{\vdenv}$), both over the same trace $\vftrc \in \ftrc$. By induction, we know that $\under{\vvmon}{\vdenv} \xRightarrow{\vftrc} \under{\yes}{\vvdenv}$ and $\under{\vvvmon}{\vdenv} \xRightarrow{\vftrc} \under{\yes}{\vvdenv}$. Then, by \Cref{prop:parallel_prod_acc}, we know that $\under{\vvmon \parprod \vvvmon}{\vdenv} \xRightarrow{\vftrc} \under{\yes}{\vvdenv}$.
      \item If $\vloc = \rec{\vpvar}{\vvmon}$, then the run is of the form $\under{\rec{\vpvar}{\vvmon}}{\vdenv} \xrightarrow{\varepsilon} \under{\vvmon}{\vdenv}$. By induction, we know that $\under{\vvmon}{\vdenv} \xRightarrow{\vftrc} \under{\yes}{\vvdenv}$. Thus, using rule $\mrecf$, $\under{\rec{\vpvar}{\vvmon}}{\vdenv} \xrightarrow[\mrecf]{\tau} \under{\vvmon}{\vdenv} \xRightarrow{\vftrc} \under{\yes}{\vvdenv}$.
      \item If $\vloc = \vpvar$, then the run is of the form $\under{\vpvar}{\vdenv} \xrightarrow{\varepsilon} \under{\recmon{\vpvar}}{\vdenv}$. By induction, we know that $\under{\recmon{\vpvar}}{\vdenv} \xRightarrow{\vftrc} \under{\yes}{\vvdenv}$. Thus, using rule $\mrecb$, $\under{\vpvar}{\vdenv} \xrightarrow[\mrecb]{\tau} \under{\recmon{\vpvar}}{\vdenv} \xRightarrow{\vftrc} \under{\yes}{\vvdenv}$.
    \end{itemize}
  \end{itemize}

  The proof of the converse implication follows similar lines.
\end{proof}

Now, monitors can conversely simulate register automata, provided they recognise a suffix-closed language (recall that monitors only recognise suffix-closed languages). The construction is essentially identical to that of~\cite[Section 4.2.1]{DBLP:journals/jlap/AcetoAFIK20}: unfold the automaton so that it is a tree with back edges to avoid edges crossing between branches (this induces an unavoidable exponential blowup), and inductively build a corresponding monitor starting from the leaves. Indeed, registers are to automata what data variables are to monitor, and the associated semantics are essentially identical.

\paragraph*{Unravelling of an Automaton}
A \emph{simple path} in a register automaton $\vregatm$ is a non-empty sequence\footnote{Note that this construction departs from that of~\cite[Section 4.2.1]{DBLP:journals/jlap/AcetoAFIK20} in that we omit the letters labelling the transitions for simplicity.} $\vloc_1 \vloc_2 \dots \vloc_k \in \locations^+$ of pairwise distinct locations such that for all $1 \leq i < k$, $\vloc_i \xrightarrow[\vregatm]{b} \vloc_{i+1}$ or $\vloc_i \xrightarrow[\vregatm]{\guess{\vreg}} \vloc_{i+1}$. To ease notations, we define the following operator $\extspath$ on simple paths:
\begin{itemize}
\item $\vloc_1 \dots \vloc_k \extspath \vloc = \vloc_1 \dots \vloc_k \vloc$ if $\vloc$ does not appear before, i.e. for all $1 \leq i \leq k$, $\vloc_i \neq \vloc$
\item $\vloc_1 \dots \vloc_i \dots \vloc_k \extspath \vloc = \vloc_1 \dots \vloc_i$ if $\vloc = \vloc_i$ (i.e. we truncate the path at the position where $\vloc$ appeared).
\end{itemize}
Note that this operator is well-defined since in the second item, $\vloc$ can only appear once before since $\extspath$ is defined on simple paths. Note moreover that it maps simple paths to simple paths.

Given a register automaton $\vregatm$, we define its \emph{unravelling} $\unravel(\vregatm) = (\locations'', \registers, \iloc, \idval, \flocs', \vtransrel')$, where:
\begin{itemize}
\item $\locations'$ is the set of simple paths of $\vregatm$
\item $\flocs'$ is the set of simple paths of $\vregatm$ that end in an accepting location
\item $\vtransrel'$ is the set of transitions of the form $\vpath \vloc \xrightarrow{e} \vpath \vloc \extspath \vloc'$ for all simple paths $\vpath \vloc$ and all transitions $\vloc \xrightarrow[\vregatm]{e} \vloc'$ (here, $e$ is either $\vbexp$ or $\guess{\vreg}$).
\end{itemize}
It is straightforward to establish that $\vregatm$ and $\unravel(\vregatm)$ recognise the same language, but the tree-like structure of $\unravel(\vregatm)$ makes it more amenable to being converted to a monitor.

In the following, we say that a register automaton is \emph{irrevocable} whenever it recognises a suffix-closed language. Without loss of generality, we can assume that an irrevocable automaton has a single accepting state which is a sink, i.e. all its outgoing transitions point to itself.

\begin{proposition}[see~{\cite[Theorem~6]{DBLP:journals/jlap/AcetoAFIK20}}]
  \label{prop:RA_to_monitor}
  Let $\vregatm$ be an irrevocable register automaton. There exists a monitor $\vmon_{\vregatm}$ which accepts the same traces as $\vregatm$, i.e. for all finite traces $\vftrc$, $\vftrc \in \lang(\vregatm)$ if and only if $\acc(\vmon_{\vregatm}, \vftrc)$.
\end{proposition}
\begin{proof}
  Let $\vregatm$ be an irrevocable register automaton. Consider $\unravel(\vregatm)$. Given $\vpath \in \locations^*$ and $\vloc \in \locations$, we recursively define $\vmon(\vpath \vloc)$ as:
  \begin{itemize}
  \item if $\vloc \in \flocs$ is accepting, then $\vmon(\vpath \vloc) = \yes$
  \item otherwise:
    \begin{itemize}
    \item if $\vloc \in \exilocs$ is existential, then
      \begin{equation}
        \label{eq:def_mon_reg_exi}
        \vmon(\vpath \vloc) = \rec{\vrvar_{\vpath}}{
          \left(
            \begin{array}{c}
              \bigoplus \left\{ \prf{(\vbexp)}{\vmon(\vpath \vloc \vloc')} \mid \vloc \xrightarrow[\vregatm]{\vbexp} \vloc' \text{ and } \vloc' \notin \vpath \vloc \right\} \\
              \oplus \\
              \bigoplus \left\{ \prf{\guess{\vreg}}{\vmon(\vpath \vloc \vloc')} \mid \vloc \xrightarrow[\vregatm]{\guess{\vreg}} \vloc' \text{ and } \vloc' \notin \vpath \vloc \right\} \\
              \oplus \\
              \bigoplus \left\{ \prf{(\vbexp)}{\vrvar_{\vpath \vloc \extspath \vloc'}} \mid \vloc \xrightarrow[\vregatm]{\vbexp} \vloc' \text{ and } \vloc' \in \vpath \vloc \right\} \\
              \oplus \\
              \bigoplus \left\{ \prf{\guess{\vreg}}{\vrvar_{\vpath \vloc \extspath \vloc'}} \mid \vloc \xrightarrow[\vregatm]{\guess{\vreg}} \vloc' \text{ and } \vloc' \in \vpath \vloc \right\}
              \end{array}
        \right)
        }
        \end{equation}
    \item if $\vloc \in \unilocs$ is universal, then\footnote{Note the absence of universal guessing since it has no counterpart in monitors.}
      \begin{equation}
        \label{eq:def_mon_reg_uni}
        \vmon(\vpath \vloc) = \rec{\vrvar_{\vpath}}{
          \left(
            \begin{array}{c}
              \bigotimes \left\{ \prf{(\vbexp)}{\vmon(\vpath \vloc \vloc')} \mid \vloc \xrightarrow[\vregatm]{\vbexp} \vloc' \text{ and } \vloc' \notin \vpath \vloc \right\} \\
              \otimes \\
              \bigotimes \left\{ \prf{(\vbexp)}{\vrvar_{\vpath \vloc \extspath \vloc'}} \mid \vloc \xrightarrow[\vregatm]{\vbexp} \vloc' \text{ and } \vloc' \in \vpath \vloc \right\}
              \end{array}
            \right)
        }
        \end{equation}
    \end{itemize}
  \end{itemize}
  Observe that the recursive definition is well-founded since we only consider simple paths.

  Then, let us show that for all finite traces $\vftrc$ and all data valuations $\vdenv \in \denv$, $\under{\vmon(\iloc)}{\idval} \xRightarrow{\vftrc} \under{\yes}{\vdenv}$ if and only if $\vftrc \in \lang(\vregatm)$. By the earlier observation that $\lang(\vregatm) = \lang(\unravel(\vregatm))$, this is equivalent to showing that $\under{\vmon(\iloc)}{\idval} \xRightarrow[LS]{\vftrc} \under{\yes}{\vdenv}$ if and only if $\vftrc \in \lang(\unravel(\vregatm))$. We show a more general result, namely that $\under{\vmon(\vloc)}{\vdval} \xRightarrow[LS]{\vftrc} \under{\yes}{\vvdenv}$ for some $\vvdenv \in \denv$ if and only if $\unravel(\vregatm)$ has a final run from state $\regcfg{\vloc}{\vdval}$.

  First, assume that $\under{\vmon(\vpath \vloc)}{\vdval} \xRightarrow[LS]{\vftrc} \under{\yes}{\vdenv}$. We show the result by induction on the length of the derivation:
  \begin{itemize}
  \item If it is of length 0, this means that  $\vftrc = \varepsilon$ and $\vmon(\vloc) = \yes$. The latter happens whenever $\vloc \in \flocs$, and by definition this implies that $\vregatm$ has a final run over $\vftrc = \varepsilon$.
  \item Otherwise, the derivation is of length $n \geq 1$ and starts from some $\under{\vmon(\vpath \vloc)}{\vdval}$. We distinguish cases based on $\vloc$:
    \begin{itemize}
    \item If $\vloc \in \exilocs$, then $\vmon(\vpath \vloc) = \rec{\vrvar_{\vpath}}{S}$, where $S$ is the (big) sum of \Cref{eq:def_mon_reg_exi}. Necessarily, the first transition follows rule $\mrec$ and we have $\under{\vmon(\vpath \vloc)}{\vdenv} \xrightarrow{\tau} \under{S}{\vdenv}$. By repeatedly applying \Cref{prop:parallel_sum_acc}, we know that $\under{S}{\vdenv} \xRightarrow{\vftrc} \under{\yes}{\vvdenv}$ if and only if one component of $S$ accepts $\vftrc$.

      If this component is of the form $\prf{e}{\vmon(\vpath \vloc \vloc')}$ (where $e$ is either $\vbexp$ or $\guess{\vreg}$), i.e. if $\vloc \xrightarrow[\vregatm]{e} \vloc'$ and $\vloc' \notin \vpath \vloc$, then we have that $\prf{e}{\vmon(\vpath \vloc \vloc')} \xrightarrow{\mu} \under{\vmon(\vpath \vloc \vloc')}{\vvvdenv} \xRightarrow{\vvftrc} \under{\yes}{\vvdenv}$. By the induction hypothesis, $\unravel{\vregatm}$ has a final run from state $\regcfg{\vpath \vloc \vloc'}{\vvvdenv}$, so, since $\regcfg{\vpath \vloc}{\vvvdenv} \xrightarrow[\unravel{\vregatm}]{\mu} \regcfg{\vpath \vloc}{\vvvdenv}$ and $\vloc$ is existential, we get that $\unravel{\vregatm}$ has a final run over $\vftrc = \mu \vvftrc$ from state $\regcfg{\vpath \vloc}{\vdenv}$.

      If the component is instend of the form $\prf{e}{\vrvar_{\vpath \vloc \extspath \vloc'}}$, i.e. if the target state already appeared in $\vpath \vloc$, then a similar reasoning applies, except that there is an additional transition $\under{\vrvar_{\vpath \vloc \extspath \vloc'}}{\vvvdenv} \xrightarrow{\tau} \vmon(\vpath \vloc \extspath \vloc')$.
    \item If $\vloc \in \unilocs$, then $\vmon(\vpath \vloc) = \rec{\vrvar_{\vpath}}{P}$, where $P$ is the product of \Cref{eq:def_mon_reg_uni}. By repeatedly applying \Cref{prop:parallel_prod_acc}, we get that each component of the product must have an accepting run. By the same reasoning as above, we obtain by induction that for all successor state $\regcfg{\vloc'}{\vvdenv}$, there is an accepting run of $\unravel{\vregatm}$ over the remaining trace $\vftrc'$. By definition, this means that $\unravel{\vregatm}$ has a final run from the universal state $\regcfg{\vloc}{\vdval}$.
    \end{itemize}
  \end{itemize}

  Conversely, assume that $\unravel(\vregatm)$ has a final run from state $\regcfg{\vloc}{\vdval}$. Let us show that then, $\under{\vmon(\vpath \vloc)}{\vdval} \xRightarrow[LS]{\vftrc} \yes$. This is done by induction on the definition of final run:
  \begin{itemize}
  \item If there is a final run because $\vftrc = \varepsilon$ and $\vloc \in \flocs$, then $\vmon(\vpath \vloc) = \yes$ and the result holds
  \item If $\vloc \in \exilocs$, then by definition there exists a successor state $\regcfg{\vloc}{\vdenv} \xrightarrow{\mu} \regcfg{\vvloc}{\vvdenv}$ such that $\vregatm$ has a final run from $\regcfg{\vvloc}{\vvdenv}$ on the finite trace $\vvftrc$ such that $\vftrc = \mu \vvftrc$. By induction, this means that $\under{\vmon(\vpath \vloc \extspath \vvloc)}{\vvdenv} \xRightarrow[LS]{\vvftrc} \yes$. By a simple case analysis similar to what has been done above, this means that one of the components of the big sum in \Cref{eq:def_mon_reg_exi} accepts $\vvftrc$, which implies that $\acc(\vmon(\vpath \vloc), \vdenv, \vftrc)$.
  \item If $\vloc \in \unilocs$, then all successor states have a final run. By a similar reasoning as above, we get that $\acc(\vmon(\vpath \vloc), \vdenv, \vftrc)$. \qedhere
  \end{itemize}
\end{proof}

As a consequence of Propositions~\ref{prop:monitor_to_RA} and~\ref{prop:RA_to_monitor}, we get that our model of monitors has the same expressive power as alternating register automata with existential guessing:
\monitorsRegisterAutomata*

\subsection{Comparing Expressiveness of \rechmld{} Fragments}
\label{app:sec:comparing_expressiveness}

To formally state our result, we introduce the following notation: we denote $\hmld(\mathcal{C})$ to specify the fragment of \rechmld{} with the constructs in $\mathcal{C}$. For instance, $\disjhmld = \hmld(\tru, \pos{}{}, \exists, \lor, \mathsf{min})$. Additionally, we introduce the two following semantic restrictions:
  \begin{itemize}
    \item \textsc{match}: occurrences of $\exists$ are necessarily of the form $\exi{\vdvar}{\pos{\vdvar = \star \land b}{\vfvar}}$: any introduced data variable is immediately matched with the current element of the trace.
    \item \textsc{det}: the above restriction holds, and additionally disjunctions are of the form $\bigvee_{i \in I} \exi{\binder{\vdvar}}{\pos{\star = \vdvar \cand \vbexp_i}{\vfvar}}$, where for all $i, j \in I$ such that $i \neq j$, the expression $\vbexp_i \wedge \vbexp_j$ is not satisfiable. Intuitively, the formula is deterministic in the sense that its associated monitor is.
  \end{itemize}

By \Cref{thm:monitors_register_automata}, along with the separation results for register automata~\cite[Section~1.5]{Bojanczyk19}, the following holds (recall that $\vclass \sqsubseteq \vvclass$ means that $\vclass$ can be expressed in $\vvclass$, as formally defined on page~\pageref{def:expressiveness}):
\begin{theorem}
  \label{thm:hierarchy_rechmld}
  $\hmld(\exists, \pos{}{}, \lor, \textsc{det}) \sqsubsetneq \hmld(\mathsf{min}, \exists, \pos{}{}, \lor, \textsc{det}) \sqsubsetneq \hmld(\mathsf{min}, \exists, \pos{}{}, \lor, \textsc{match}) \sqsubsetneq \hmld(\mathsf{min}, \exists, \pos{}{}, \lor) \sqsubsetneq \hmld(\mathsf{min}, \exists, \pos{}{}, \lor, \land) \sqsubsetneq \hmld(\mathsf{min}, \mathsf{max}, \exists, \forall, \pos{}{}, \nec{}{}, \lor, \land) = \rechmld{}$.
\end{theorem}

\subsection{Missing Proofs of \Cref{sec:optimal_monitors}}
\label{app:proof:sec:optimal_monitors}
\optimalMonitorsDisjhmld*
\begin{proof}
Let $\vfvar \in \disjhmld$. We know that there exists a non-deterministic register automaton with guessing $A_{\vfvar}$ such that $L(A_{\vfvar}) = \sem{\vfvar}$. Since emptiness is decidable, we can assume without loss of generality that all states of $A_{\vfvar}$ accept at least one word (we also assume that each state stores the type of its valuation). Then, one can complete $A_{\vfvar}$ with a single sink rejecting state $\lightning$. By dualising $A_{\vfvar}$ and setting $\lightning$ as its only accepting state, one gets a universal register automaton which accepts exactly the bad prefixes of $\vfvar$, i.e. a violation-optimal monitor for $\vfvar$.
\end{proof}
\optimalMonitorscHMLdUndecidable*
\begin{proof}
By~\Cref{thm:satisfiability_chmld}, satisfiability for \chmld{} is undecidable, and by~\Cref{cor:sat-chml-re}, it is recursively enumerable.
Thus, unsatisfiability is not recursively enumerable.
Let $\vfvar \in \chmld{}$.
Towards a contradiction, assume that we can effectively construct a violation-optimal monitor $\vmon$ for $\vfvar$. By definition, a formula $\vfvar$ is unsatisfiable iff all traces violate $\vfvar$ iff on any trace the monitor $\vmon$ rejects. Consider the tree that includes all rejecting prefixes up to renaming. As in the proof of \Cref{prop:CM_HML}, since there are finitely many types, that tree is finitely branching and every branch is finite.
We can enumerate all these trees and check if the monitor rejects all branches, so unsatisfiability is recursively enumerable, which is a contradiction.
\end{proof}

\subsection{Missing Proofs from \Cref{sec:beyond_chmld}}
\label{app:beyond_chml}
\langfsefblocksNotcHMLd*
\begin{proof}[Proof of \Cref{prop:langfsefblocks_not_chmld}]
  Given a monitor $m$, define $\parc(m)$, the number of parallel components of $m$ recursively on $m$: if $m$ is of the form $m_1 \parany  m_2$, then $\parc(m) = \parc(m_1) + \parc(m_2)$; and $\parc(m) = 1$, otherwise.
  The parallel submonitors $\psub(m)$ of $m$ are also defined recursively on $m$:
  if $m$ is of the form $m_1 \parany  m_2$, then $\psub(m) = \{m\} \cup \psub(m_1) \cup \psub(m_2)$; and $\psub(m) = \{m\}$ otherwise.

  If  there exists a formula $\vfvar \in \chmld$ such that $\sem{\vfvar} = \langfsefblocks$, then by \Cref{thm:mon_satisfactions_cHMLd}, there exists a sound monitor $m$ for $\langfsefblocks$ that recognises all satisfactions of the property. Let $k$ be its number of data variables. Intuitively, on reading $\$$, $m$ can only remember that many distinct data values, so if the first blocks contains more values, we can trick $m$ into not accepting inputs it should accept.
  %
  Consider a run $m,\delta_0 \xRightarrow{d_0 d_1 \dots d_{k}\$} m',\delta$ of the monitor
  on input $d_0 d_1 \dots d_{k}\$ $,
  and
  let
  $s$ be a finite trace that has $\parc(m')$ distinct data values $c_1,\ldots,c_{\parc(m')}$ that are also distinct from all the $d_i$.
  Let $S = \{s'\# d^\omega \mid s'$ results from $s$ by replacing all of the $c_i$'s except one by values among $d_0,\ldots,d_{k}\}$.
  We prove the following claim that directly implies that any such $m$ is either unsound or incomplete, yielding the proposition.

  \emph{Claim:}
   Let $n \in \psub(m')$, and $p = \parc(n)$.
  If $n,\delta$ does not accept
  at least $2^{p-1}$ traces in $S$, then
  $n,\delta $ does not accept some trace that results from $s\# d^\omega$ by replacing all of the $c_i$'s in $s$ by values among $d_0,\ldots,d_{k}$.

  The proof of the claim is by induction on $p$.
  For the case of $p = 1$,
  let $d_i$ be such that $0\leq i \leq k$ and for any register $x$ in $n$, $\delta(x) \neq d_i$.
  From our assumptions, there is some $s_0 \# d^\omega \in S$ that $n,\delta$ does not accept, where, without loss of generality, $s_0$ has exactly one data value $c_1$ that is distinct from $d_0,\ldots,d_k$.
  Then, by \Cref{prop:invariance_renaming_monitors}, $n,\delta $ does not accept $s'\# d^\omega$, where $s'$ results from $s_0$ by replacing $c_1$ by $d_i$.
  For the case of $n = n_1 \parprod n_2$, if
  $n,\delta$ does not accept
  at least
  $2^{p-1}$ traces in $S$, then it is straightforward to see that
  for at least one of $j=1,2$,
  $n_j,\delta$ does not accept at least $2^{\parc(n_j)-1}$ traces in $S$.
  The claim then follows.
  The case of $n = n_1 \parsum n_2$ is similar and therefore the inductive proof of the claim is complete.
\end{proof}

\subsection{Proof of \Cref{thm:guarded-annotation-semantics}}
\label{app:proof:thm:guarded-annotation-semantics}
To demonstrate \Cref{thm:guarded-annotation-semantics}, we show the following more precise statement:
\begin{proposition}\label{prop:guarded-annotation-semantics_pf}
    For every closed $\vfvar \in \minhmldg$, $\vdenv \in \denv$, and $\vtrc \in \trc$,
    the following are equivalent:
    \begin{enumerate}
      \item $(\vfvar, \vdenv,\vtrc)$ has an annotation;
      \item $(\vfvar, \vdenv,\vtrc)$ has a  guarded-branching annotation;
       \item $(\vfvar, \vdenv,\vtrc)$ has a finite guarded-branching annotation; and
        \item $ \vtrc \in \sem{\vfvar,\vdenv} $.
    \end{enumerate}
\end{proposition}

We can extend the notation $\vdenv[x \mapsto d]$ to sets of variables, such that for $J = \{ x_1, x_2,\ldots ,x_k \}$,
$\vdenv[J \mapsto d] = \vdenv[x_1 \mapsto d][x_2 \mapsto d] \cdots [x_k \mapsto d]$.

The following, \Cref{lem:GBA-of-in-V-any-will-do,lemma:guarded-x-not-in-D}, demonstrate that, for a guarded-branching annotation, the specific values of formulas that are in $V$ or that are not encountered by the annotation, do not affect the evaluation of the formula.

\begin{lemma}\label{lem:GBA-of-in-V-any-will-do}
  Let $\vvfvar \in \minhmldg[V,F]$ be a subformula of a closed formula $\vfvar\in \minhmldg[\emptyset,\emptyset]$.
    Let $(\vvfvar, \vdenv, \vvtrc)$ have a finite guarded-branching annotation $(A,\dep{})$, $ J \subseteq V$, and $d,d'\in \dat$ be such that
    \begin{itemize}
        \item for every $\vvtrc',\vvdenv,\vvvfvar$
         it is not the case that
        $(
    \vvfvar,\vdenv,\vvtrc
        ) \dep{}^* (\vvvfvar,\vvdenv,\vdat\vvtrc')$,
        \item for every $x \in J$ and $y \in F \setminus J$, $\vdenv(x) = d'$ and $\vdenv(y) \neq d, d'$.
    \end{itemize}
        Then,
    $(\vvfvar, \vdenv[J \mapsto d],
     \vvtrc
    )$ has a finite guarded-branching annotation  that has at most the same size as $(A,\dep{})$.
\end{lemma}

\begin{proof}
  We assume that $(A,\dep{})$ is minimal and we decorate $A$ by assigning to each $a \in A$ a pair of sets of variables $H,G$, so that if $a = (\vvvfvar,\vvdenv,\vvtrc)$, then $\vvvfvar \in \minhmldg[H,G]$. We do so in the following recursive way:
  we assign $(V,F)$ to $(\vvfvar, \vdenv, \vvtrc)$, and if $a = (\vvvfvar,\vvdenv,\vvtrc) \dep{} c$ and $a$ is assigned with $G,H$, then we take cases for $\vvvfvar$:
  \begin{description}
      \item[It is not possible that $\vvvfvar = \fls$ or $\tru$.]
      \item[If $\vvvfvar = \pos{\vbexp}{\vvvfvar'} $,
      ] then by
      the annotation conditions, $\evalrel{\vbexp\vdenv[\star\mapsto\vact]}{\ltru}$ and $c = (\vvvfvar', \vvdenv, \vvvtrc) \in A$, where $\vvtrc = \vact\vvvtrc$. We assign $G,H$ to $c$.
      \item[If $\vvvfvar = \exi{\binder{\vdvar}}{(\vdvar \neq V'
      \land \vfvar_1) \lor (
      x \sim V'
      \land \vfvar_2)}$],
      then $V' = G$ and
      either $c = (\vfvar_1,\vvdenv,\vvtrc)$ or $c = (\vfvar_2,\vvdenv,\vvtrc)$.
      In the first case, we assign $G \bar{\vdvar} ,H \vdvar$ to $c$, and in the second case we assign $G \vdvar ,H \vdvar$ to $c$.
      \item[If $\vvvfvar = \gforall{\vguard}{F'} \vdvar.\vvvfvar'$], then $F' = H$; let $D$ and $\vdat_*$ be as in the universal quantifier condition. We assign $G\vdvar,H\vdvar$ to $(\vguard,\vvdenv[\vdvar \mapsto \vdat_*],\vtrc)$ and $G\bar{\vdvar},H\vdvar$ to $(\vvvfvar',\vvdenv[\vdvar \mapsto \vdat_D],\vtrc)$, for every $\vdat_D \in D$.
      \item[In all other cases,] we assign $G,H$  to $c$.
  \end{description}

  By straightforward induction on the number of $\dep{}$-steps that we need to reach $a = (\vvvfvar,\vvdenv,\vvvtrc)$  from $(\vvfvar,\vdenv,\vvtrc)$, we can see that if $a$ is assigned $G,H$, then for every $x,y \in H$, if $x \in G$ and $\vvdenv(x) = \vvdenv(y)$, then $y \in G$.

  Since $(A,\dep{})$ is finite and loop-free, we can proceed by induction on the longest $\dep{}$-path from $a \in A$ to prove that
  for every $a = (\vvvfvar,\vvdenv,\vvvtrc)$ that is assigned with $G,H$,
  for every $J'\subseteq G \subseteq H$,
  if for every $x \in J'$ and $y \in H$, $y \in J'$ if and only if $\vvdenv(x) = \vvdenv(y)$, then
  $ (\vvvfvar,\vvdenv[J' \mapsto \vdat],\vvvtrc)$ has a finite guarded-branching annotation, which yields the statement of the lemma.
%
  The base case of $\vvvfvar = \tru$ is straightforward.
  For the inductive cases, we only need to treat the cases of diamonds and quantifiers:
  \begin{description}
    \item[if $\vvvfvar = \pos{\vbexp}{\vvvfvar'} $
    ,] then
    $\evalrel{\vbexp\vdenv[\star\mapsto\vact]}{\ltru}$ and $a \dep{} c = (\vvvfvar', \vvdenv, \vvvtrc) \in A$, where $\vvtrc = \vact\vvvtrc$. From our definition of our annotation decoration, $c$ is assigned with  $G,H$.
    By the inductive hypothesis,
    $(\vvvfvar', \vvdenv[J' \mapsto d], \vvvtrc)$ has a minimal, finite guarded-branching annotation $(A',\dep{}')$.
    If $ (\vvvfvar,\vvdenv[J' \mapsto \vdat],\vvvtrc) \in A'$, then we are done, and therefore we will assume that $ (\vvvfvar,\vvdenv[J' \mapsto \vdat],\vvvtrc) \notin A'$.


    Let $(B,\dep{}_{x})$ be such that
    \[ B = A' \cup \{ (\pos{\vbexp}{\vvfvar'}, \vdenv[J' \mapsto d], \vact\vtrc) \}
    ~~~~\text{ and}
    \]
    \[ \dep{}_{x} = \dep{}' \cup \{( (\pos{\vbexp}{\vvfvar'}, \vdenv[J' \mapsto d], \vact\vtrc) ~,~ (\vvfvar', \vdenv[J' \mapsto d], \vtrc)) \} .\]
    From our assumption that $(\vvfvar, \vdenv[J' \mapsto d], \vvtrc) \notin A'$, $\dep{}_{x}$ remains acyclic.
    To verify that $(B,\dep{}_{x})$ is a finite guarded-branching annotation, it suffices to verify the condition for $(\pos{\vbexp}{\vvfvar'}, \vdenv[J' \mapsto d], \vact\vtrc)$.
    Since $J' \subseteq G$ and $\vvvfvar \in \minhmldg[G,H]$,
    $\vbexp$ is of the form $\star \neq G \land \vbexp'(H)$, so $\vact \neq d,d'$, and by straightforward induction on $\vbexp'$, we can see that $\evalrel{\vbexp\vdenv[J' \mapsto d][\star\mapsto\vact]}{\ltru}$.
    Furthermore, by the definition of $\dep{}_{x}$,
    $(\pos{\vbexp}{\vvvfvar'}, \vdenv[J' \mapsto d], \vact\vtrc) \dep{}_{x} (\vvvfvar', \vdenv[J' \mapsto d], \vtrc))$, which concludes this case.


    \item[if $\vvvfvar = \exi{\binder{\vdvar}}{(\vdvar \neq V'
    \land \vfvar_1) \lor (
    x \sim V'
    \land \vfvar_2)}$],
    then $V' = G$ and
    $a \dep{} c$, where either $c = (\vfvar_1,\vvdenv[\vdvar \mapsto \vdat_e],\vvtrc)$ or $c = (\vfvar_2,\vvdenv[\vdvar \mapsto \vdat_e],\vvtrc)$.
    We now take the following cases:
    \begin{itemize}
      \item
    If 
    $\vdat_e = \vvdenv(\vvdvar)$ for some $\vvdvar \in J'\bar{\vdvar}$, then from $J'\bar{\vdvar} \subseteq V'\bar{\vdvar}$ and
    by the conditions of the guarded-branching annotations for the existential quantifier, $c = (\vfvar_2,\vvdenv[\vdvar \mapsto \vdat_e],\vvtrc)$, and we have assigned $G \vdvar ,H \vdvar$ to $c$.
    Therefore, we can use the inductive hypothesis for $J'' = J' \cup \{x\}$ to conclude that $(\vfvar_2,\vvdenv[J' \mapsto \vdat][\vdvar \mapsto \vdat],\vvtrc) = (\vfvar_2,\vvdenv[\vdvar \mapsto \vdat_e][J'' \mapsto \vdat],\vvtrc)$ has a finite guarded-branching annotation, and
    we can proceed to prove that
    so does $(\vvvfvar, \vvdenv[J' \mapsto d], \vvtrc)$, similarly to the diamond case.
    \item  If $J' = \{\vdvar\}$ and $\vdat_e = \vvdenv(\vdvar)$, then
    $\vdat_e \neq \vvdenv(\vvdvar)$ for every $\vvdvar \in V'\bar{\vdvar}$. Therefore,
    by the conditions of the guarded-branching annotations for the existential quantifier, $c = (\vfvar_1,\vvdenv[\vdvar \mapsto \vdat_e],\vvtrc)$, and we have assigned $G \bar{\vdvar} ,H \vdvar$ to $c$.
    Then, we see that
    $J'' = J' \setminus \{x\} = \emptyset$ 
    and
    $(\vfvar_1,\vvdenv[J' \mapsto \vdat][\vdvar \mapsto \vdat_e],\vvtrc) = (\vfvar_1,\vvdenv[\vdvar \mapsto \vdat_e],\vvtrc) = c \in A$,
    and therefore $(\vvvfvar, \vvdenv[J' \mapsto d], \vvtrc)$ has a finite guarded-branching annotation.
    \item If $\vdat_e \neq \vvdenv(\vvdvar)$ for every $\vvdvar \in J'$ and $\vdat_e = \vvdenv(\vvdvar)$ for some $\vvdvar \in V'\bar{\vdvar}$, then
    by the conditions of the guarded-branching annotations for the existential quantifier, $c = (\vfvar_2,\vvdenv[\vdvar \mapsto \vdat_e],\vvtrc)$, and we have assigned $G \vdvar ,H \vdvar$ to $c$.
    From
    the inductive hypothesis, for
    $J'' =   J' \setminus \{x\} $,
    $(\vfvar_2,\vvdenv[J' \mapsto \vdat][\vdvar \mapsto \vdat_e],\vvtrc) = (\vfvar_2,\vvdenv[\vdvar \mapsto \vdat_e][J'' \mapsto \vdat],\vvtrc)$ has a finite guarded-branching annotation, and
    we can proceed to prove that
    so does $(\vvvfvar, \vvdenv[J' \mapsto d], \vvtrc)$, similarly to the diamond case.
    \item If
    $\vdat_e \neq \vvdenv(\vvdvar)$ for every $\vvdvar \in V'\bar{\vdvar}$, then
    by the conditions of the guarded-branching annotations for the existential quantifier, $c = (\vfvar_1,\vvdenv[\vdvar \mapsto \vdat_e],\vvtrc)$, and we have assigned $G \bar{\vdvar} ,H \vdvar$ to $c$.
    Let $J'' = J' \setminus \{x\} \subseteq G \bar{\vdvar}$.
    We can use the inductive hypothesis
    to conclude that $(\vfvar_1,\vvdenv[J' \mapsto \vdat][\vdvar \mapsto \vdat_e],\vvtrc) = (\vfvar_1,\vvdenv[\vdvar \mapsto \vdat_e][J'' \mapsto \vdat],\vvtrc)$ has a finite guarded-branching annotation, and therefore so does $(\vvvfvar, \vvdenv[J' \mapsto d], \vvtrc)$.
    \end{itemize}
%
    \item[if $\vvvfvar = \gforall{\vguard}{F'} \vdvar.\vvvfvar'$], then $F' = H$;
    there is some finite $D \cup \{ \vdat_* \} \subseteq \dat$, such that $\vdat_* \notin D$, and:
    \begin{enumerate}
      \item for every $c \in D \cup \{ \vdat_* \}$,
    $(\vguard \lor \vfvar, \vdenv[x\mapsto  c ], \vvtrc ) \in A$ and $a \dep{} (\vguard \lor \vfvar, \vdenv[x\mapsto  c ], \vvtrc )$; 

      \item $ c  \in D$, for every $(\exi{\binder{\vdvar}}{\vguard},\vdenv,\vvtrc) \dep{}^* (\vguard',\vvdenv, c \vvvtrc)$;
    and
      \item $\{ \vdenv(y) \mid y \in F\bar{x} \cap \dvar\} \cup ( F \cap \dat) \subseteq D$.
    \end{enumerate}
    Due to the minimality of $(A,\dep{})$, we can assume that $D$ is also minimal, and therefore $d \notin D$.
    If $J' = \{ x \}$, then by
    the inductive hypothesis,  for every $c \in D \cup \{ \vdat_* \}$,
    $(\vguard \lor \vfvar, \vdenv[x\mapsto  c ], \vvtrc )$
    has a finite guarded-branching annotation $(A_c,\dep{}_{c})$.
    Let
    $(B,\dep{}_{x})$ be such that
    \[
      B = \{(\gforall{\vguard}{F} \vdvar.\vvfvar, \vdenv[J' \mapsto d], \vvtrc)\} \cup \bigcup_{c\in D \cup \{ \vdat_* \}} A_c
      ~~~~\text{ and }
    \]
    \[
      \dep{}_{x} = \bigcup_{c\in D \cup \{ \vdat_* \}} \dep{}_{c}  \cup ~
      \{
      (
        (\gforall{\vguard}{F} \vdvar.\vvfvar, \vdenv[J' \mapsto d], \vvtrc)
        ~,~
        (\vguard \lor \vfvar, [x\mapsto \vact(c) ], \vvtrc )
      )
      \} .
    \]
    It is straightforward to confirm that $(B,\dep{}_{x})$ is a guarded-branching annotation.

    If $J' \neq \{ x \}$, then let $y \in J'\bar{x}$.
    Observe that $\vdenv(y) \in D$, and therefore $\vdenv(y) \neq d_*$.
    For each $c\in D \cup \{ \vdat_* \}$,
    if $\vdenv(y) = c$, then let $J(c) = J'x$; and otherwise, let $J(c) = J'\bar{x}$.


    For each $c \in D \cup \{ \vdat_* \}$, let $\vact(c) = c$, if $c \neq d,\vdenv(y)$;
    $\vact(c) = d$, if $c = \vdenv(y)$; and $\vact(c) = d''$, if $c = d$ ($= d_*$), where
    $d'' \notin D \cup \{ \vdvar_*, d, \vdenv(y)\}$
    and
    for every $\vvtrc',\vvdenv,\vvvfvar$ it is not the case that
    $(\vvfvar,\vdenv,\vvtrc ) \dep{}^* (\vvvfvar,\vvdenv,d''\vvtrc')$.
    Similarly to the the other cases for
    $\vvvfvar$,
    we can use the inductive hypothesis to prove that for every $c \in D \cup \{ \vdat_* \}$,
    \[(\vguard \lor \vfvar, \vdenv[J' \mapsto d][x\mapsto  \vact(c) ], \vvtrc ) = (\vguard \lor \vfvar, \vdenv[x\mapsto  c ][\{x\}\mapsto  \vact(c) ][J(c) \mapsto d], \vvtrc )\]
    has a finite guarded-branching annotation $(A_c,\dep{}_{c})$.
    Let $(B,\dep{}_{x})$ be such that
    \[
      B = \{(\gforall{\vguard}{F} \vdvar.\vvfvar, \vdenv[J' \mapsto d], \vvtrc)\} \cup \bigcup_{c\in D \cup \{ \vdat_* \}} A_c
      ~~~~\text{ and }
    \]
    \[
      \dep{}_{x} = \bigcup_{c\in D \cup \{ \vdat_* \}} \dep{}_{c}  \cup ~
      \{
      (
        (\gforall{\vguard}{F} \vdvar.\vvfvar, \vdenv[J' \mapsto d], \vvtrc)
        ~,~
        (\vguard \lor \vfvar, \vdenv[J' \mapsto d][x\mapsto \vact(c) ], \vvtrc )
      )
      \}
    \]

    Let $D' = \{ \vact(c) \mid c \in D \}$.
    From $\vdenv(y) \neq \vdat_*$, we get that
    $\vact(\vdat_*) = \vdat_* \neq d, \vdenv(y), d''$, or $\vdat_* = d$ and $\vact(\vdat_*) = d'' \neq d, \vdenv(y)$.
    In the first case it is easy to see that
    $\vact(\vdat_*) \notin D$.
    In the second case, $d = \vdat_* \notin D$, and therefore $\vact(\vdat_*) = d'' \notin D'$.
    Therefore, we can conclude that
    $(B,\dep{}_{x})$ is a finite guarded-branching annotation for
    $(\gforall{\vguard}{F} \vdvar.\vvfvar, \vdenv[J' \mapsto d], \vvtrc)$.
%
\end{description}
  The above completes the inductive proof. To complete the proof of the lemma, notice that our construction above never introduces more triples in the annotation than what was already in $A$, and only affects the data environment $\vdenv$.
\end{proof}

\begin{lemma}\label{lemma:guarded-x-not-in-D}
    Let $\vvfvar \in \minhmldg[V,F]$ be a subformula of a closed formula $\vfvar$.
    Let $(\vvfvar, \vdenv, \vvtrc
    )
    $ have a finite guarded-branching annotation $(A,\dep{})$, $J \subseteq F$, and $d,d'\in \dat \setminus F$ be such that
    \begin{itemize}
        \item for every $\vvtrc',\vvdenv,\vvvfvar$ it is neither the case that
    $(
    \vvfvar,\vdenv,\vvtrc
        ) \dep{}^* (\vvvfvar,\vvdenv,\vdat\vvtrc')$
        nor
        $(
    \vvfvar,\vdenv,\vvtrc
        ) \dep{}^* (\vvvfvar,\vvdenv,\vdat'\vvtrc')$,
        \item for every $x \in J$ and $y \in F \setminus J$, $\vdenv(x) = d'$ and $\vdenv(y) \neq d, d'$. 
    \end{itemize}
        Then,
    $(\vvfvar, \vdenv[J \mapsto d],
     \vvtrc
    )$ has a finite guarded-branching annotation that has at most the same size as $(A,\dep{})$.
\end{lemma}


\begin{proof}
  We assume that $(A,\dep{})$ is a minimal finite guarded-branching annotation for $(\vvfvar, \vdenv, \vvtrc)$.
  Notice that if $(\vvfvar, \vdenv[J \mapsto d], \vvtrc) \in A$ or $d = d'$, then
  $(\vvfvar, \vdenv[J \mapsto d], \vvtrc)$ has a finite guarded-branching annotation and the proof is complete. Therefore, we will assume that $(\vvfvar, \vdenv[J \mapsto d], \vvtrc) \notin A$ and $d \neq d'$.
  The proof is by induction on the size of $(A,\dep{})$.

  For the base case, if $(\vvfvar, \vdenv, \vvtrc)$ is the only reachable triple from $(\vvfvar, \vdenv, \vvtrc)$, then $\vvfvar = \tru$ and the lemma follows.
  Otherwise, we take cases for $\vvfvar$:
  \begin{description}
    \item[if $(\vvfvar, \vdenv, \vvtrc) = (\pos{\vbexp}{\vvfvar'}, \vdenv, \vact\vtrc) \dep{} (\vvfvar', \vdenv, \vtrc) \in A$,] then, by the inductive hypothesis on the restriction of $(A,\dep{})$ on the triples reachable from
    $(\vvfvar', \vdenv, \vtrc)$, $(\vvfvar', \vdenv[J \mapsto d], \vtrc)$ has a
    minimal finite guarded-branching annotation $(A',\dep{}')$.
    Let $(B,\dep{}_{x})$ be such that
    \[ B = A' \cup \{ (\pos{\vbexp}{\vvfvar'}, \vdenv[J \mapsto d], \vact\vtrc) \}
    ~~~~\text{ and}
    \]
    \[ \dep{}_{x} = \dep{}' \cup \{( (\pos{\vbexp}{\vvfvar'}, \vdenv[J \mapsto d], \vact\vtrc) ~,~ (\vvfvar', \vdenv[J \mapsto d], \vtrc)) \} .\]
    From our assumption that $(\vvfvar, \vdenv[J \mapsto d], \vvtrc) \notin A$, $\dep{}_{x}$ remains acyclic.
    To verify that $(B,\dep{}_{x})$ is a finite guarded-branching annotation, it suffices to verify the condition for $(\pos{\vbexp}{\vvfvar'}, \vdenv[J \mapsto d], \vact\vtrc)$.
    From $\vvfvar \in \minhmldg[V,F]$, we get that $\vbexp = \vbexp'(F) \land \bigwedge_{y \in V} y \neq \star$, where $\vbexp'$ uses only the variables in $F$.
    From the lemma's assumptions, $d \neq \vact$, and
    since $(A,\dep{})$ is an annotation,
    $\evalrel{\vbexp\vdenv[\star\mapsto\vact]}{\ltru}$,
    and therefore
    $\evalrel{\big(\bigwedge_{y \in V} y \neq \star\big)\vdenv[\star\mapsto\vact]}{\ltru}$.
    This, together with the lemma's assumptions, yield that
    $\vdenv(x) \neq \vact$, $d \neq \vact$, and $d \neq \vdenv(y) \neq \vdenv(x)$ for every $x \in J$ and $y \in F \setminus J$.
    By straightforward induction on $\vbexp'$, we conclude that
    $\evalrel{\vbexp'\vdenv[J \mapsto d][\star\mapsto\vact]}{\ltru}$, and therefore
    $\evalrel{\vbexp\vdenv[J \mapsto d][\star\mapsto\vact]}{\ltru}$.
    Furthermore, by the definition of $\dep{}_{x}$,
    $(\pos{\vbexp}{\vvfvar'}, \vdenv[J \mapsto d], \vact\vtrc) \dep{}_{x} (\vvfvar', \vdenv[J \mapsto d], \vtrc))$, which concludes this case.
    \item[if
     $(\vvfvar, \vdenv, \vvtrc) = (\exi{\binder{\vdvar}}{\vvfvar'}, \vdenv, \vvtrc ) \dep{} (\vvfvar', \vdenv {[ x \mapsto \vact ]} , \vvtrc)$
    for some
    $\vact \in \dat$,
    ] then
    we distinguish the following cases:
    $\vact = d$;
    $\vact = d'$;
    or
    $\vact \neq d, d'$.

    \emph{For the case of $\vact \neq d,d'$, }
    $(\vvfvar', \vdenv[J \mapsto d][x \mapsto \vact], \vvtrc) = (\vvfvar', \vdenv[x \mapsto \vact][J\bar{x} \mapsto d], \vvtrc)$, and
    by the inductive hypothesis on the restriction of $(A,\dep{})$ on
    the triples reachable from
    $(\vvfvar', \vdenv[x \mapsto \vact], \vvtrc)$,
    $(\vvfvar', \vdenv[x \mapsto \vact][J\bar{x} \mapsto d], \vvtrc)$ has a
    minimal finite guarded-branching annotation $(A',\dep{}')$; let $\vact' = \vact$.

    \emph{For the case of $\vact = d$, }
    let $J' = \{ y \in F \mid \vdenv(y) = d \} \cup \{x\}$.
    By the inductive hypothesis on the restriction of $(A,\dep{})$ on
    the triples reachable from
    $(\vvfvar', \vdenv[x \mapsto d], \vvtrc)$, $(\vvfvar', \vdenv[x \mapsto d][J' \mapsto d''], \vvtrc)$ has a
    minimal finite guarded-branching annotation, where $d'' \neq d, d'$ is such that
    for every $y \in F$, $\vdenv(y) \neq d''$, and
    for every $\vvtrc',\vvdenv,\vvvfvar$ it is not the case that
    $(\vvfvar,\vdenv,\vvtrc ) \dep{}^* (\vvvfvar,\vvdenv,d''\vvtrc')$.
    We can now
    proceed as with the case of $\vact \neq d,d'$
    to prove that $
    (\vvfvar', \vdenv[J \mapsto d][x \mapsto d''], \vvtrc) =
    (\vvfvar', \vdenv[x \mapsto d''][J\bar{x} \mapsto d], \vvtrc)$ has a
    minimal finite guarded-branching annotation $(A',\dep{}')$; let $\vact' = d''$.

    \emph{For the case of $\vact = d'$, }
    $(\vvfvar', \vdenv[J \mapsto d][x \mapsto d], \vvtrc) = (\vvfvar', \vdenv[J x \mapsto d], \vvtrc)$, and
    by the inductive hypothesis on the restriction of $(A,\dep{})$ on
    the triples reachable from
    $(\vvfvar', \vdenv[x \mapsto d'], \vvtrc)$, $(\vvfvar', \vdenv[J x \mapsto d], \vvtrc) = (\vvfvar', \vdenv[J \mapsto d][x \mapsto d], \vvtrc) = (\vvfvar', \vdenv[x \mapsto d'][J x \mapsto d], \vvtrc) $ has a
    minimal finite guarded-branching annotation $(A',\dep{}')$; let $\vact' = d$.

    For all three cases, let $(B,\dep{}_{x})$ be such that
      \[
      B = A \cup \{ (\exi{\binder{\vdvar}}{\vvfvar'}, \vdenv[J\mapsto d], \vvtrc) \} ~~~\text{ and}
      \]
      \[\dep{}_{x} = \dep{}' \cup \{((\exi{\binder{\vdvar}}{\vvfvar'}, \vdenv[J\mapsto d], \vvtrc) ~,~
      (\vvfvar', \vdenv[J \mapsto d][\vdvar \mapsto \vact'], \vvtrc)
      )\}.
      \]
      From our assumption that $(\vvfvar, \vdenv[J \mapsto d], \vvtrc) \notin A$, $\dep{}_{x}$ remains acyclic.
      It is now clear that $(B,\dep{}_{x})$ is  a finite guarded-branching annotation for $(\exi{\binder{\vdvar}}{\vvfvar'}, \vdenv[J\mapsto d], \vvtrc)$.
    \item[if $(\vvfvar, \vdenv, \vvtrc) = (\gforall{\vguard}{F} \vdvar.\vvfvar, \vdenv, \vvtrc)$,] then we know that
    there is some finite $D \cup \{ \vdat_* \} \subseteq \dat$, such that $\vdat_* \notin D$, and:
    \begin{enumerate}

      \item for every $c \in D \cup \{ \vdat_* \}$,
    $(\vguard \lor \vfvar, \vdenv[x\mapsto  c ], \vvtrc ) \in A$ and $\alpha \dep{} (\vguard \lor \vfvar, \vdenv[x\mapsto  c ], \vvtrc )$; 

      \item $ c  \in D$, for every $(\exi{\binder{\vdvar}}{\vguard},\vdenv,\vvtrc) \dep{}^* (\vvfvar,\vvdenv, c \vvvtrc)$;
    and
      \item $\{ \vdenv(x) \mid x \in F \cap \dvar\} \cup ( F \cap \dat) \subseteq D$.
    \end{enumerate}
    Due to the minimality of $(A,\dep{})$, we can assume that $D$ is also minimal, and therefore $d \notin D$.

    For each $c \in D \cup \{ \vdat_* \}$, let
    \[
      \vact(c) =
      \begin{cases}
        c, &\text{ if } c \neq d,d' ;\\
        d, &\text{ if } c = d' ; \text{ and} \\
        d'', &\text{ if } c = d,
      \end{cases}
    \]
    where
    $d'' \notin D \cup \{ \vdvar_*, d, d'\}$
    and
    for every $\vvtrc',\vvdenv,\vvvfvar$ it is not the case that
    $(\vvfvar,\vdenv,\vvtrc ) \dep{}^* (\vvvfvar,\vvdenv,d''\vvtrc')$.
    Similarly to the cases for $(\vvfvar, \vdenv, \vvtrc) = (\exi{\binder{\vdvar}}{\vvfvar'}, \vdenv, \vvtrc )$,
    we can use the inductive hypothesis to prove that for every $c \in D \cup \{ \vdat_* \}$,
    $(\vguard \lor \vfvar, \vdenv[J \mapsto d][x\mapsto  \vact(c) ], \vvtrc )$
    has a finite guarded-branching annotation $(A_c,\dep{}_{c})$.
    Let $(B,\dep{}_{x})$ be such that
    \[
      B = \{(\gforall{\vguard}{F} \vdvar.\vvfvar, \vdenv[J \mapsto d], \vvtrc)\} \cup \bigcup_{c\in D \cup \{ \vdat_* \}} A_c
      ~~~~\text{ and }
    \]
    \[
      \dep{}_{x} = \bigcup_{c\in D \cup \{ \vdat_* \}} \dep{}_{c}  \cup ~
      \{
      (
        (\gforall{\vguard}{F} \vdvar.\vvfvar, \vdenv[J \mapsto d], \vvtrc)
        ~,~
        (\vguard \lor \vfvar, \vdenv[J \mapsto d][x\mapsto \vact(c) ], \vvtrc )
      )
      \}
    \]

    Let $D' = \{ \vact(c) \mid c \in D \}$.
    Notice that since $d' \in \{ \vdenv(x) \mid x \in F\} \subseteq D$, $\vdat_* \neq d'$, and therefore $\vact(\vdat_*) = \vdat_* \neq d, d', d''$, or $\vdat_* = d$ and $\vact(\vdat_*) = d'' \neq d, d'$.
    In the first case it is easy to see that
    $\vact(\vdat_*) \notin D$.
    In the second case, $d = \vdat_* \notin D$, and therefore $\vact(\vdat_*) = d'' \notin D'$.
    Therefore, we can conclude that
    $(B,\dep{}_{x})$ is a finite guarded-branching annotation for
    $(\gforall{\vguard}{F} \vdvar.\vvfvar, \vdenv[J \mapsto d], \vvtrc)$.

    \item[The remaining cases] are straightforward, and the induction is complete.
\end{description}
To complete the proof of the lemma, notice that our construction above never introduces more triples in the annotation than what was already in $A$, and only affects the data environment $\vdenv$.
%
\end{proof}

\begin{corollary}\label{cor:GBA-gamma-inf-values}
  Let $(A\dep{})$ be a guarded-branching annotation,
$$a = (\gforall{\vguard}{F} \vdvar.\vfvar, \vdenv, \vtrc) \in A,$$ and
let some finite $D \cup \{ \vdat_* \} \subseteq \dat$ be
as in the conditions for the universal quantifiers for guarded-branching annotations.
  Then, $(\vguard, \vdenv[\vdvar \mapsto \vvdat], \vtrc)$ has a finite guarded-branching annotation for every $\vvdat \notin D$, which has at most the same size as the sub-annotation of $(A,\dep{})$ on $(\vguard, \vdenv[\vdvar \mapsto \vdat_*], \vtrc)$.
\end{corollary}

We are now ready to prove \Cref{prop:guarded-annotation-semantics_pf}.

\begin{proof}[Proof of \Cref{prop:guarded-annotation-semantics_pf}]
  The equivalence of statement 1 with statement 4 was established by \Cref{prop:annotation-semantics}. Statement 3 trivially implies statement 2; and the implication from statement 2 to 3 results, similarly to \Cref{prop:annotation-chmld}, from observing that any minimal guarded-branching annotation is finitely-branching.

  It suffices to prove that statement 1 implies statement 2, and statement 3 implies statement 1,
  for every closed $\vfvar \in \minhmld$, $\vdenv \in \denv$, and $\vtrc \in \trc$.

    If $(\vfvar, \vdenv,\vtrc)$ has a finite guarded-branching annotation $(A,\dep{})$, then it is straightforward to extend $(A,\dep{})$ to an annotation for $(\vfvar, \vdenv,\vtrc)$, using \Cref{cor:GBA-gamma-inf-values} and recursion on the longest $\dep{}$-path from $(\vfvar, \vdenv,\vtrc)$.

    On the other hand,
    we can also construct a guarded-branching annotation from an annotation $(A,\dep{})$ for $(\vfvar, \vdenv,\vtrc)$.
    Let $a \dep{}' c$ if and only if $a \dep{} c$ and if $a = (\gforall{\vguard}{F} \vdvar.\vfvar, \vvdenv, \vvtrc)$, then $c = (\exists x.(x \neq F \land \vguard), \vvdenv, \vvtrc)$.
    Observe that $\dep{}'$ is finitely branching and every branch of $\dep{}'$ is finite. Therefore, $\dep{}'$ is finite.
    It is now straightforward to construct a guarded-branching annotation from $(A,\dep{})$, by induction on the number of triples reachable by $\dep{}'$. The interesting case is that of
    $a = (\gforall{\vguard}{F} \vdvar.\vvfvar, \vvdenv, \vvtrc)$, where we can define $D = \{ d \in \dat \mid d \in F, \text{ or } \exists y\in F. \vvdenv(y)=d,  , \text{ or } a {\dep{}'}^* (\vvvfvar,\vvvdenv,d\vvvtrc) \text{ for some }(\vvvfvar,\vvvdenv,d\vvvtrc) \in A \}$ and $d_* \in \dat \setminus D$. Then, by the inductive hypothesis, there is some $d_*' \in D$, such that for every $y \in F \cap \dvar$ and $c \in F \cap \dat$, $\vvdenv(y) \neq d_*'$ and $d_*' \neq c$, and $(\vguard, \vvdenv[x \mapsto d_*'], \vvtrc)$ has a finite guarded-branching annotation. By \Cref{lem:GBA-of-in-V-any-will-do}, $(\vguard, \vvdenv[x \mapsto d_*], \vvtrc)$ has a finite guarded-branching annotation.
    Also by the inductive hypothesis, for every $d\in D$,
    $(\vvfvar, \vvdenv[x \mapsto d], \vvtrc)$ has a finite guarded-branching annotation or $(\vguard, \vvdenv[x \mapsto d], \vvtrc)$ has a finite guarded-branching annotation.
    Therefore, we can construct a finite guarded-branching annotation for $a$, completing the inductive argument.
\end{proof}

We now proceed to prove \Cref{thm:opt-eff-mon,cor:maximality-of-guarded-in-min}

We can straightforwardly lift a function $f : \data \to \data$ to traces, by defining $f(\vact_1\vact_2\cdots) = f(\vact_1)f(\vact_2)\cdots$.

\begin{lemma}\label{cor:replace-data-all-ok}
  Let $P$ be an infinite subset of $\data$,  $f: P \to \data$ be one-to-one and onto [and preserves constants], and $\vfvar \in \minhmldg$.
  Then, for every $t \in P^\omega \cap \sem{\vfvar}$,
  $f(t) \in \sem{\vfvar}$.
\end{lemma}

\begin{proof}
  Let $t \in P^\omega \cap \sem{\vfvar}$ and $C = \dat \setminus P$.
  \Cref{prop:guarded-annotation-semantics_pf} yields that $(\vfvar,\delta_0,t)$ has a finite guarded-branching annotation, where $\delta_0$ is the empty assignment. Therefore, it suffices to prove that for every subformula $\vvfvar$ of $\vfvar$, $\vvtrc$, and $\vdenv$, if $(\vvfvar,\vdenv,\vvtrc)$ has a finite guarded-branching annotation and for every $x$ in the domain of $\vdenv$, $\vdenv(x) \in P$, then
  $(\vvfvar,\vvdenv,f(\vvtrc))$ has a finite guarded-branching annotation,
  where
  $\vvdenv = f \circ \vdenv$.

  The proof of the statement is by induction on the longest branch of the guarded-branching annotation, taking cases for $\vvfvar$.
  The only interesting cases are those of the quantified formulas $\vvfvar = \forall x.\vvfvar'$ or $\vvfvar = \exists x.\vvfvar'$, where
  $(\vvfvar,\vvdenv,\vvtrc) \dep{} (\vvvfvar,\vvvdenv,\vvvtrc)$ and $\vvvdenv(x) \in C$.
  Then, $\vvdenv(x) \neq \vvdenv(y)$ for every $y \neq x$, and therefore we can use \Cref{lemma:guarded-x-not-in-D} to reduce to the case where $\vvvdenv(x) \in P$.
\end{proof}


\begin{proposition}\label{propo:guard-is-consequence}
  For every $\vfvar \in \minhmld$, $\sem{\guard(\vfvar)} \subseteq \sem{\vfvar}$.
\end{proposition}

\begin{proof}
  It suffices to prove that
  if
  $(\guard(\vfvar),\vdenv,\vtrc)$ has an annotation, then so does $(\vfvar,\vdenv,\vtrc)$.
  Let $(A,\dep{})$ be an annotation for $(\guard(\vfvar),\vdenv,\vtrc)$.
  Let $A' = \{ (\vvfvar,\vvdenv,\vvtrc) \mid (\guard(\vvfvar,V,F,\Pi),\vvdenv,\vvtrc) \in A \text{ for some } V,F,\Pi \}$, and
  $(\vvfvar_1,\vvdenv_1,\vvtrc_1)\dep{}'(\vvfvar_2,\vvdenv_2,\vvtrc_2)$ if and only if
  \begin{itemize}
    \item
  $(\guard(\vvfvar_1,V_1,F_1,\Pi_1),\vvdenv_1,\vvtrc_1)\dep{}^+ (\guard(\vvfvar_2,V_2,F_2,\Pi_2),\vvdenv_2,\vvtrc_2)$ for some $V_1,V_2,F_1,F_2,\Pi_1,\Pi_2$, or
  \item
  $\vvfvar_1 = \vpvar$, $\vvfvar_2 = \vfvar_\vpvar$, $\vvdenv_1 = \vvdenv_2$, and $\vvtrc_1 = \vvtrc_2$.
  \end{itemize}
  It is now straightforward to confirm that $(A',\dep{}')$ is an annotation by verifying each annotation condition.
\end{proof}

\begin{proposition}\label{prop:guard-preserves-good-prefixes}
  For every $\vfvar \in \minhmld$ and $\vftrc \in \data^*$, if $\vftrc \cdot \data^\omega \subseteq \sem{\vfvar}$, then $\vftrc \cdot \data^\omega \subseteq \sem{\guard(\vfvar)}$.
\end{proposition}

\begin{proof}
    Let $\vftrc \cdot \data^\omega \subseteq \sem{\vfvar}$ and let $\vftrc \prec \vtrc$. We prove that $\vtrc \in \sem{\guard(\vfvar)}$.
    Let $P \cup C = \data$, where $P \cap C = \emptyset$, $P,C$ are infinite, and $P_0 \subseteq P$, where $P_0$ is the (finite) set of data values in $\vftrc$.
    Let $f:P \xrightarrow[onto]{1-1} \data$, such that for every $\vdat \in P_0$, $f(\vdat) = \vdat$.
    Let $\vvtrc = f^{-1}(\vtrc) \in \vftrc\cdot P^\omega$;  $\vvtrc$ is an extension of $\vftrc$ that only uses data in $P$.
    By \Cref{cor:replace-data-all-ok}, it suffices to prove that $\vvtrc \in \sem{\guard(\vfvar)}$.
    From our assumptions, $(\vvtrc \in \sem{\vfvar})$, and therefore, by \Cref{prop:annotation-semantics}, $(\vfvar,\vdvar_0,\vvtrc)$ has an annotation $(A,\dep{})$, which we assume is minimal.
    By \Cref{prop:guarded-annotation-semantics_pf}, it suffices to prove that $(\guard(\vfvar),\vdvar_0,\vvtrc)$ has a guarded-branching annotation.

    We note that in the context of formula $\guard(\vfvar)$, each variable $X_{V,F}$ is in the scope of a unique $\min X_{V,F}.\vfvar_{X_{V,F}} = \guard(\vfvar_X,V,F,\Pi)$ for some $\Pi$, which we will denote as $\Pi_{X_{V,F}}$.


    For each $a = (\vvfvar,\vdenv,\vvvtrc) \in A$ and every finite $V \subseteq F \subseteq Vr(\vfvar)$ and $\Pi \subseteq (2^{Vr(\vfvar)})^2$, such that $\vdenv[V] \subseteq C$, we define a finite $A_a(V,F,\Pi),\dep{}_a(V,F,\Pi)$ that satisfies all conditions for the guarded-branching annotations, and we do so by induction on the
    $\dep{}$-branches from $a$.
    Let $a = (\vvfvar,\vdenv,\vvvtrc)$, finite $V \subseteq F \subseteq Vr(\vfvar)$ and $\Pi \subseteq (2^{Vr(\vfvar)})^2$, such that $\vdenv[V] \subseteq C$.
%
    \begin{description}
      \item[If $\vvfvar = \tru$,] then $A_a(V,F,\Pi) = \{ a \}$ and $\dep{}_a(V,F,\Pi) = \emptyset $.
      \item[If $\vvfvar = \vrvar$ and $(V,F)\in \Pi$,] then $a \dep{} (\vfvar_X,\vdenv,\vvvtrc) \in A$. Let
      \[A_a(V,F,\Pi) = \{ (X_{V,F},\vdenv,\vvvtrc) \} \cup A_{(\vfvar_X,\vdenv,\vvvtrc)}(V,F,\Pi_{X_{V,F}}), \text{ and}\]
      \[\dep{}_a(V,F,\Pi) = \{ ((X_{V,F},\vdenv,\vvvtrc) , (\vfvar_{X_{V,F}},\vdenv,\vvvtrc) ) \} \cup \dep{}_{(\vfvar_X,\vdenv,\vvvtrc)}(V,F,\Pi_{X_{V,F}}) .\]
    \item[If $\vvfvar = \vrvar$ and $(V,F)\notin \Pi$,] then $a \dep{} (\vfvar_X,\vdenv,\vvvtrc) \in A$. Let
      \[A_a(V,F,\Pi) = \{ (\fx{X_{V,F}},\vdenv,\vvvtrc) \} \cup A_{(\vfvar_X,\vdenv,\vvvtrc)}(V,F,\Pi_{X_{V,F}}), \text{ and}\]
      \[\dep{}_a(V,F,\Pi) = \{ ((\fx{X_{V,F}},\vdenv,\vvvtrc) , (\vfvar_{X_{V,F}},\vdenv,\vvvtrc) ) \} \cup \dep{}_{(\vfvar_X,\vdenv,\vvvtrc)}(V,F,\Pi_{X_{V,F}}) .\]
    \item[If $\vvfvar = \min \vrvar.\vfvar_\vrvar$,] then $a \dep{} (\vfvar_X,\vdenv,\vvvtrc) \in A$. Let
      \[A_a(V,F,\Pi) = \{ (\fx{X_{V,F}},\vdenv,\vvvtrc) \} \cup A_{(\vfvar_X,\vdenv,\vvvtrc)}(V,F,\Pi \cup \{ (V,F)\}), \text{ and}\]
      \[\dep{}_a(V,F,\Pi) = \{ ((\fx{X_{V,F}},\vdenv,\vvvtrc) , (\vfvar_{X_{V,F}},\vdenv,\vvvtrc) ) \} \cup \dep{}_{(\vfvar_X,\vdenv,\vvvtrc)}(V,F,\Pi \cup \{ (V,F)\}) .\]
    \item[If $\vvfvar = \forall \vdvar.\vvvfvar$,] then
    let
    $c \in C$, such that $c \neq \vdenv(y)$ for every $y$.
    We know that
    $a \dep{} (\vvvfvar,\vdenv[\vdvar \mapsto c],\vvvtrc) \in A$.
    Let $D = \{ \vdat \in P \mid \exists (\vvvfvar,\vvdenv,\vdat\vvvtrc') \in A_{(\vvvfvar,\vdenv[\vdvar \mapsto c],\vvvtrc)}(V\vdvar,F\vdvar,\Pi) \}$ and let for each $\vdat \in D \cup \{c\}$, $a_\vdat = (\vvvfvar,\vdenv[\vdvar \mapsto \vdat],\vvvtrc)$. Then, we define
    \begin{align*}
      A_a(V,F,\Pi) = \{ (\guard(\vvfvar,V,F,\Pi),\vdenv,\vvvtrc) \} \cup A_{a_c}(V\vdvar,F\vdvar,\Pi) \cup \bigcup_{\vdat \in D} A_{a_\vdat}(V\bar{\vdvar},F\vdvar,\Pi), \text{ and}
    \end{align*}
    \begin{align*}
      \dep{}_a(V,F,\Pi) = & \{ ((\guard(\vvfvar,V,F,\Pi),\vdenv,\vvvtrc)~,~(\guard(\vvvfvar,V,F,\Pi),\vdenv[\vdvar \mapsto \vdat],\vvvtrc) ) \mid \vdat \in D \cup\{ c \} \} \\ & \quad
      \cup
      \dep{}_{a_c}(V\vdvar,F\vdvar,\Pi) \cup
      \bigcup_{\vdat \in D} A_{a_\vdat}(V\bar{\vdvar},F\vdvar,\Pi).
    \end{align*}
    \item[If $\vvfvar = \exists \vdvar.\vvvfvar$,] then
    let $\vdat \in \data$, such that
    $a \dep{} a_\vdat = (\vvvfvar,\vdenv[\vdvar \mapsto \vdat],\vvvtrc) \in A$.
    If $\vdat = \vdenv(y)$ for some $y \in V$, then
    we define
    \begin{align*}
      A_a(V,F,\Pi) = \{ (\guard(\vvfvar,V,F,\Pi),\vdenv,\vvvtrc) \} \cup A_{a_\vdat}(V\vdvar,F\vdvar,\Pi), \text{ and}
    \end{align*}
    \begin{align*}
      \dep{}_a(V,F,\Pi) = &\{ ((\guard(\vvfvar,V,F,\Pi),\vdenv,\vvvtrc)~,~(\guard(\vvvfvar,V\vdvar,F\vdvar,\Pi),\vdenv[\vdvar\mapsto \vdat],\vvvtrc) ) \}
      \cup
      \dep{}_{a_\vdat}(V\vdvar,F\vdvar,\Pi).
    \end{align*}
    On the other hand, if If $\vdat \neq \vdenv(y)$ for every $y \in V$, then
    we define
    \begin{align*}
      A_a(V,F,\Pi) = \{ (\guard(\vvfvar,V,F,\Pi),\vdenv,\vvvtrc) \} \cup A_{a_\vdat}(V\bar{\vdvar},F\vdvar,\Pi), \text{ and}
    \end{align*}
    \begin{align*}
      \dep{}_a(V,F,\Pi) = & \{ ((\guard(\vvfvar,V,F,\Pi),\vdenv,\vvvtrc)~,~(\guard(\vvvfvar,V\bar{\vdvar},F\vdvar,\Pi),\vdenv[\vdvar\mapsto \vdat],\vvvtrc) ) \} \\ & \quad
      \cup
      \dep{}_{a_\vdat}(V\bar{\vdvar},F\vdvar,\Pi).
    \end{align*}
    \item[If $\vvfvar = \pos{\vbexp}{\vvvfvar}$,] then $\evalrel{\vbexp\vdenv}{\ltru} $, and there exists some
      $a \dep{} (\vvvfvar,\vdenv,\vvvtrc')$. Let
      \begin{align*}
      A_a(V,F,\Pi) = \{ (\guard(\vvfvar,V,F,\Pi),\vdenv,\vvvtrc) \} \cup A_{(\vvvfvar,\vdenv,\vvvtrc')}(V,F,\Pi), \text{ and}
    \end{align*}
    \begin{align*}
      \dep{}_a(V,F,\Pi) = \{ ((\guard(\vvfvar,V,F,\Pi),\vdenv,\vvvtrc)~,~(\guard(\vvvfvar,V,F,\Pi),\vdenv,\vvvtrc') ) \}
      \cup
      \dep{}_{(\vvvfvar,\vdenv,\vvvtrc')}(V,F,\Pi).
    \end{align*}
    We note that our condition that $\vdenv[V] \subseteq C$ yields $\evalrel{(* \neq V)\vdenv}{\ltru} $, and therefore $\evalrel{(* \neq V) \land \vbexp\vdenv}{\ltru} $.
    \item[If  $\vvfvar = \vvfvar_1 \land \vvfvar_2$,] then
    $a \dep{} a_1 = (\vvfvar_1,\vdenv,\vvvtrc)$ and
    $a \dep{} a_2 = (\vvfvar_2,\vdenv,\vvvtrc)$. Let
    \begin{align*}
      A_a(V,F,\Pi) = \{ (\guard(\vvfvar,V,F,\Pi),\vdenv,\vvvtrc) \} \cup A_{a_1} \cup A_{a_2} , \text{ and}
    \end{align*}
    \begin{align*}
      \dep{}_a(V,F,\Pi) = & \left\{
        \begin{Array}{l}
        \Big((\guard(\vvfvar,V,F,\Pi),\vdenv,\vvvtrc)~,~(\guard(\vvfvar_1,V,F,\Pi),\vdenv,\vvvtrc) \Big), \\
        \Big((\guard(\vvfvar,V,F,\Pi),\vdenv,\vvvtrc)~,~(\guard(\vvfvar_2,V,F,\Pi),\vdenv,\vvvtrc) \Big)
        \end{Array}
        \right\}
      \\ & \quad
      \cup
      \dep{}_{a_1}(V,F,\Pi) \cup
      \dep{}_{a_2}(V,F,\Pi).
    \end{align*}
    \item[the case of $\vvfvar = \vvfvar_1 \lor \vvfvar_2$] is similar to the previous one.
    \end{description}
    It is now straightforward to see that $A_{(\vfvar,\vdenv_0,\vvtrc)}(\emptyset,\emptyset,\emptyset)$ is a guarded-branching annotation for $(\guard(\vfvar),\vdenv_0,\vvtrc)$.
\end{proof}


We can now prove \Cref{thm:opt-eff-mon}, the first part of the main result of this subsection:

\begin{proposition}\label{thm:opt-eff-mon}
    Every formula $\vfvar \in \minhmld$ is optimally effectively monitorable.
\end{proposition}

\begin{proof}
  By \Cref{prop:guard-preserves-good-prefixes,propo:guard-is-consequence}, it suffices to see that $\guard(\vfvar)$ is  effectively monitorable for satisfactions, which results from \Cref{cor:min-effective-mon}.
\end{proof}

\begin{corollary}
   Let $\vfvar \in \minhmld$. If $\vfvar$ is monitorable for satisfactions, then
   $\sem{\vfvar} = \sem{\guard(\vfvar)}$.
\end{corollary}

\Cref{cor:maximality-of-guarded-in-min} is then a direct result.

\begin{proof}[Proof of \Cref{lem:2formulae-for-halting}]

  For the first statement of the lemma, notice that in the proof of \Cref{thm:validity_disjhmld}, presented in \Cref{app:proof_thm_validity_disjhmld},
  if we omit $\psi_{\bar{h}}$ from the definition of $\varphi_{M}$,
  the resulting formula is satisfied exactly when the trace encodes the run of $M$ (on 0
  ).

  For the second statement of the lemma, again, notice that in the proof of \Cref{thm:validity_disjhmld}, presented in \Cref{app:proof_thm_validity_disjhmld},
  we can replace in $\varphi_M$, subformula $\psi_6$, which also ensures that the first configuration only uses one tape position, by the formula
  $[x^\#]\recmin{X}{[\star \neq x^\#]X } \land \psi'_6$, where
  \begin{align*}
    \psi'_6 = [x^\#]
    \forall x.
      \recmax{Y}{[\star \neq x ]
    Y \land [x^\#][\_]
                   &\recmax{Z}{[\star \neq x ][\_] Z \land \\
    &[x][\_][\star \neq x^\#]\fls}
      }
      ,
    \end{align*}
  we still ensure that the second block is finite and thus encodes a configuration, but we allow that configuration to be any initial configuration.
  Let's call the resulting formula $\varphi_{M}^i$. Then,
  $\varphi_{M}^i$ is satisfied by the  traces that encode nonterminating runs of $M$ on any input.
  As we argued for the proof of the first statement, by removing the
  $\psi_{\bar{h}}$ formula from $\varphi_{M}^i$, the resulting formula $\varphi_{M}^r$ is satisfied by the  traces that encode terminating or nonterminating runs of $M$ on any input.
  Furthermore, let $\varphi_{M}^p = \varphi_{N}^r$, where $N$ is a Turing machine that halts on its initial configuration. Therefore, $\varphi_{M}^p$ expresses that the trace has a prefix that encodes a non-empty prefix of a run of $M$.
  We can then define $\vvfvar_M^{\neg{H}} = \varphi_{M}^p \land (\varphi_{M}^i \lor \neg \varphi_{M}^r)$, where $\neg \varphi_{M}^r$ is a \rechmld\ formula equivalent to the negation of $\varphi_{M}^r$.
\end{proof}

\subsection{Effective Monitorability}
\label{app:effective_monitorability}
\begin{definition}\label{def:effective-monitorability}
    A property $P \in 2^\trc$ is effectively monitorable for satisfactions (\resp\ violations) if there is a Turing machine such that for every $\vtrc \in \trc$, $\vtrc \in P$ (\resp\ $\vtrc \notin P$) if and only if there exists a finite prefix of $\vtrc$ that the Turing machine accepts.
\end{definition}


\end{document}